\newcommand{\cL}{{\cal L}}
\newcommand{\cO}{{\cal O}}
\newcommand{\nt}{\notag\\}

\newcommand{\cA}{{\cal A}}

\renewcommand{\a}{{\alpha}}

\newcommand{\be}{\begin{equation}}
\newcommand{\ee}{\end{equation}}
\newcommand{\bea}{\begin{eqnarray}}
\newcommand{\eaa}{\end{eqnarray}}

\newcommand{\cF}{{\cal F}}

\renewcommand{\a}{\alpha}

\newcommand{\cN}{{\cal N}}

\newcommand{\cM}{{\cal M}}
\newcommand{\tr}{{\rm tr}}

\newcommand{\p}[1]{(\ref{#1})}
\newcommand{\bt}[1]{{\bar t}}

\newcommand \vev [1] {\langle{#1}\rangle}

\newcommand\lr[1]{{\left({#1}\right)}}
\newcommand{\ft}[2]{{\textstyle\frac{#1}{#2}}}
\newcommand\re[1]{(\ref{#1})}
\def \qqquad {\qquad\quad}
\def \qqqquad {\qquad\qquad}

\documentclass[12pt]{article}
\usepackage{epsfig,amssymb,amsmath,psfrag,subfigure}

\catcode`\@=11
\def \tr {\mathop{\rm tr}\nolimits}

\def\numberbysection{\@addtoreset{equation}{section}
                     \def\theequation{\thesection.\arabic{equation}}}

\numberbysection
 
\begin{document}
  
\thispagestyle{empty}


\null\vskip-53pt \hfill
\begin{minipage}[t]{50mm}
CERN-PH-TH/2012-014 \\
DCPT-12/03 \\
HU-EP-12/03\\
HU-MATH-2012-27\\
IPhT--T12/005 \\
LAPTH-005/12 
\end{minipage}

\vskip1.1truecm
\begin{center}
\vskip 0.0truecm

 {\Large\bf
  Constructing the correlation function of four  stress-tensor \\[1mm] multiplets and the four-particle  amplitude in  $\cN=4$ SYM\\
  }
\vskip 1.0truecm

{\bf    Burkhard Eden$^{a}$, Paul Heslop$^{b}$, Gregory P. Korchemsky$^{c}$, 
Emery Sokatchev$^{d,e,f}$ \\
}

\vskip 0.4truecm
$^{a}$ {\it Institut f\"ur Mathematik, Humboldt-Universit\"at zu Berlin,
\\
Rudower Chaussee 25, Johann von Neumann-Haus, 12489 Berlin}
 \\
  \vskip .2truecm
$^{b}$ {\it  Mathematics Department, Durham University, 
Science Laboratories,
 \\
South Rd, Durham DH1 3LE,
United Kingdom \\
 \vskip .2truecm
$^{c}$ Institut de Physique Th\'eorique\,\footnote{Unit\'e de Recherche Associ\'ee au CNRS URA 2306},
CEA Saclay, \\
91191 Gif-sur-Yvette Cedex, France\\
\vskip .2truecm $^{d}$ Physics Department, Theory Unit, CERN,\\ CH -1211, Geneva 23, Switzerland \\
\vskip .2truecm $^{e}$ Institut Universitaire de France, \\103, bd Saint-Michel
F-75005 Paris, France \\
\vskip .2truecm $^{f}$ LAPTH\,\footnote[2]{Laboratoire d'Annecy-le-Vieux de Physique Th\'{e}orique, UMR 5108},   Universit\'{e} de Savoie, CNRS, \\
B.P. 110,  F-74941 Annecy-le-Vieux, France
                       } \\
\end{center}

\vskip .3truecm

\centerline{\bf Abstract} 
\medskip
\noindent

We present a construction of the integrand of the correlation function of four stress-tensor multiplets in {$\cN=4$}
SYM at weak coupling. It does not rely on Feynman diagrams and makes use of the recently
discovered symmetry of the integrand under permutations of external and integration points.
This symmetry holds for any gauge group, so it can be used
to predict the integrand   both in the planar and non-planar sectors.
We demonstrate the great efficiency of graph-theoretical tools in the systematic study of the possible
permutation symmetric integrands. We formulate a general
ansatz for the correlation function as a linear combination of all relevant graph topologies, with arbitrary coefficients.
Powerful restrictions on the coefficients come from the analysis of the logarithmic divergences of the correlation function in two singular regimes: Euclidean short-distance and Minkowski light-cone limits.
We demonstrate that the planar integrand is completely fixed by the procedure up to six loops and probably beyond. In the non-planar sector, we show the absence of non-planar corrections at three loops and we reduce the freedom at four loops to just four constants. Finally,  the correlation function/amplitude duality allows us to show the  complete agreement of our results with the four-particle planar amplitude in $\cN=4$ SYM.

\newpage

\thispagestyle{empty}

{\small \tableofcontents}

\newpage
\setcounter{page}{1}\setcounter{footnote}{0}



\section{Introduction and summary of the results}

In the present paper we continue the investigation of the four-point
correlation function of the stress-tensor multiplet in $\cN=4$ super-Yang-Mills (SYM) theory initiated in \cite{Eden:2011we}. This correlation function is a very interesting object, from various points of view. In the past it has been extensively studied in the context of the AdS/CFT correspondence,  both at strong coupling (in AdS supergravity) \cite{ADS} and at weak coupling (see, e.g., \cite{partialNonRen,Heslop:2002hp,Dolan:2001tt} and references therein). {The asymptotic behavior of the correlation function at short distances contains information about the anomalous dimensions of a large variety of Wilson operators. The spectrum of anomalous dimensions of the theory can alternatively be calculated by diagonalization of the dilatation operator 
which is believed to be integrable in planar $\cN=4$ SYM (for a recent review see~\cite{Beisert:2010jr}).} It is then natural to expect that the four-point
correlation function itself should have some non-trivial hidden structure.

More recently, renewed interest in this type of correlation function
arose from the observation that  they, if considered in  the planar
limit  and restricted to the light cone, are dual to scattering
amplitudes \cite{EKS2,EKS3,Eden:2011yp,Eden:2011ku,Adamo:2011dq}.  
The scattering amplitudes have a
 number of intriguing properties (see, e.g. \cite{ampreview} for a review).  
  Again, we should expect that the rich structure of the four-particle amplitude should be reflected in the properties of the (dual) four-point correlation function in $\cN=4$ SYM. 

Indeed, one unexpected property was discovered  in the recent paper \cite{Eden:2011we}. As was shown there, to any loop level in the weak coupling expansion
and for an arbitrary gauge group,
 the integrand of the correlation function has full $S_{4+\ell}$ permutation symmetry with respect to {the exchange of} the four external points and the $\ell$ integration points.  This property, combined with the (super)conformal symmetry of  the correlation function, puts strong restrictions on the structure of  its integrand. Using the
permutation symmetry and some input from the duality with the amplitude,  the precise form of the planar correlation function in terms of a set of conformal four-point integrals was established at three and four loops. This was done without even mentioning Feynman diagrams, whose number and complexity render a conventional perturbative calculation at this level very hard.  

We would like to emphasize that the power of the new permutation symmetry is that it holds regardless of the 
rank of the gauge group and, therefore, it can be used to predict the integrand of the correlation function 
both in the planar and non-planar sectors. In the present paper we show this explicitly by constructing 
the four-point correlation function in the planar limit up to six loops and in the non-planar sector up to 
four loops.

We demonstrate the great efficiency of the graph-theoretical tools \cite{nauty,Sage} in the systematic study of the possible permutation symmetric integrands,  including both planar and non-planar topologies. They
allow us to identify all relevant graph topologies and to formulate a general ansatz for the correlation function
as a linear combination of all such topologies, with arbitrary coefficients. Further powerful restrictions  come from the analysis  of the expected logarithmic divergences of the correlation function in various singular limits. 

We consider two such limits: the Euclidean double short-distance limit in which the four points coincide pairwise, and the Minkowski light-cone limit in which the four points become sequentially light-like separated. Based on the properties of the operator product expansion in the two regimes, we require that the correlation function should
have the correct singular behavior in the two limits. The origin of the singularities is traced back to the properties of the integrand of the logarithm
of the correlation function. In particular, its numerator should vanish in  the singular regimes, to prevent the integral from developing a more severe divergence than required. These simple criteria yield strong restrictions on the values of the coefficients of the various topologies.  

We show that the Minkowski criterion is powerful enough to completely
fix the form of the planar correlation function up to six loops and we
conjecture the same to be true to all loops.  The same criterion was
recently proposed in Ref.~\cite{Bourjaily:2011hi} in the context of
the four-particle planar amplitude written down in a basis of dual conformal
integrals, and  led to its determination up to seven loops. Our
  result for the six-loop four-point  correlation function
  gives, via the correlation function/amplitude duality a six-loop
  four-particle amplitude which
fully agrees with that of
Refs.~\cite{CarrascoJohansson,Bourjaily:2011hi}. This is a new, very
non-trivial confirmation of the duality.  We would like to point out that our analysis is not restricted to the planar limit of the correlation function (and hence to the possibility to compare  with the amplitude), but equally well applies to the non-planar sector. Thus, we are able to show that at three loops the correlation function does not receive any non-planar contributions. At four loops we reduce the freedom in the non-planar sector to just four arbitrary coefficients. To fix the value of these coefficients, we will need more
detailed information about the properties of the correlation function in the singular limit.

The Euclidean criterion turns out to be somewhat weaker, leaving some freedom in the planar correlation function already at four loops. Nevertheless,  the Euclidean double short-distance limit  provides us with the simplest and most efficient way of computing the anomalous dimension of the Konishi  operator. In a separate publication \cite{toapp}  we will show how to obtain its five-loop value, in full agreement with the prediction of Ref.~\cite{Bajnok:2008bm}, without making use of the conventional Feynman diagram technique. 

We would like to emphasize that our study of the correlation function in this paper does not require any knowledge about the scattering amplitude. We can obtain the form of the integrand using only the properties of the correlation function. At the end our results can be used to predict the dual scattering amplitude, and not the other way round, as was done in Ref.~\cite{Eden:2011we}. Still, the duality with the amplitude can be very helpful in improving the efficiency of our method. Just using the very fact that such a duality exists, but no explicit information about the amplitude, we derive a recursion procedure for determining the contribution to the correlation function 
from a large subclass of planar graphs. This procedure is the exact analog of the so-called ``rung rule" for the scattering amplitudes  \cite{Bern:1997nh}. This rule not only generates new higher-loop topologies from the lower-loop ones, but it also fixes their coefficients. Thus, after using the rung rule, we are left with a very small number of coefficients to determine.  For example, at four and five loops there remains a single non-rung-rule coefficient to fix. {At six loops we have 23 topologies of the rung-rule type and 13 non-rung-rule, with only 3 of the latter having non-zero coefficients. This should be compared with the analogous counting  in the approach of Ref.~\cite{Bourjaily:2011hi}, where the numbers are 245, 29 and 13, respectively. }This shows that our method can prove very efficient at even higher loops both for the correlation function and for the dual scattering amplitude.

The paper is organized as follows. 

In Section 2, we summarize the properties of the four-point correlation function and formulate our general 
ansatz for the integrand as a sum over all relevant topologies with arbitrary coefficients. In Section 3, we 
explain how these coefficients can be fixed to two loops from the known singular behavior of the correlation 
function in the Euclidean short-distance and in the Minkowski light-cone limits. In Section 4, we repeat the same
analysis at three loops and demonstrate that, unlike the two-loop case, the singular behavior of the three-loop
correlation function fixes all but one (color-dependent) coefficient. However, the integrand accompanying
this coefficient is proportional to the Gram determinant and, therefore, vanishes in four dimensions. The
resulting expression for the three-loop correlation function is unique and it does not receive non-planar 
corrections. In Section 5, we extend our analysis to four loops. We show that the four-point correlation 
function is uniquely fixed at four loops in the planar sector only, whereas in the non-planar sector 
the obtained expression depends on four arbitrary constants. In Sections 6 and 7, we construct the unique four-point
correlation function in the planar limit at five  and six loops, respectively. We also use the duality relation between
the correlation function and scattering amplitude to obtain the expression for the planar four-particle
amplitude up to six loops and we observe perfect agreement with the known results.
Some technical details of our calculation are described in four appendices.

\section{Symmetries of four-point correlation functions}

In this paper, we study four-point correlation functions of the simplest representative of the class of
half-BPS operators in the $\mathcal{N}=4$ SYM theory.  These are bilinear gauge invariant operators built from the six real scalars $\Phi^I$ (with $I=1,\ldots,6$ being an $SO(6)$ index) in the adjoint representation of the gauge group $SU(N_c)$:
\begin{align}\label{O-IJ}
  \mathcal{O}_{\mathbf{20'}}^{IJ} &= \tr\left(\Phi^I \Phi^J\right) - \frac16 \delta^{IJ} \tr\left(\Phi^K \Phi^K\right)\,.
\end{align}
The operator $\mathcal{O}^{IJ}$ belongs to the representation 
$\mathbf{20'}$ of the $R$ symmetry group $SO(6)\sim SU(4)$ and is the lowest-weight state 
of the so-called $\cN=4$ stress-tensor supermultiplet. The latter contains, among others, the stress-tensor (hence the name) and the (on-shell) Lagrangian of the theory. 
An important property of the operators \p{O-IJ} is that their scaling dimension is protected 
from perturbative corrections. The same is true for the two- and three-point correlation
functions of the operators $\mathcal{O}^{IJ}$ and it is only starting from four points that the correlation functions receive perturbative corrections in the $\mathcal{N}=4$ SYM theory
\cite{Penati:1999ba}. 

The four-point correlation function of the operators \p{O-IJ} have a non-trivial  $SO(6)$ tensor structure. A convenient way to keep track of it is to introduce auxiliary $SO(6)$ harmonic variables $Y_I$, defined as a (complex)
null vector, $Y^2\equiv Y_I Y_I=0$, and project the indices of  $\mathcal{O}^{IJ}$ as follows:
\begin{align}\label{defO}
{\mathcal O}(x,y) \equiv Y_I \, Y_J\, \mathcal{O}_{\mathbf{20'}}^{IJ}(x) =Y_I \, Y_J\,  \tr\left(\Phi^I(x) \Phi^J(x)\right)  \,, 
\end{align}
where $y$ denotes the dependence on the $Y-$variables. The main subject of this paper is the four-point correlation function of the  operators (\ref{defO}) :
\begin{equation}
G_4=\vev{ {\mathcal O}(x_1,y_1) {\mathcal O}(x_2,y_2) {\mathcal O}(x_3,y_3)    {\mathcal O}(x_4,y_4)}   =   \sum_{\ell=0}^\infty a^{\ell}
 \, G^{(\ell)}_{4}(1,2,3,4)\,.
\label{cor4loop}
\end{equation}
On the right-hand side $G_4$ is  expanded in powers of the `t Hooft coupling $a=g^2N_c/(4\pi^2)$, with $G^{(\ell)}_{4}$  denoting the perturbative correction at $\ell$ loops.
Notice that here we do not assume the planar limit and allow $G^{(\ell)}_{4}$ to have a  non-trivial dependence  on
$N_c$.

At tree level,  i.e. for $\ell=0$, the correlation function \p{cor4loop} reduces to a product of free scalar propagators:
\begin{align}\label{eq:1}
G^{(0)}_4(1,2,3,4) 
& =   \frac{(N_c^2-1)^2}{4 \, (4 \pi^2)^4}
\bigg[\lr{\frac{y^2_{12}}{x^2_{12}} \frac{y^2_{34}}{x^2_{34}}}^2+
 \lr{\frac{y^2_{13}}{x^2_{13}} \frac{y^2_{24}}{x^2_{24}}}^2 +
 \lr{\frac{y^2_{41}}{x^2_{41}} \frac{y^2_{23}}{x^2_{23}}}^2 \bigg] \\
& +   \frac{N_c^2 -1}{(4 \pi^2)^4} \,\
\left( \frac{y_{12}^2}{x_{12}^2} \frac{y_{23}^2}{x_{23}^2}  
\frac{y_{34}^2}{x_{34}^2} \frac{y_{41}^2}{x_{41}^2} +
 \frac{y_{12}^2}{x_{12}^2} \frac{y_{24}^2}{x_{24}^2}  
\frac{y_{34}^2}{x_{34}^2} \frac{y_{13}^2}{x_{13}^2} +
 \frac{y_{13}^2}{x_{13}^2} \frac{y_{23}^2}{x_{23}^2}  
 \frac{y_{24}^2}{x_{24}^2} \frac{y_{41}^2}{x_{41}^2} \right), \nonumber
\end{align}
where the notation is introduced for $x_{ij}^2=(x_i-x_j)^2$
and $y_{ij}^2=(Y_i\cdot Y_j)$. Here the first and second lines describe the disconnected 
and connected contributions, respectively, which explains their different scaling at large 
$N_c$. The right-hand side of \p{eq:1} contains six different $SO(6)$ harmonic $y-$structures. This is in agreement with the presence of six $SU(4)$ channels of the tensor product $\mathbf{20'}\times\mathbf{20'} = \mathbf{1} + \mathbf{15} + \mathbf{20'} + \mathbf{84} + \mathbf{105} +\mathbf{175}$. 
 
The superconformal symmetry of the $\mathcal{N}=4$ SYM theory, combined with the Lagrangian insertion mechanism, impose tight restrictions on the possible form of the loop corrections to the four-point correlation function, known as ``partial non-renormalization" \cite{partialNonRen,Eden:2011we}. Namely, the loop corrections to $G_4$ take the following factorized form:
\begin{align}\label{intriLoops}
  G_{4}^{(\ell)}(1,2,3,4)= \frac{2 \, (N_c^2-1)}{(4\pi^2)^{4}} \times R(1,2,3,4)   \times  F^{(\ell)}(x_i) \qquad \mbox{(for $\ell \ge 1$)} \,,
\end{align} 
where the first factor on the right-hand side is chosen for later convenience. Here
$R(1,2,3,4)$ is a universal, $\ell-$independent
rational function of the space-time, $x_i$, and harmonic, $y_i$, coordinates at the four external points $1,2,3,4$:
 \begin{align}\label{eq:7}
 R(1,2,3,4) &= \frac{y^2_{12}y^2_{23}y^2_{34}y^2_{14}}{x^2_{12}x^2_{23}x^2_{34} x^2_{14}}(x_{13}^2 x_{24}^2-x^2_{12} x^2_{34}-x^2_{14} x^2_{23})\notag
 \\ &
+\frac{ y^2_{12}y^2_{13}y^2_{24}y^2_{34}}{x^2_{12}x^2_{13}x^2_{24} x^2_{34}}(x^2_{14} x^2_{23}-x^2_{12} x^2_{34}-x_{13}^2 x_{24}^2)  \notag
\\
& +\frac{y^2_{13}y^2_{14}y^2_{23}y^2_{24}}{x^2_{13}x^2_{14}x^2_{23} x^2_{24}}(x^2_{12} x^2_{34}-x^2_{14} x^2_{23}-x_{13}^2 x_{24}^2) \notag 
 \\ &
+
\frac{y^4_{12} y^4_{34}}{x^2_{12}x^2_{34}}   + \frac{y^4_{13} y^4_{24}}{x^2_{13}
x^2_{24}} + \frac{y^4_{14} y^4_{23}}{x^2_{14}x^2_{23}} \,,
\end{align}
while $F^{(\ell)}(x_i)$ is a function of $x_i$ only (with $i=1,2,3,4$). Then, it follows from \p{intriLoops} that
the loop corrections to the four-point correlation function are determined by a single
function with the perturbative expansion
\begin{align}\label{loop-cor}
F(x_i; a)= \sum_{\ell \geq 1} a^\ell F^{(\ell)}(x_i)  \,.
\end{align} 
As before, we do not impose the planar limit. For the gauge group $SU(N_c)$, 
the expansion of $F(x_i; a)$ in powers of $1/N_c$ takes the form
\begin{align}\label{top}
F(x_i; a) = F_{{\rm g}=0} (x_i;a)+ \frac1{N_c^2} F_{{\rm g}=1}(x_i;a)  +\ldots\ ,
\end{align}
where $F_{\rm g}(x_i;a)$ is associated with color graphs of { (maximal)} genus ${\rm g}$, that is, a sphere for 
${\rm g}=0$, a torus for ${\rm g}=1$, etc.  

{It is easy to see that at one and two loops there exist no non-planar Feynman graphs contributing to the correlation function, therefore non-planar corrections to $F$ may appear
only starting from three loops, }  
\begin{align}\label{top1}
 F_{{\rm g}=1}(x_i;a)=O(a^3)\,.
\end{align}
Later in the paper we shall demonstrate that, in fact, $F_{{\rm g}=1}(x_i;a)$ vanishes at three loops and can receive  non-trivial contributions only starting from four loops (see Sects.~\ref{3lcondi} and \ref{Nps}).

What can we tell about the function $F(x_i)$ based on the symmetries and other basic properties of the correlation function? 

\subsection{Conformal symmetry and OPE constraints}\label{Sect:OPE}

Let us start with the corollaries of conformal symmetry. Each operator ${\mathcal O}(x_i,y_i)$ carries  
conformal weight $(+2)$. Examining the expression on the right-hand side of \p{intriLoops}, 
we find that $F(x_i)$ should be a conformally covariant function of $x_i$ with conformal weight $(+1)$ at each point. The remaining conformal weight $(+1)$ is provided by the rational factor $R(1,2,3,4)$. This implies, in particular, that $F(x_i)$ transforms under inversions
$x_i^\mu \to x_i^\mu/x_i^2$
as
\begin{align}
F(x_i) \to (x_1^2 x_2^2 x_3^2 x_4^2) F(x_i) \,.
\end{align}
This allows us to write the loop correction functions in \p{loop-cor} in the following manifestly conformal form:
\begin{align}\label{phil}
F^{(\ell)}(x_i)=\frac1{x^2_{13}x^2_{24}} \Phi^{(\ell)}(u,v)\,,
\end{align}
where $u$ and $v$ are the two conformally invariant cross-ratios made of the four points $x_i$,
\begin{align}\label{cr}
u=\frac{x^2_{12}x^2_{34}}{x^2_{13}x^2_{24}}\,, \qquad
v=\frac{x^2_{14}x^2_{23}}{x^2_{13}x^2_{24}} \,.
\end{align}

Another set of powerful constraints comes from the analysis of the four-point correlation 
function in the limit where a  pair of operators become null separated, $x_{ij}^2\to 0$. In this
limit, we can apply the operator product expansion (OPE) to find the asymptotic
behavior of the function $F(x_i)$. Notice that the limit $x_{ij}^2\to 0$ has different meanings in Euclidean and Minkowski  
kinematics. In the former case, the two points $x_i$ and $x_j$ coincide, while in the latter
case they can be separated along a light-ray direction. This allows us to distinguish two
different limits:
\begin{align}\label{lim1}
& \bullet \text{ Euclidean double short-distance limit:} && \text{$x_1\to x_2$,  $x_3\to x_4$;  }
\\[3mm] \label{lim2}
& \bullet \text{ Minkowski light-cone limit:} && \text{$x_{12}^2\,,x_{23}^2\,,x_{34}^2\,,x_{41}^2\to 0$.}
\end{align}
{In terms of the conformal cross-ratios, the two limits correspond to $u\to 0,v\to 1$ and $u,v\to 0$,
respectively.}

Let us consider the first limit. For $x_1\to x_2$ we have the OPE of the operators \p{defO},
\begin{align}\label{ope}
 {\mathcal O}(x_1, y_1)  \; {\mathcal O}(x_2, y_2)  & =     c_{\mathcal I}  \frac{y_{12}^4}{x_{12}^4} \; \mathcal I 
+ c_{\mathcal O}   \frac{y_{12}^2}{x_{12}^2} Y_{1I} Y_{2J}\, {\mathcal O}_{\mathbf{20'}}^{IJ}(x_1)  + c_{\mathcal K}(a)   \frac{y_{12}^4}{(x_{12}^2)^{1-{\gamma_{\mathcal K}}/{2}}} \;
{\mathcal K}(x_1)    + \ldots\,,
\end{align}
where the ellipsis denotes the  subleading contribution for $x_1\to x_2$. Here,
the dominant $O(1/x_{12}^4)$ contribution comes from the identity operator $\mathcal I $ and the first
subleading correction comes from the half-BPS operators ${\mathcal O}_{\mathbf{20'}}^{IJ}$
defined in \p{O-IJ}  and the Konishi operator ${\mathcal K} = \tr(\Phi^I\Phi^I)$. The
coefficients $c_{\mathcal I}= \lr{N_c^2 -1}/\lr{32 \pi^4}$ and $c_{\mathcal O}=1/(2\pi^2)$ do not depend on the coupling constant. Unlike the  operators ${\mathcal I}$ and ${\mathcal O}_{\mathbf{20'}}^{IJ}$, the Konishi operator ${\mathcal K}$ is not protected, so it acquires
an anomalous dimension $\gamma_{\mathcal K}(a)$ and the coefficient $c_{\mathcal K}(a)$ on the right-hand side of \p{ope} depends on the coupling
constant. 

Applying the OPE \p{ope} to the two pairs of operators at points $1,2$ and $3,4$, we can obtain the asymptotic behavior of the four-point
correlation function in the double short-distance limit $x_1\to x_2$, $x_3\to x_4$.  
Then, matching this behavior with the expected form of the correlation function \p{intriLoops}, we find the following  OPE limit for the function $F(x_i)$  as $x_1\to x_2$ and $x_3\to x_4$:
\begin{align}\label{cons}
  \sum_{\ell\ge 1} a^\ell F^{(\ell)}(x_i)   \
\stackrel{u\to 0 \atop v\to 1}{\longrightarrow} \ 
\frac1{6x_{13}^4}
\left[ c_{\mathcal K}^2(a) 
u^{{\gamma_{\mathcal K}(a)}/{2}} -1 \right] +\dots \,,
\end{align}
where the ellipsis denotes subleading terms {and the factor $1/6$ comes from  the
two-point correlation function of the Konishi operator.} For our 
purposes it is convenient to rewrite the OPE limit as
\begin{align}\label{SD}
 \ln  \lr{1+6x_{13}^4\sum_{\ell\ge 1} a^\ell F^{(\ell)}(x_i)} \
\stackrel{u\to 0 \atop v\to 1}{\longrightarrow} \  \frac12 \gamma_{\mathcal K}(a) \ln u+ O(u^0)\,.
\end{align}
This relation has the following meaning.
According to \p{cons}, the perturbative corrections $F^{(\ell)}(x_i)$ are expanded in powers of $\ln u$ with the maximal power equal to the number of loops $\ell$. The relation \re{SD}
implies that, at each perturbative level $a^\ell$, the particular combination of functions $F^{(\ell')}(x_i)$ with $\ell' \leq \ell$, arising from the expansion of the left-hand side of \p{SD} in powers of $a$, should
contain a single power of $\ln u$, independently of the loop order. This is a very powerful constraint which relates the $\ell-$loop correction $F^{(\ell)}(x_i)$ to the lower-loop terms  $F^{(\ell')}(x_i)$ with $1\leq \ell' < \ell$.

We would like to emphasize that the relation \p{SD} arose from the analysis of the
contribution of just one unprotected operator, the Konishi operator, to the OPE \p{ope}. Examining the subleading terms on the right-hand side of \p{ope} (denoted by an ellipsis there) we can single out the twist-two operators ${\mathcal O}_{\mu_1\ldots\mu_S}(x)$ carrying Lorentz spin $S$ and scaling dimension $\Delta=2+S+\gamma_{S}(a)$.~%
\footnote{It is worth mentioning that the Konishi operator $\mathcal K$ is the twist-two operator
with spin $S=0$.}
Their contribution to the right-hand side of \p{ope} is accompanied by the coefficient function
$x_{12}^{\mu_1} \ldots x_{12}^{\mu_S}/(x_{12}^2)^{1-\gamma_{S}(a)/2}$ and, therefore,
it is suppressed in the Euclidean short distance limit $x_{1}\to x_2$ by a factor of $(x_{12}^2)^{S/2}$. The situation changes, however, in the light-cone (Minkowski) limit, $x_{12}^2\to 0$ with $x_{12}^\mu\neq 0$. In this case the twist-two operators with arbitrary spin $S$ produce equally important contributions to the OPE \p{ope}. As a result, to find the asymptotic behavior of the four-point correlation function in the light-cone limit, we have to take into account the contributions of an {\it infinite} number of twist-two operators \cite{Belitsky:2003ys}. 

{This problem has been studied in Ref.~\cite{Alday:2010zy} where it was 
shown that in the multiple light-cone limit, $x_{12}^2,x_{23}^2,x_{34}^2,x_{41}^2\to 0$, 
the leading asymptotic behavior of the four-point correlation function is controlled
by the contribution of the twist-two operators with large spin $S$. According to \p{cr}, both
conformal cross-ratios vanish in the light-cone limit, $u,v\to 0$. As a consequence, the perturbative
corrections to the correlation function diverge as powers of $\ln u$
and $\ln v$.

{
More precisely, 
one finds that in the light-cone limit the perturbative expansion of $G_4/G_4^{(0)}$ takes the following form:
\begin{align}\label{Sud}
G_4/G_4^{(0)}\ \ \stackrel{u,v\to 0}{\longrightarrow} \ \ 1+ 2 x_{13}^2 x_{24}^2 \sum_{\ell\ge 1} a^\ell F^{(\ell)}(x_i) = 
1+\sum_{\ell\ge 1} a^\ell \sum_{m,n=0}^\ell c^{(\ell)}_{m,n}  (\ln u)^m(\ln
v)^n + \dots\,,
\end{align}
where $c_{m,n}^{(\ell)}$ are (coupling-independent) coefficients and the
ellipsis denotes finite terms. So, at
$\ell-$loop order, the expression for
$F^{(\ell)}(x_i)$ diverges like $ (\ln u)^m(\ln
v)^n$, where the total power  $m+n$ of the logs does  not exceed $2\ell$. 

All singular terms on the right-hand side of \p{Sud} with $n+m\ge \ell$ can be resummed  
to all orders in the coupling using the approach developed in Ref.~\cite{Mueller:1979ih}.
This leads to the following   (Sudakov) asymptotic behavior of the {\it logarithm} of the correlation function ~\cite{Alday:2010zy}
\begin{align}\label{LC}
  \ln \lr{1+ 2 x_{13}^2 x_{24}^2 \sum_{\ell\ge 1} a^\ell F^{(\ell)}(x_i)}  \stackrel{u,v\to 0}{\longrightarrow}  \lr{-\frac12 a+\frac14 a^2 \zeta_2} \ln u \ln v  +  \sum_{\ell\ge 2} c_\ell \big[ (a\ln u)^\ell + (a\ln v)^\ell \big] + \ldots\,,
\end{align}
with the coefficients $c_\ell = (-1)^\ell \zeta_\ell /( 2^{\ell-1}\, \ell)$.
The terms on the right-hand side of \p{LC} are ordered according to the maximal power
of logarithms per loop. For example, the $O(a)$ contribution to \p{LC} involves a double-log while  the $O(a^2)$ correction produces terms of the form
$a^2\ln u\ln v$, $(a\ln u)^2$ and $(a\ln v)^2$, in which the total power 
of the logs equals the loop order. The ellipsis on the right-hand side of \p{LC} denotes subleading
terms with a smaller number of logarithms at each loop order, {\it e.g.} $a^3 \ln u \ln v$, $a^3 (\ln u)^2$, $a^3\ln u$, etc.

Notice that the perturbative expansion \p{LC} diverges at $\ell>2$ loops like $ (\ln u)^\ell+(\ln
v)^\ell$.  This should be compared with the right-hand side of
\p{Sud} where the total power of the logs can reach $2\ell$.  In other words, the 
logarithm of the correlation function has a much softer light-cone singularity than the correlation function itself.~\footnote{Substituting \p{LC} into \p{Sud} we can compute the coefficients $c_{m,n}^\ell$ with $\ell\le m+n\le 2\ell$.}
Similarly to the short-distance limit, this fact once again implies that the very special combination of $\ell-$loop integrals $F^{(\ell')}$ with $1 \leq \ell' \leq \ell$ appearing in the left-hand side of \p{LC} at each perturbative level 
should be less singular in the light-cone limit than each individual integral  $F^{(\ell)}$. }

We would like to stress that the asymptotics \p{LC} does not rely on the planar limit and it holds for arbitrary $N_c$. The fact that the right-hand side of \p{LC} 
does not involve the  $1/N_c-$dependence implies that the non-planar corrections 
only contribute to terms denoted by an ellipsis in \p{LC}. In other
words, at  any loop
level $\ell$, the non-planar corrections to the logarithm of the correlation function should 
involve $\ln u$ and $\ln v$ to the total power $\le \ell-1$.

Comparing \p{SD} and \p{LC}, we observe that the two limits, the double short-distance
and the light-cone limits, lead to two different conditions on the function $F(x_i)$. As was explained above, the relation \p{LC} effectively takes into 
account the contribution of an infinite number of twist-two operators and, therefore, we expect it to 
be more restrictive than \p{SD}. Later on in the paper we demonstrate that this is indeed the case, starting at four loops.\footnote{A similar criterion applied to the integrand of the four-gluon amplitude, written in a basis of dual conformal integrals, allows one to determine its form up to seven loops \cite{Bourjaily:2011hi}. The special role played by the logarithm of the integrand of the amplitude was also pointed out in \cite{Drummond:2010mb}.} 

\subsection{Permutation symmetry}

As was shown in Ref.~\cite{Eden:2011we}, the four-point correlation function 
in the ${\mathcal N}=4$ SYM
theory has a new symmetry. To formulate this symmetry, we represent the $\ell-$loop correction to the function $F^{(\ell)}(x_i)$ in the following
form
\begin{align}\label{integg}
  F^{(\ell)}(x_1,x_2,x_3,x_4)={x_{12}^2 x_{13}^2 x_{14}^2 x_{23}^2 
  x_{24}^2 x_{34}^2\over \ell!\,(-4\pi^2)^\ell} \int d^4x_5 \dots
  d^4x_{4+\ell} \,  f^{(\ell)}(x_1,  \dots , x_{4+\ell})\,,
\end{align}
where $f^{(\ell)}$ is some function depending on the four external coordinates $x_1,\ldots,x_4$
and the $\ell$ additional (internal) coordinates $x_5,\ldots,x_{4+\ell}$ giving the positions of  the interaction vertices. Defined in this way, the function $f^{(\ell)}$ has the meaning
of the $\ell-$loop {\it integrand}.

The particular form of \p{integg} is related to computing the loop corrections to
the correlation function by Lagrangian insertions. In this method, the four-point function $F^{(\ell)}$ is determined by a $(4+\ell)-$point correlation function involving $\ell$ 
additional insertions of the ${\mathcal N}=4$ SYM Lagrangian, whose positions are integrated over.
Then, the function $f^{(\ell)}$ naturally appears as the Born-level approximation to this $(4+\ell)-$point correlation function. As was shown in Ref.~\cite{Eden:2011we}, in $\mathcal{N}=4$ SYM with gauge group $SU(N_c)$ the function $f^{(\ell)}$ has the following
remarkable properties:
\begin{itemize}
\item[(i)] it is symmetric under the exchange of any pair of points (both external and internal)
\begin{align}
 f^{(\ell)}(\ldots,x_i,\ldots,x_j,\ldots) =  f^{(\ell)}(\ldots,x_j,\ldots,x_i,\ldots)\,;
\end{align}

\item[(ii)] it is a rational function of the distances $x_{ij}^2$ with only simple poles at the coincident points
\begin{align}
 f^{(\ell)}(x_i) \sim \frac1{x_{ij}^2} \,,\qquad \text{for $x_i\to x_j$}\,;
\end{align}

\item[(iii)] it has conformal weight $(+4)$ at all $(4+\ell)$ points and transforms under inversions $x_i^\mu \to x_i^\mu/x_i^2$ as
\begin{align}\label{f-inv}
  f^{(\ell)}(x_i)  \to (x_1^2 \ldots x_{4+\ell}^2)^4\, f^{(\ell)}(x_i) \,.
\end{align}
\end{itemize}
Here the least obvious first condition is ultimately related to the fact  that the half-BPS operator $ \mathcal{O}_{\mathbf{20'}}^{IJ}(x)$ belongs to the same short superconformal multiplet as the ${\mathcal N}=4$ SYM Lagrangian $\cL$. The second condition follows from the analysis of the OPE of  the operators $\cO$  and $\cL$. Finally, the last condition ensures that $F^{(\ell)}(x_i)$, as defined by the integral in  \p{integg},  has the required conformal weight $(+1)$ at the external points $x_1,\ldots,x_4$.
 
We can easily satisfy the above three conditions by writing $f^{(\ell)}$ in the
following form:
\begin{align}
   \label{eq:10}
   f^{(\ell)}(x_1, \dots, x_{4+\ell})= { P^{(\ell)}(x_1, \dots ,
   x_{4+\ell}) \over \prod_{1\leq i<j \leq 4+\ell}   x_{ij}^2}\ ,
 \end{align} 
where the denominator contains the product of all distances between the $(4+\ell)$ points
and $P^{(\ell)}$ is a homogeneous polynomial in $x_{ij}^2$
needed to restore the correct conformal weights of the function $f^{(\ell)}$. Since the
denominator in \p{eq:10} is explicitly symmetric under the exchange of any pair of points,
the polynomial $P^{(\ell)}$ should also be invariant under the $S_{4+\ell}$ permutations of $x_1,\ldots,x_{4+\ell}$. Further, in order to verify \p{f-inv} the polynomial $P^{(\ell)}$ should
have uniform conformal weight $-(\ell-1)$ at all points. This implies in particular that $P^{(\ell)}$ is a homogeneous polynomial in $x_{ij}^2$ of degree $(\ell-1)(\ell+4)/2$ and the distance
between, say, point $x_1$ and another point $x_i$ appears in the expansion of  $P^{(\ell)}$ exactly $(\ell-1)$ times.

As an example, let us consider \p{eq:10} in the two simplest cases $\ell=1$ and $\ell=2$.
At one loop, for $\ell=1$, the degree of $P^{(1)}$ is zero, reducing it to a constant:
\begin{align} \notag 
& P^{(1)} = 1\,,
\\
\label{f1}
& f^{(1)}(x_1, \dots, x_5)={c^{(1)}  \over \prod_{1\leq i<j \leq 5}   x_{ij}^2}\,,
\end{align}
with $c^{(1)}$ arbitrary. Here and in all the subsequent expressions for $P^{(\ell)}$ we choose to normalize the basis polynomials so that each monomial appears with unit weight. 

At two loops, for $\ell=2$,  $P^{(2)}$ is a polynomial in $x_{ij}^2$ of degree $3$ given by the $S_6$ invariant sum of terms, each linear in $x_{1i}^2$  
\begin{align} \notag 
&P^{(2)} =  \ft1{48} \sum_{\sigma\in S_6} x_{\sigma_1\sigma_2}^2x_{\sigma_3\sigma_4}^2x_{\sigma_5\sigma_6}^2 = x_{12}^2x_{34}^2x_{56}^2 + \dots \,,
\\
\label{f2}
&f^{(2)}(x_1, \dots, x_6)={c^{(2)} P^{(2)} \over \prod_{1\leq i<j \leq 6}   x_{ij}^2}\,,
\end{align}
with $c^{(2)}$ arbitrary. In the first relation in \p{f2} the sum runs over all permutations $(\sigma_1,\ldots,\sigma_6)$ of the points $(1,\ldots,6)$. The coefficient $1/48$ ensures that each term appears with unit weight (the ellipsis 
denotes the terms needed to restore the $S_6$ permutation symmetry of $P^{(2)}$). 
Combining the two relations in  \p{f2}, we get
\begin{align}\label{f-2loop}
 f^{(2)}(x_1, \dots, x_6)= \frac{c^{(2)}}{x_{13}^2x_{14}^2x_{15}^2 x_{16}^2x_{23}^2x_{24}^2x_{25}^2 x_{26}^2x_{35}^2x_{36}^2x_{45}^2x_{46}^2} +\text{ $S_6$ permutations}\ .
\end{align}

In summary, the construction of the correlation function at a given loop level consists of two major steps:
\begin{itemize}
\item Find the most general expression for the permutation symmetric integrand ($f-$function) or, equivalently, for its numerator ($P-$function) having the properties listed above. In general, this step produces a number of inequivalent integrand topologies, some of them planar, the others non-planar. This step is most efficiently done by using graph-theoretic tools, as explained in Sect.~\ref{GTR}. 
\item Try to fix the coefficients of the various topologies by imposing the OPE constraints from Sect.~\ref{Sect:OPE}. Later in the paper we will show that this is sufficient to completely determine the correlation function in the planar limit up to six loops {(with no obstruction for extending analysis to higher loops), while in the non-planar sector
some freedom is left in the value of a few coefficients, starting at four loops.  }

\end{itemize}

\subsection{Graph-theoretical interpretation}\label{GTR}

As mentioned above, to classify the general functions $f^{(\ell)}$ and $P^{(\ell)}$ at higher
loops, it proves convenient to use their graphical representation. To this end, we 
identify each point $x_1, \dots, x_{4+\ell}$ as a vertex of a graph and denote each
factor $x_{ij}^2$ in the expression for the functions by a line connecting the
vertices $i$ and $j$. In this way, we can associate  two sets of graphs  with the  functions $f^{(\ell)}$ and $P^{(\ell)}$ that we shall call $f-$ and $P-$graphs. In principle, the two graphs are complementary -- knowing one allows us to fix the other. However, each type of graphs has its own features, so we will discuss them separately.  

\subsubsection{Numerator or $P-$graphs}\label{Npg}

Let us first examine the properties of the $P-$graphs. By construction, they contain $(4+\ell)$ vertices. The permutation symmetry $S_{4+\ell}$ allows us not to label each vertex 
and treat them on an equal footing. 
The conformal properties of the polynomial  $P^{(\ell)}(x_1, \dots,x_{4+\ell})$  imply that
the corresponding $P-$graph should have $(\ell-1)$ edges attached to each vertex, so that
the total number of edges is $(\ell-1)(\ell+4)/2$, in agreement with the total degree of the polynomial. In general, the $P-$graph is made of several connected subgraphs.
For instance, for $\ell=2$, from \p{f2} we read that the corresponding graph
contains three components, each consisting of a single edge. Also, the polynomial $P^{(\ell)}$ may contain higher powers of the distances $(x_{ij}^2)^n$, in which
case the vertices $i,j$ are connected by $n$ edges.%
\footnote{In graph theory, for $n=1$  such graphs are called ``regular" and for $n>1$ they are called  ``multigraphs".}
At the same time, the $P-$graph
cannot contain loops (edges attached to the same vertex)\footnote{A ``loop" should be distinguished from a ``cycle", i.e. a subset of vertices connected by a closed cycle of edges.} since their contribution to the $P-$polynomial vanishes due to $x_{ii}^2=0$. Summarizing these
properties, we conclude that
\begin{align}\label{P-cond}
\fbox{\textit{ $P-$graph  =  loop-less multigraph with $(4+\ell)$ vertices of degree $(\ell-1)$}}
\end{align}
The main advantage of the $P-$graphs is that each vertex has the same degree. This makes it possible to determine their number at each loop level, using standard graph-theoretical tools from Refs.~\cite{nauty,Sage}.  

Let us denote by $n_\ell$ the number of isomorphism classes of loop-less multigraphs of order $(4+\ell)$ and of degree  $(\ell-1)$. Up to six loops, $n_\ell$ takes the following
values:\\
 
\begin{table}[h!]\begin{center}
  \begin{tabular}{ | c | c | c | c | c | c |c | }
    \hline
  $\ell$ & 1 & 2 & 3 & 4 & 5 & 6   \\ \hline
   $n_\ell$ & 1 & 1 & 4 & 32 & 930 & 189341  \\ \hline
   $n^{\rm planar}_\ell$ & 1 & 1 & 1 & 3 & 7 & 36  \\ \hline
   $n^{\rm rung-rule}_\ell$ & - & 1 & 1 & 2 & 6 & 23  \\ \hline
   Ref. & \phantom{A12941}  &  \phantom{A12941} & \phantom{A12941}  & A129416 & A129418 & A129420  \\
    \hline
  \end{tabular}\end{center}
  \caption{Numbers of  $P-$graphs up to six loops (second row), of planar integrand  topologies (third row) and of iterative (``rung-rule") planar topologies (fourth row). The bottom row contains references to  \cite{Ency}.
  The first three cases of  $P-$graphs from this table are displayed in Fig.~\ref{Pgr}.}
  \label{Table1}
\end{table}  
\noindent
As can be seen from Table \ref{Table1}, the number $n_\ell$ of $P-$graphs grows very rapidly with the loop order
$\ell$. However, as we will show later in the paper, the vast majority of such graphs produce non-planar corrections
to the correlation function, leaving a much smaller number $n^{\rm planar}_\ell$ of planar topologies. Moreover, the latter have an interesting iterative structure,
the so-called ``rung rule" (see Sect.~\ref{sec:rung-rule}), which allows us to determine the contribution of the 
majority of the planar graphs (listed as    $n^{\rm rung-rule}_\ell$ in Table \ref{Table1}) in the planar limit from lower loop data. 
  Only a small number of  relevant planar graphs cannot be obtained by the rung rule and require a different approach. 

\begin{figure}[th!]%
\vspace*{5mm}
\psfrag{a}[cc][cc]{$P^{(1)}$}
\psfrag{b}[cc][cc]{$P^{(2)}$}
\psfrag{c}[cc][cc]{$P^{(3)}_1$}
\psfrag{d}[cc][cc]{$P^{(3)}_2$}
\psfrag{e}[cc][cc]{$P^{(3)}_3$}
\psfrag{f}[cc][cc]{$P^{(3)}_4$}
\centerline{\includegraphics[width=0.95\textwidth]{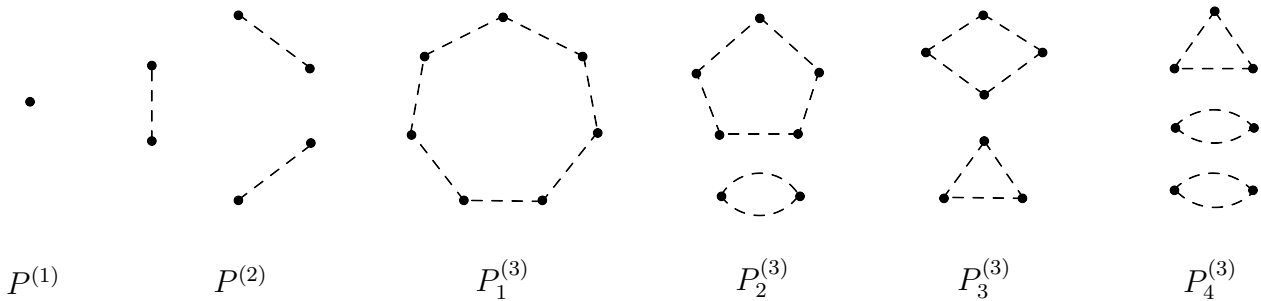}}
\caption{$P-$graphs up to three loops. The one-loop graph $P^{(1)}$ shrinks to a point corresponding to the constant numerator in \p{f1}. The two-loop graph $P^{(2)}$ consists of three disconnected segments. At three loops there are four possible graphs, one connected $P^{(3)}_1$ and three disconnected $P^{(3)}_2$, $P^{(3)}_3$, $P^{(3)}_4$. }
\label{Pgr} 
\end{figure}

As explained, $P-$graphs provide a straightforward way to obtain a
general basis for the full non-planar correlation function. However, it also
proves useful to consider 
 $f-$graphs described below. These are much closer to more standard Feynman
 integral graphs, with edges corresponding to propagators. 
 

\subsubsection{Denominator or $f-$graphs}

If the $P-$graphs represent the {\it numerator} of the $f-$function \p{eq:10}, the $f-$graphs are constructed from its {\it denominator}. Here we have to distinguish the effective denominator of the integrand of the function $F^{(\ell)}$ from the universal one shown on the right-hand side of \p{eq:10}.  The latter defines the so-called ``complete graph" $K_{4+\ell}$, that is the graph in which each pair of the $(4+\ell)$ vertices are connected by an edge. The polynomial $P^{(\ell)}$ contains positive powers of $x^2_{ij}$, so that it removes some of the factors from the denominator in \p{eq:10}. In graphical terms  this amounts to deleting the corresponding edges in the complete graph  $K_{4+\ell}$. The remaining factors  of $x^2_{ij}$  in the denominator constitute the edges of what we call the ``$f-$graph".    

If the polynomial $P^{(\ell)}$ 
does not contain factors of $(x_{ij}^2)^n$ with power $n > 1$, or equivalently, if the corresponding
$P-$graph is regular (and not a multigraph), then the effective numerator of $f^{(\ell)}$ is a constant (see, e.g., Eq.~\p{f-2loop}) and the corresponding $f-$graph coincides with the complement of the $P-$graph. Namely, it is obtained from the complete graph $K_{4+\ell}$ by removing all edges of the $P-$graph. In what follows we shall refer to such $f-$graphs
as ``{pure}". By construction, a pure $f-$graph is a regular {\it connected} \footnote{It is possible to have potential $f-$graphs which are disconnected, but satisfy all other properties. The first example occurs at six loops in the non-planar theory and is given by $K_5 \times K_5$. We believe such $f-$graphs cannot contribute to the correlation function and thus should come with coefficient zero. This is an additional restriction on the basis.} graph with
$(4+\ell)$ vertices and exactly four edges attached to each vertex. The latter number
matches the conformal weight of the function $f^{(\ell)}$ at each point. 

However, if the polynomial $P^{(\ell)}$ contains a factor of $(x_{ij}^2)^n$ with $n>1$, then, after cancellation with the denominator,  the numerator of $f^{(\ell)}$ still involves some factor of $(x_{ij}^2)^{n-1}$. This factor induces a negative conformal weight  at the points $x_i$ and $x_j$ in the numerator and it has to be balanced by an excess of   
conformal weight in the denominator. In terms of the $f-$graph this means that
the two vertices should have $(3+n) > 4$ edges attached to each of them. 

To summarize,
\begin{align}\label{f-cond}
\fbox{\textit{$f-$graph = connected  graph with $(4+\ell)$ vertices of degree $\ge 4$}}
\end{align}
The pure $f-$graphs are the special case where all vertices have the same minimal degree $4$.
As an example,  in Fig.~\ref{Fgr} we show the $f-$graphs corresponding to the one-loop and two-loop expressions, Eqs.~\p{f1} and \p{f-2loop}, respectively, as well as the three-loop graphs discussed in Sect.~\ref{cftl}.

 \begin{figure}[h!]
\vspace*{5mm}
\psfrag{a}[cc][cc]{$f^{(1)}$}
\psfrag{b}[cc][cc]{$f^{(2)}$}
\psfrag{c}[cc][cc]{$f^{(3)}_1$}
\psfrag{d}[cc][cc]{$f^{(3)}_2$}
\psfrag{e}[cc][cc]{$f^{(3)}_3$}
\psfrag{f}[cc][cc]{$f^{(3)}_4$}
\centerline{\includegraphics[width=\textwidth]{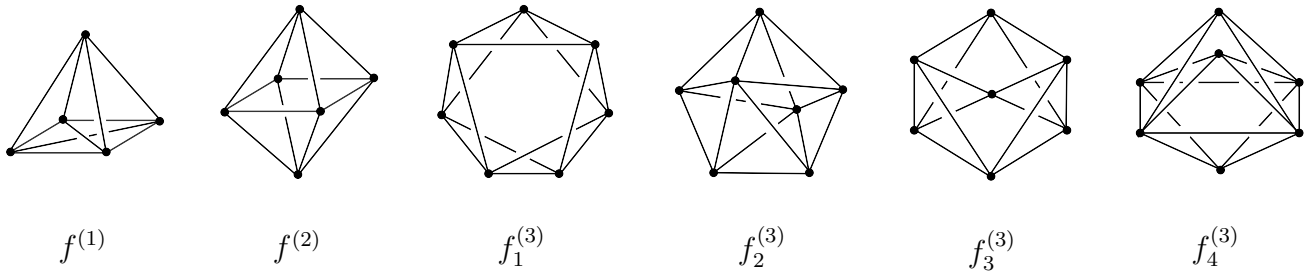}}
\caption{ $f-$graphs up to three loops obtained from the $P-$graphs in Fig.~\ref{Pgr}. The one-loop graph $f^{(1)}$ is non-planar, but it becomes planar after multiplying it with the prefactor  in \p{integg}. The two-loop  graph $f^{(2)}$ and one of the tree-loop graphs, $f^{(3)}_2$, are planar. The remaining three-loop graphs $f^{(3)}_1$, $f^{(3)}_3$, $f^{(3)}_4$ are non-planar, even including  the prefactor  in \p{integg}.}
\label{Fgr} 
\end{figure}

The main advantage of dealing with graphs 
is that they allow us to replace the (difficult) problem of finding the most
general expression for the functions $f^{(\ell)}$ and/or $P^{(\ell)}$ by the much simpler problem of classifying all (non-isomorphic) $f-$ and/or $P-$graphs satisfying
the conditions \p{f-cond} and \p{P-cond}, respectively. We find both $P-$graphs and $f-$graphs useful depending on the context.  The $P-$graphs allow a clean and systematic procedure for obtaining the full basis of all possible terms that can occur in the full non-planar theory, whereas for the $f-$graphs one has to check that it is possible to attach   suitable numerators to the graph in order to ensure that all vertices have overall conformal weight four, without canceling a propagator. 
The $f-$graphs are however much more closely related to standard Feynman integral graphs.  Moreover, importantly, in the planar theory we may restrict our basis to planar $f-$graphs, which are easier to classify than the corresponding $P-$graphs.  As we shall see shortly, restricting the scope to planar $f-$graphs allows a huge further reduction in the basis.

\subsubsection{General form of the $f-$function}

Now, suppose that we have found all such graphs.
For each $f-$ and/or $P-$graph topology we can easily reconstruct the corresponding contribution to the function $f^{(\ell)}$, 
following the rules formulated above. Then the general form of $f^{(\ell)}$ is given by a  linear combination of all topologies with {\it arbitrary coefficients}
$c^{(\ell)}_{\alpha}$ 
\begin{align}\label{f-sum}
f^{(\ell)}(x) = \sum_{\alpha=1}^{n_\ell} c^{(\ell)}_{\alpha} \, f_\alpha^{(\ell)}(x_1,\ldots,x_{4+\ell}) 
= \sum_{\alpha=1}^{n_\ell} c^{(\ell)}_{\alpha}\, \frac{P^{(\ell)}_\alpha (x_1,\ldots,x_{4+\ell})}{\prod_{1\le i< j \le 4+\ell} x_{ij}^2} \,.
\end{align}
Here the sum runs over the $n_\ell$ non-isomorphic graphs satisfying  \p{f-cond} and \p{P-cond}. Their numbers for $\ell \leq 6$ are listed in Table~\ref{Table1}.
 
We recall that the function $f^{(\ell)}(x)$ defines the integrand of the $\ell-$loop correction $F^{(\ell)}(1,2,3,4)$ in \p{integg}. Substituting  the general ansatz \p{f-sum} into 
\p{integg}, we conclude that the permutation symmetry of the integrand allows us to reduce the freedom in the loop
corrections to the four-point correlation function to the set of
coefficients $c^{(\ell)}_{\alpha}$. {As was already mentioned, the
  majority of terms on the right-hand side of \p{f-sum} {give 
  non-planar integrands and their coefficients are expected to vanish in the limit $N_c \to \infty$.} 
Moreover, most of the coefficients $c^{(\ell)}_\a$ of the integrands
in \p{f-sum} {which survive in the planar limit}, are inherited from lower loops by the ``rung rule" (see Sect.~\ref{sec:rung-rule}). As a result, to determine the correlation function
in the planar limit, we will have to fix a much smaller number of coefficients, for example, only 13 coefficients at six loops with only three of them being different from zero (see Sect.~\ref{Sl}).    }

In order to determine the values of the coefficients $c^{(\ell)}_\a$, we have to take into account  the following additional information about 
 the correlation function. Firstly, as was shown in Sect.~\ref{Sect:OPE}, the
function $F^{(\ell)}$ has to satisfy the relations \p{SD} and \p{LC} that follow from the OPE
analysis of the correlation function. Combined with \p{integg}, these relations lead to very powerful
restrictions on the coefficients $c^{(\ell)}_{\alpha}$. Secondly, the (non-planar) correlation function admits an
expansion \p{top} in powers of $1/N_c^2$. In our ansatz for the integrand,
Eqs.~\p{integg} and \p{f-sum}, the dependence on $N_c$ is contained in the coefficients 
$c^{(\ell)}_{\alpha}$. Therefore, {it is natural to conjecture that $c^{(\ell)}_{\alpha}$ admit a similar expansion in $1/N_c^2$,
with the leading power  related to the genus of the
corresponding $f-$graph.\footnote{The connection between the genus of the $f-$graphs and that of the actual components of the correlation function \p{intriLoops}   is discussed in Appendix~\ref{app:genus}.}} 
Finally, the polynomials $P_\alpha^{(\ell)}(x_i)$ in the numerators in \p{f-sum} depend on $(4+\ell)$ four-dimensional
vectors $x_i^\mu$. Since any five vectors are linearly dependent in four dimensions, this
leads to additional relations between these polynomials which take the form of vanishing Gram determinant 
conditions. Taking them into account, we can reduce the number of independent terms on the
right-hand side of \p{f-sum}.

\subsection{Duality with the planar four-particle scattering amplitude} 
 
In the planar $\mathcal{N}=4$ SYM theory there exists a remarkable relation between the {\it planar} four-point
correlation function $G_4$ and the {\it planar} four-particle MHV amplitude $\cA_4$:~\footnote{This duality between $G_n$ and $\cA^{\rm MHV}_n$   with an arbitrary number of points $n$ was first proposed in \cite{EKS2,EKS3}. Later on it has been extended to the super-correlation functions of stress-tensor multiplets and non-MHV super-amplitudes~\cite{Eden:2011yp,Eden:2011ku,Adamo:2011dq}.}
\begin{align}\label{G=A}
\lim_{x_{i,i+1}^2\to 0}  (G_4/G_4^{(0)})(x_1,x_2,x_3,x_4) =     [(\cA_4/\cA_4^{(0)})(p_1,p_2,p_3,p_4)]^2 \,.
\end{align}
Here the superscript `$(0)$' denotes the tree-level approximation and the on-shell momenta of the scattered 
particles $p_i^\mu$ are related to the (dual) coordinates $x_i^\mu$ by
\begin{align}\label{duco}
p_i^\mu=x_i^\mu-x_{i+1}^\mu\,,\qqquad p_i^2=0\,.
\end{align}
Notice that the relation \p{G=A} is formulated in terms of the {\it integrands} of the two objects, and not in terms of the corresponding Feynman integrals. The latter diverge in the light-cone limit (for the correlation function) and for massless particles with $p^2_i=0$ (for the amplitude), and hence require a regularization, say dimensional regularization in $D=4-2\epsilon$ dimensions. What appears on the right-hand side of  \p{G=A} is the {\it four-dimensional} integrand of the amplitude, which is a rational function of the momenta. This rational function, rewritten in terms of dual coordinates according to \p{duco}, is then compared to the rational integrand of the correlation function. The latter is conformally covariant by construction, while the integrand of the amplitude is known to  have dual conformal invariance \cite{Drummond:2006rz, Bern:2010tq, Bern:2007ct, Carrasco:2011mn}.  

We would like to emphasize once again that the duality \p{G=A} only applies to the planar limit of the two objects. Indeed, the correlation function is known not to have non-planar corrections at one and two loops \cite{Eden:2000mv,Bianchi:2000hn},~\footnote{In this paper we show that  the  non-planar corrections do not appear below four loops, see Section~\ref{3lcondi}.} while the four-particle amplitude starts having non-planar contributions already at two loops \cite{Bern:1997nh,Bern:1998ug,Bern:2010tq}. 

The duality relation \p{G=A} involves the ratio of the correlation functions defined in the kinematical configuration
in which two neighboring operators are light-like separated, $x_{i,i+1}^2=0$. In this limit the left-hand side of \p{G=A} becomes (see Eqs.~\p{cor4loop}--\p{intriLoops})~\footnote{In what follows, to simplify the formulae, we will drop the explicit mentioning of the integrand, but it will always be implied.}  
\begin{align}\label{G/G0}
\lim_{x_{i,i+1}^2\to 0}  (G_4/G_4^{(0)}) =     1+ 2  \sum_{\ell\ge 1} a^\ell \cF^{(\ell)}(x_i)\,,
\end{align}
where the notation was introduced for
\begin{align}\label{F-hat}
\cF^{(\ell)} = \lim_{x_{i,i+1}^2\to 0} x_{13}^2 x_{24}^2 F_{{\rm g}=0}^{(\ell)}(x_i)
\end{align}
and the subscript ${\rm g}=0$ indicates the planar limit. 
Similarly, the perturbative corrections to the scattering amplitude $\cA_4$ take the form
\begin{align}\label{M}
\cA_4/\cA_4^{(0)} = 1+\sum_{\ell\ge 1} a^\ell  \cM^{(\ell)}(p_i)\,,
\end{align}
where the loop corrections $\cM^{(\ell)}(p_i)$ are given by linear combinations of the  integrands of scalar $\ell-$loop Feynman integrals. With this notation, the duality relation \p{G=A} becomes
\begin{align}\label{F=AA}
  {1+ 2  \sum_{\ell\ge 1} a^\ell \cF^{(\ell)}(x_i)} =  \Big({1+\sum_{\ell\ge 1} a^\ell  \cM^{(\ell)}(p_i)}\Big)^2\,.
\end{align}
Expanded in powers of the coupling, Eq.~\p{F=AA} leads to 
\begin{align}\notag
& \cF^{(1)} = \cM^{(1)}\,,\qquad \cF^{(2)} = \cM^{(2)}+\frac12\lr{\cM^{(1)}}^{2}
\,,\qquad \cF^{(3)} = \cM^{(1)}\cM^{(2)}+\cM^{(3)}
\\ \label{F-M}
& \cF^{(4)} = \cM^{(1)}\cM^{(3)}+\cM^{(4)}+\frac12\lr{\cM^{(2)}}^{2}
\,,\qquad \cF^{(5)} = \cM^{(1)}\cM^{(4)}+\cM^{(2)}\cM^{(3)}+\cM^{(5)}\,,\ \ldots
\end{align}
and, conversely, 
\begin{align}\notag
& \cM^{(1)} = \cF^{(1)}\,,\qquad  \cM^{(2)} = \cF^{(2)}-\frac12\big(\cF^{(1)}\big)^{2}
\,,\qquad \cM^{(3)}=\cF^{(3)} - \cF^{(1)}\cF^{(2)} +\frac12 \big(\cF^{(1)}\big)^3\,,
\\ \label{M-F}
& \cM^{(4)}= \cF^{(4)}-\cF^{(1)}\,\cF^{(3)} -\frac12\big({\cF^{(2)}}\big)^{2}+\frac32\,\cF^{(2)}\big({\cF^{(1)}}\big)^{2}-\frac58\big({\cF^{(1)}}\big)^{4}\,,\ \ldots\ .
\end{align} 
Making use of \p{M-F}, we can apply our 
results for the correlation function to obtain the four-dimensional integrands of the four-particle 
scattering amplitude in the planar limit.

\subsection{Rung rule}\label{sec:rung-rule}

The amplitude/correlation function duality outlined in the previous subsection
provides a means of inducing  $\ell$-loop terms in the correlation
function (and hence amplitude) from lower loops. To see how this works, note  that in the light-cone limit the correlation function at any loop order always contains a contribution of the  form (see Eq.~\p{F-M})~\footnote{At two loops there is an additional  factor of 1/2 in front of this term due to the additional symmetry but this does not affect the argument which follows.}
\begin{equation}\label{eq:2}
\cF^{(\ell)}= {\cM^{(1)}} \times {\cM^{(\ell-1)}} +\ldots\, .
\end{equation}
This describes part of the integrand of the correlation function in the light-like limit, which is directly determined by iterating lower loop information. However, if we now  
relax the light-cone condition and make use of the $S_{4+\ell}$ permutation symmetry of the integrand, we can 
reconstruct an even bigger part of the correlation function, exploiting this simple fact. 

It is most straightforward to consider the implications of this observation geometrically in terms of $f-$graphs. In essence, the iteration consists in gluing the one-loop graph (a square pyramid) to any planar $f^{(\ell-1)}-$ loop graph, thus  producing a subset of all the planar $f^{(\ell)}-$loop graphs. The $f-$graphs generated in this way up to four loops are illustrated in Fig.~\ref{fig:bodies}.
\begin{figure}[h!]
  \centering
  \includegraphics[width=13cm]{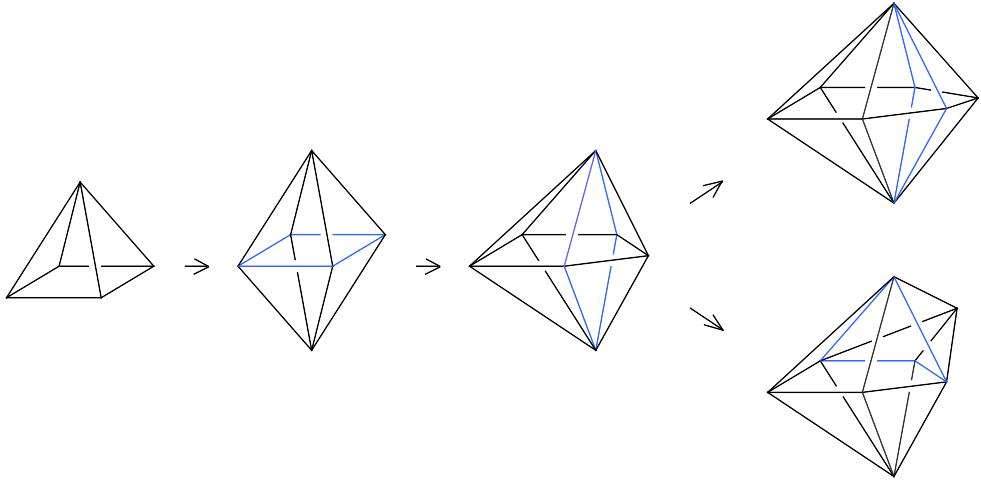}
  \caption{Planar $f-$graphs at  one, two, three and four loops. One can see
   how applying the rung rule (i.e. gluing a pyramid across rectangle  face shown
   in blue) to a lower-loop
   $f-$graph produces higher-loop $f-$graphs. }
\label{fig:bodies}
\end{figure}
 We explain the idea of this graphical construction with the example of the two-loop correlation function in Sect.~\ref{rr}, after which we formulate the general rule for any loop order. 

The above procedure is reminiscent of the so-called ``rung rule'' \cite{Bern:1997nh} for iteration of loop integrals 
for the four-particle amplitude in planar $\cN=4$ SYM. This empirical rule allows one to 
generate new planar $\ell-$loop integrals for   $\cM^{(\ell)}$, starting from the known $(\ell-1)-$loop integral topologies in  $\cM^{(\ell-1)}$  and adding a ``rung" (avoiding bubbles and triangles). 
The validity of the rung rule can be verified by considering the unitarity cuts of the amplitude.  
In fact, what we observe here  is more than just an analogy, the rung rule is {\em
  precisely} the dual momentum space description of the graphical iteration procedure
mentioned above. The factor $\cM^{(1)}$ in  \p{eq:2}  plays the role of the added
rung (see Sect.~\ref{rr}).

We would like to emphasize that our induction mechanism not only allows us to predict the new ``rung-rule'' topologies, but
also to fix their coefficients. They are inherited from the lower-loop coefficients via the cross-term in  \p{eq:2}.  Going away from the light-cone limit in  \p{eq:2} and using the permutation symmetry of the integrand,  we can thus reconstruct a substantial part of the correlation function. After that, we can feed this information back into Eq.~\p{M-F} and predict part of  the amplitude itself.

It is well known  \cite{Bern:2006ew} that the rung rule fails to generate all the relevant topologies of the four-particle planar amplitude
from four loops on. Then, we should expect that the
cross-terms in the duality relation  \p{eq:2} will not be sufficient
to predict all the {\it planar} topologies appearing in the integrand
$f^{(\ell)}$ for $\ell \geq 4$. Indeed, this is the case already at four loops: the function
$f^{(4)}$ contains two planar topologies of the rung-rule type (i.e.,
which talk to the cross-terms in the duality relation) and one of a
different type (see Sect.~\ref{Pls}).

\section{Correlation function to two loops}\label{Coftl}

In this section we wish to illustrate the general procedure for constructing the four-point correlation function $G_4$, outlined in the preceding section, by the simplest examples of the one- and two-loop corrections. Although very simple, the two-loop case exhibits many of the key features.

\subsection{One loop}

\subsubsection{The one-loop integrand}

We recall that the loop corrections to the four-point correlation function  \p{intriLoops} are described by scalar functions $F^{(\ell)}(x_i)$ given by $\ell-$fold integrals, see \p{integg}. At one loop 
the relation \p{integg} reads
\begin{align}\label{F-1loop} 
  F^{(1)}(x_1,x_2,x_3,x_4)={x_{12}^2 x_{13}^2 x_{14}^2 x_{23}^2 
  x_{24}^2 x_{34}^2\over (-4\pi^2)} \int  d^4x_5  \,  f^{(1)}(x_1,  \dots , x_{5})\,,
\end{align}
where the integrand ($f-$function) is given by \p{f1}:
\begin{align} 
 f^{(1)}(x_1,\ldots,x_5) &= {c^{(1)} \over \prod_{1\leq i<j \leq 5}   x_{ij}^2} \,.
\label{2.1}
\end{align}

In Eq.~\p{2.1}, the product of distances in the denominator has the required conformal
weight  $(+4)$ at each point, whereas the numerator has weight zero, so it is a constant $c^{(1)}$. This is the only 
constant in our whole construction whose value we cannot fix without an explicit Feynman 
graph calculation of the correlation function, or comparison with known values of anomalous 
dimensions from the OPE data. Substituting \p{2.1} into \p{F-1loop} we notice that in the 
expression for the correlation function \p{cor4loop} the constant $c^{(1)}$ is accompanied 
by the coupling constant $a c^{(1)}$. Therefore, this constant can be 
absorbed into the definition of the coupling constant. For this reason we are free to make 
the choice
\begin{align}\label{2.2}
c^{(1)}=1\,.
\end{align} 

The $f-$function \p{2.1} is depicted  in Fig.~\ref{Fgr} as a graph,
in which each vertex is connected to the remaining four vertices by four lines.  
What this graph represents is just the product of propagators connecting five points (the four external points $x_1,\ldots,x_4$ and the single integration point $x_5$). It is easy
to see that the graph $f^{(1)}$  in Fig.~\ref{Fgr} is non-planar (in fact, it has genus
${\rm g}=1$ according to the Euler-Poincar\'e  formula). This may seem surprising because the one-loop
Feynman diagrams contributing to the correlation function and, hence, to the function
$F^{(1)}(x_i)$ are all planar. We should keep in mind however that the integral
defined by the one-loop $f-$graph,
$ \int  d^4x_5  \,  f^{(1)}(x_1,  \dots , x_{5})$ is not yet the Feynman integral appearing in the one-loop correlation function. According to \p{F-1loop}, 
the latter is obtained after multiplication with the universal prefactor 
$x_{12}^2x_{13}^2 x_{14}^2 x_{23}^2x_{24}^2 x_{34}^2$. This factor removes the six lines in 
the graph $f^{(1)}$ in Fig.~\ref{Fgr} connecting the four external points, leaving us with the familiar one-loop ``cross" integral:
\begin{align}\label{2.3} 
F^{(1)}(x_1,x_2,x_3,x_4) \equiv g(1,2,3,4)  
 = - \frac1{4\pi^2} \int \frac{d^4x_5}{x^2_{15}x^2_{25}x^2_{35}x^2_{45}} \,,
 \end{align}  
where the notation $g(1,2,3,4)$ is introduced for later convenience. 
Its explicit expression in a manifestly conformal form is \cite{davussladder} 
\begin{align}\label{2.3'}
F^{(1)}(x_1,x_2,x_3,x_4) =  \frac{1}{4 x^2_{13} x^2_{24}} \int_0^1 d\xi\ \frac{\ln(v/u) + 2
\ln\xi}{v   \xi^2 + (1 - u - v)   \xi + u}\,.
\end{align}
The relation \p{2.3} is in agreement with the well-known result \cite{GonzalezRey:1998tk,Eden:1998hh} for the one-loop correction 
to the four-point correlation function of half-BPS operators. 

We would like to emphasize that the one-loop case is exceptional in the sense that the 
topology of the $f-$graph is non-planar while its contribution to the correlation function is 
planar. As we argue in Appendix~\ref{app:genus},   the planar limit at higher loops
always gives planar $f-$graphs. The latter constitute a very small subset of all possible 
$f-$graph topologies, providing an extremely useful reduction of the basis in the planar theory.

\subsubsection{Logarithmic singularities}\label{Los}

The integral in \p{2.3} is well defined and conformally covariant provided that the external points are kept apart, $x^2_{ij}\neq 0$ (with $i,j=1,\ldots,4$). However, the integral develops logarithmic singularities when this condition is not satisfied. Understanding the origin of these singularities provides us with key information for constructing the correlation function at higher loops. This is why we wish to illustrate  the phenomenon first at one loop. 

One way to create logarithmic singularities of the integral \p{2.3}
 is to go to the Euclidean double short-distance limit $x_1\to x_2$ and $x_3\to x_4$,
 {or equivalently $u\to 0$ and $v\to 1$}.  As was explained in Sect.~\ref{Sect:OPE}, in this regime the correlation function should have the asymptotic behavior \p{SD}. At one loop,  \p{SD} becomes
\begin{align}\label{2.4}
  x_{13}^4   F^{(1)}(x_i) \xrightarrow{u\to 0 \atop v\to 1} \frac1{12} \gamma^{(1)}_\mathcal{K}\ln u +O(u^0) \,, 
\end{align}
where $\gamma^{(1)}_\mathcal{K}$ is the one-loop anomalous dimension of the Konishi operator. Indeed, using \p{2.3'}, it is easy to obtain 
\begin{align}\label{davphiE}
 x_{13}^4 F^{(1)}(x_i) \ \stackrel{u\to 0 \atop   v \to 1}{\longrightarrow }   \   \frac{1}{4} \ln u    +   O(u^0)\,. 
\end{align}
Comparing the coefficient in front of the logarithm, we can 
extract the one-loop Konishi anomalous dimension, $\gamma^{(1)}_\mathcal{K}=3$.

In practice, we don't need to  know the explicit form \p{2.3'} of the function $F^{(1)}(x_i)$  in order to see the origin of the logarithmic singularity in the one-loop integral  \p{2.3}.  {Let us the examine expression for $x_{13}^4 F^{(1)}(x_i)$ in the limit $x_1\to x_2$ and $x_3\to x_4$:} 
\begin{align}\label{2.5}
\lim_{x_1\to x_2 \atop x_3\to x_4}\int \frac{d^4x_5\,(x_{13}^2)^2}{x^2_{15}x^2_{25}x^2_{35}x^2_{45}} & \sim \int \frac{d^4x_5\,(x_{13}^2)^2}{(x^2_{15})^2 (x^2_{35})^2}  
 \sim \int_{x_{15}^2\ll x_{13}^2} \frac{d^4x_5}{(x^2_{15})^2}+\int_{x_{35}^2\ll x_{13}^2} \frac{d^4x_5}{(x^2_{35})^2} \,.  
\end{align}
{Here in the first relation we interchanged the integration with taking the limit. This 
renders the integral divergent thus allowing us to identify the integration regions producing
logarithmic singularities}. Indeed, it is easy to see that the logarithmic singularities of the resulting integral 
originate from the two regions where the integration point $x_5$ approaches one of the 
coincident outer points, $x_5\to x_1$ and $x_5\to x_3$. In the last relation in \p{2.5}
we separated the contribution of these two regions. It is straightforward to verify that 
the sum of the two integrals on the right-hand side of \p{2.5} has the required simple log divergence \p{2.4}. 

An alternative singular regime is obtained by considering the correlation function \p{2.3} in Minkowski space-time and by taking the light-cone limit, in which the external points become sequentially light-like separated, $x_{12}^2,x_{23}^2,x_{34}^2,x_{41}^2\to 0$. In this case, for $u,v\to 0$, the log of the correlation function is expected to have the logarithmic scaling \p{LC} to all loops. Indeed, in the light-cone limit $u,v \to 0$, the one-loop function  \p{2.3'} behaves as follows:
\begin{align}\label{davphiM}
 x_{13}^2 x_{24}^2  F^{(1)}(x_i)   \ \stackrel{u,v\to 0}{\longrightarrow }   \     - \frac{1}{4} \ln u \ln v   -
\frac{\pi^2}{12}   + O(u)+O(v) \,,
\end{align}
in agreement with \p{LC} at the lowest
order in the coupling. 

Let us examine more closely the origin of this logarithmic singularity (following Ref.~\cite{EKS2}). In the light-cone limit,  the four external points $x_1,\ldots,x_4$ define the vertices
of a light-like rectangle. Then, the singularities of the integral in \p{2.3} originate 
from the $x_5-$integration over the region, in which  $x_5$ approaches the light-like edges 
$[x_i,x_{i+1}]$ of the rectangle. For instance, if $x_5$ approaches the light-like segment $[x_1,x_2]$,
\begin{align}\label{lc-iden}
x_5^\mu \to (1-\alpha) x_1^\mu+ \alpha x_{2}^\mu \qquad \Rightarrow\qquad x_{i5}^2 \to (1-\alpha) x_{1i}^2 + \alpha x_{2i}^2\,,
\end{align}
then two of the factors in the denominator in \p{2.3} vanish simultaneously, $x_{15}^2,x_{25}^2\to 0$ for arbitrary $\alpha$, thus producing a single logarithmic asymptotic behavior $\sim \ln x^2_{12}$. 
Moreover, since   $x_{45}^2\to \alpha x_{24}^2$,  an additional singularity $\sim \ln x^2_{41}$ appears for  $\alpha\to0$ and the integral acquires the double logarithmic singularity $\sim (\ln x^2_{12} \ln x^2_{41})$. In a similar
manner, the double logarithmic singularity $\sim (\ln x^2_{12} \ln x^2_{23})$ arises from integration around $\alpha=1$. These double-logs appear in the term  $\ln u \ln v$  in \p{davphiM}.

Another way to see this is to rewrite the integral in \p{2.3} in terms of dual momentum variables
 $p_i=x_i-x_{i+1}$ (with $i=1,\ldots,4 \ {\rm mod} \ 4$) and  $k=x_{15}$:
\begin{align}\label{box}
 F^{(1)}(x_i) = -\frac1{4\pi^2}\int \frac{d^4 k}{k^2(k-p_1)^2(k-p_1-p_2)^2(k+p_4)^2}\,.
\end{align}
In the light-cone limit, $x_{i,i+1}^2\to 0$, the dual momenta become light-like, $p_i^2=0$,
and the integral can be identified (after modifying the integration measure
$d^4 k \to d^D k$) with the one-loop box integral defining the one-loop correction to the four-particle amplitude $\cM_4$, Eq.~\p{M}. The coincidence is not accidental, of course. It
is a manifestation of the duality between scattering amplitudes and correlation function 
in the light-cone limit, Eq.~\p{F-M}. 
We recall that the duality is understood  at the level of the finite and rational four-dimensional {\it integrands} rather than the divergent {\it integrals}. 

Coming back to the integral \p{box}, we observe that the light-cone divergence in question can be assimilated to an infrared singularity for the dual $k-$momentum. It is well known that the infrared divergences of the massless scalar box originate from integration over the loop momentum $k$ collinear with the light-like momenta $p_i$ of the external legs. For instance, for 
$k$ collinear with $p_1$ we have $k^\mu=\alpha p_1^\mu$.  In terms of  the $x-$variables, this corresponds to the limit where the integration point $x_5$ approaches  the light-like segment $[x_1,x_2]$, that is $x_5^\mu\to (1-\alpha)x_1^\mu+\alpha x_2^\mu$, in agreement with the discussion above.

\subsection{From one to two loops}

\subsubsection{The two-loop integrand}

According to \p{integg}, the two-loop correction to the correlation function takes the form
\begin{align}\label{F-2loop} 
  F^{(2)}(x_1,x_2,x_3,x_4)={x_{12}^2 x_{13}^2 x_{14}^2 x_{23}^2 
  x_{24}^2 x_{34}^2\over 2!\,(-4\pi^2)^2} \int d^4x_5   d^4x_6 \,  f^{(2)}(x_1,  \dots , x_{6})\,.
\end{align}
Here the function $f^{(2)}(x_1,  \dots , x_{6})$ is invariant under $S_6$ permutations of the
six points and is given by \p{f-2loop}. It is depicted in Fig.~\ref{Fgr} as a connected octahedron graph with $6$ vertices of degree $4$  (the number of lines leaving each vertex).
Notice that, in distinction with the one-loop $f-$graph shown in Fig.~\ref{Fgr}, the 
two-loop octahedron $f-$graph is planar (i.e., of genus 0). {As
  already mentioned, this continues to hold for all higher loops according to
  the argument in Appendix~\ref{app:genus}.}
 
Replacing the function $f^{(2)}$ in \p{F-2loop} by its explicit expression \p{f-2loop},
we can expand $F^{(2)}$ into a sum of conformally covariant scalar two-loop integrals:
\begin{align}\label{2.9}
 F^{(2)}   = c^{(2)}\bigg(&   h(1,2;3,4) + h(3,4;1,2) + h(1,4;2,3)+ h(2,3;1,4)    
 \nt  + & h(1,3;2,4) +h(2,4;1,3) +
 \frac12 \lr{x_{12}^2x_{34}^2+ x_{13}^2 x_{24}^2+x_{14}^2x_{23}^2}  [g(1,2,3,4)]^2  \bigg) \,,
\end{align}
where the one-loop integral $g(1,2,3,4)$ was defined in \p{2.3} and the notation
\begin{align}\label{eq:g+h}
h(1,2;3,4)  & =  \frac{x^2_{34}}{(4 \pi^2)^2}
\int \frac{d^4x_5 \, d^4x_6}{(x_{15}^2 x_{35}^2 x_{45}^2) x_{56}^2
(x_{26}^2 x_{36}^2 x_{46}^2)}\nt
& =  -\frac{1}{32\; x^2_{13} x^2_{24}} \int_0^1 d\xi\ \frac{(\ln(v/u) + 2
\ln\xi)(\ln(v/u) +  
\ln\xi) \ln\xi}{v   \xi^2 + (1 - u - v)   \xi + u}  
\end{align}
 was introduced 
for the two-loop ladder integral \cite{davussladder}.
The remaining $h-$integrals in \p{2.9} are obtained from $h(1,2;3,4)$ by permuting the indices of 
the external points.

\subsubsection{Fixing the constant from the logarithmic singularities}\label{ficolo}

What remains undetermined by the symmetry properties is the overall normalization $c^{(2)}$ on the right-hand side of \p{2.9}. We need some additional information to fix it. As was already mentioned in Sect.~2,  this information comes from the analysis of the correlation function in either of the two  singular limits considered in Sect.~\ref{Sect:OPE}. 

Let us start with the double short-distance limit \p{lim1}. Expanding the expression in the left-hand side of \p{SD} to terms of order $a^2$, we find
\begin{align}\label{2.10}
  x_{13}^4\left[  F^{(2)} - 3 x^4_{13} (F^{(1)})^2\right] \
\stackrel{u\to 0 \atop v\to 1}{\longrightarrow} \
 \frac1{12}\gamma_{K}^{(2)} \ln u+O(u^0)\,,
\end{align}
where $\gamma_{K}^{(2)}$ is the two-loop Konishi anomalous dimension.
This relation implies that the particular combination,  $F^{(2)} - 3 x^4_{13} (F^{(1)})^2$, of 
two- and one-loop integrals should have a {\it single-log} short-distance singularity.
Replacing $F^{(1)}$ and $F^{(2)}$ by their explicit expressions, Eqs.~\p{2.3} and  \p{2.9}, 
respectively,  we
find~\footnote{Note that $h(1,2;3,4)$ does not contribute to this relation since it is proportional to $x_{34}^2$ and,
therefore, vanishes in the short-distance limit. }
\begin{align}\label{2.11}
\lim_{x_1\to x_2 \atop x_3\to x_4} \left[F^{(2)} - 3 x^4_{13} (F^{(1)})^2\right]\ \sim \ x_{13}^2 \int d^4 x_5 d^4 x_6 \frac{2c^{(2)}(x_{15}^2 x_{36}^2 + x_{16}^2 x_{35}^2) + (c^{(2)}-3) x_{13}^2 x_{56}^2}{(x_{15}^4x_{16}^4)x_{56}^2(x_{35}^4x_{36}^4) } \,,
\end{align}
where the integrand is manifestly symmetric under the exchange of the integration points $x_5$ and $x_6$. 
Like in the one-loop case \p{2.5},  {we swapped the integration and  the limit  on the right-hand side of \p{2.11}.} 
Similarly to \p{2.5},  the integral \p{2.11} develops ultraviolet (short-distance) 
divergences when the integration points $x_5$ and $x_6$ independently approach one 
of the external points $x_1$ and $x_3$.  For arbitrary $c^{(2)}$, the integration over each of the points $x_5$
and $x_6$ produces a single-log divergence, so that the right-hand side of 
\p{2.11} has a double-log divergence. In order to lower the degree of divergence of
\p{2.11} to the single log required by \p{2.10},  we have to make sure that the numerator of the integrand in \p{2.11} vanishes when, e.g., $x_5\to x_1$ and $x_6$ remains in a generic position:
\begin{align}\label{2.12}
\lim_{x_5\to x_1} [2c^{(2)}(x_{15}^2 x_{36}^2 + x_{16}^2 x_{35}^2) + (c^{(2)}-3) x_{13}^2 x_{56}^2] = 3(c^{(2)}-1)x_{13}^2 x_{16}^2 =0  \,,
\end{align}
leading to 
\begin{align}\label{valc2}
 c^{(2)}=1\,.
\end{align}
We would like to emphasize that the condition \p{2.12} only lowers, but does not completely remove the logarithmic singularity of the integral in \p{2.11}. Indeed, for $c^{(2)}=1$ the integral in \p{2.11} has a (single-log) divergence originating from the regions
where both integration points approach the external points simultaneously, $x_5,x_6\to x_1$ and $x_5,x_6\to x_3$. To see this, we examine the contribution to \p{2.11}
from the integration region $x_5,x_6\to x_1$
\begin{align}\label{resdiv}
\mbox{r.h.s. of  \p{2.11}} \sim \frac1{x^4_{13}}  \int_{x_{5}, x_{6} \to x_1} d^4 x_5 d^4 x_6 \frac{x_{51} \cdot x_{61}}{x_{51}^4 x^4_{61} x^2_{56}}\,,\end{align}
and observe that the integral clearly has a log divergence, as can be seen by power counting. 

Thus, the condition on the correlation function to have a single logarithmic asymptotic behavior in the Euclidean OPE limit, Eq.~\p{SD}, allows us to fix the only remaining two-loop constant $c^{(2)}$ in \p{2.9}. The two-loop correlation function we have constructed in this way agrees with the results of the explicit perturbative calculations \cite{Eden:2000mv,Bianchi:2000hn}. 

Alternatively, we may impose the condition for the correlation function to have a
double-log singularity \p{LC} in the light-like limit \p{lim2}. In this case we are considering the expansion of the log appearing in the left-hand side of \p{LC},  
\begin{align}\label{2.13}
\lim_{x_{i,i+1}^2\to 0} \ln (G_4/G^{(0)}_4) & \sim \ \ln\lr{1 + 2 a \cF^{(1)} + 2 a^2 \cF^{(2)} + O(a^3) } \nt 
& = 2a \cF^{(1)} + 2 a^2 [ \cF^{(2)} - (\cF^{(1)})^2] + O(a^3)\,,
\end{align}
where $\cF^{(1,2)} = \lim_{x_{i,i+1}^2\to 0} (x^2_{13} x^2_{24} F^{(1,2)})$. 
 Using the expressions  for the integrals in \p{2.3'} and \p{eq:g+h} as functions of the cross-ratios $u,v$  and performing the asymptotic expansion in the limit $u,v \to 0$, we obtain (cf. Eq.~\p{davphiM})
\begin{align}\label{2lphi} \notag
 \cF^{(2)} - (\cF^{(1)})^2 \, =\, &\frac1{48} [\pi ^2     \left(\ln u \ln
   v+\ln ^2u+\ln ^2v\right)  \\[2mm]
   & +  (c^{(2)}-1) \left( 3 \ln ^2u \ln ^2v+3
   \pi ^2 \ln u \ln v+  \pi ^2 \ln ^2u+  \pi ^2 \ln ^2v \right) ] + \ldots \ ,
\end{align}
where the dots stand for less singular terms. 
As expected, the second line in this relation contains the leading double-log singularity $(\ln u \ln v)^2 \sim (\log)^4$, proper to a generic two-loop conformal integral. However,  choosing the correct value  $c^{(2)}=1$ as in \p{valc2}  removes this leading term and lowers the degree of singularity according to the general formula  \p{LC}. 

Let us now try to identify the source of the double-log singularity at the level of the integrand. Using 
\p{2.3} and  \p{2.9}, we find in the light-cone limit  
\begin{align}\label{2.14}
    \cF^{(2)} - (\cF^{(1)})^2\sim & \int d^4 x_5 d^4 x_6 \frac{x_{13}^2 x_{24}^2}{x_{15}^2 x_{25}^2 x_{35}^2 x_{45}^2x_{16}^2 x_{26}^2 x_{36}^2 x_{46}^2 x_{56}^2}
    \\[2mm] \notag
   & \times  { [(c^{(2)}-2 ) x_{13}^2 x_{24}^2 x_{56}^2+ c^{(2)} x_{13}^2 (x_{25}^2x_{46}^2+x_{45}^2x_{26}^2) +c^{(2)} x_{24}^2(x_{15}^2x_{36}^2+x_{35}^2x_{16}^2) ]} \,,
\end{align}
with $\cF^{(\ell)} = \lim_{x_{i,i+1}^2\to 0}  x^2_{13} x^2_{24} F^{(\ell)}$.
As before, to reveal the structure of light-cone singularities, here we interchanged the integration with
taking the light-cone limit $x_{i,i+1}^2\to 0$.
As in the one-loop case, the light-cone divergences of the integral come from the region where the integration points $x_5$ and/or $x_6$ approach independently the light-like
edges of the rectangle with vertices located at the points $x_1,\ldots,x_4$. 
For arbitrary  $c^{(2)}$ the integration over this region in \p{2.14} produces a $(\log)^4$ singularity. As before, to lower this singularity down to $(\log)^2$ we require that 
the expression in the second line of \p{2.14} should vanish for $x_5^\mu \to (1-\alpha)x_1^\mu +\alpha x_2^\mu$ and $x_6$ in a general position. Applying \p{lc-iden} we find 
that this condition takes the form
\begin{align}\label{faco}
 (c^{(2)}-1) x_{13}^4 x_{24}^4 \left[ (1-\alpha) x_{16}^2  + \alpha x_{26}^2\right]=0 \,,
\end{align}
with $\alpha$ arbitrary, leading to $c^{(2)}=1$.~\footnote{The two-loop case is
  exceptional in the sense that the dependence on the single unknown
  coefficient $c^{(2)}$ factorizes  in the light-cone condition
  \p{faco}. This might suggest that it is sufficient to impose the
  constraint for just one value of $\a$, say, $\a=0$ (the integration
  point $x_5$ approaching the end point of the segment
  $[x_1,x_2]$). However, at higher loops, where we have to determine
  several coefficients, we will indeed need the full power of the constraint for
  {\it arbitrary} $\a$. }
As before, this condition only softens the light-cone 
 singularity of the integral \p{2.14}.   Indeed, for $c^{(2)}=1$ the integral in \p{2.14} develops
a (double-log) singularity when $x_5$ and $x_6$ approach the same light-like edge.
Thus, the condition for the correlation function to have the correct double log asymptotics in the Minkowski light-like limit \p{LC} leads to the same value of the constant $c^{(2)}$ as we found in the Euclidean short-distance limit above. 

We remark that the qualitative argument above only points at a natural condition for lowering the singularity of the integral. It does not guarantee, at higher loops, that the singularity will be lowered  sufficiently, in order to obtain the desired asymptotics  \p{LC}. Nevertheless, later on in the paper we will see that this simple condition turns out to  also be {\it sufficient}, at least in what concerns the planar limit of the correlation function.  

We conclude that the two conditions, Eqs.~\p{SD} and \p{LC}, are equally powerful at two loops and they allow us to completely determine the four-point correlation function, without having done any Feynman graph calculations. However, as we will show in Sect.~\ref{Pls},
 at higher loop levels the Minkowski condition turns out to be more powerful than the Euclidean one. It will allow us to fix all undetermined coefficients in the planar sector and to greatly reduce the number of constants in the non-planar sector. The Euclidean condition is somewhat weaker and leaves some arbitrary constants even in the planar sector, starting with four loops. 
 
\subsubsection{Relation to the two-loop amplitude}\label{rel2loam}

There exists a third way of fixing the two-loop coefficient $c^{(2)}$.\footnote{This third alternative was used in \cite{Eden:2011we}.}  It relies on the conjectured duality \p{F=AA} between correlation functions and amplitudes
in planar $\cN=4$ SYM. Up to two loops it reads (recall \p{M-F})
\begin{align}\label{spec1}
 \cM^{(1)} = \cF^{(1)}\,,
 \qqquad
 \cM^{(2)} = \cF^{(2)} -\frac12 \lr{\cF^{(1)}}^2\,,
 \end{align}   
 where $ \cM^{(1,2)}$  denote the two-loop corrections to the four-particle
 amplitude \p{M} and $\cF^{(1,2)}$ are obtained from the two-loop correlation function
 in the light-cone limit \p{F-hat}. 

Substituting \p{2.3} and \p{2.9} into \p{F-hat} we get from the duality relation \p{spec1}
the following prediction for the amplitude
\begin{align}\notag
 \cM^{(1)} &= x_{13}^2 x_{24}^2\ g(1,2,3,4)\,,
 \\[2mm] \label{spec2}
  \cM^{(2)} &=  x_{13}^2 x_{24}^2\,[ h(1,3;2,4) + h(2,4;1,3) ]\,.
\end{align}
Notice that the term $[g(1,2,3,4)]^2$ cancels in the expression for $\cM^{(2)}$ only for $c^{(2)}=1$.
We recall that these relations should be understood at the level of the four-dimensional
{\it integrands}. Indeed, Eq.~\p{spec2} is in perfect agreement with the known one- and two-loop results for the
planar four-particle scattering amplitude in $\mathcal{N}=4$ SYM \cite{Green:1982sw,Bern:1997nh,Bern:1998ug}.

Now, we can turn the duality relations \p{spec1} around and use them to derive $c^{(2)}=1$. To this end, we rewrite 
the second relation in \p{spec1} as (recall \p{F-M}) 
\begin{align}\label{rung-2loop}
\cF^{(2)}=\frac12 \lr{\cF^{(1)}}^2+\cM^{(2)}\,.
\end{align}
We start with the important observation that, in virtue of the $S_6-$permutation
symmetry of the two-loop integrand $f^{(2)}(x_1,\ldots,x_6)$, it is sufficient to know the coefficient
of just one of the terms on the right-hand side of \p{f-2loop}, in order to be able to reconstruct the
whole expression. For this purpose, it is convenient to choose the term $\frac12 \lr{\cF^{(1)}}^2$.
Its integrand does not contain a factor $x_{56}^2$ depending on the two integration points 
simultaneously, so its contribution to the two-loop correlation function \p{F-2loop} is factorized
into the square of the one-loop contribution $F^{(1)}$, Eq.~\p{2.3}. Going to the light-cone limit
and making use of \p{F-hat}, we find that the contribution of this term to the two-loop
correlation function takes the form $\cF^{(2)}= c^{(2)} \lr{\cF^{(1)}}^2/2+\ldots$. Here the dots
denote the remaining terms on the right-hand side of \p{f-2loop} whose contribution
is not factorizable into a product of one-loop integrals. Comparing
the last relation with \p{rung-2loop}, we immediately conclude that $c^{(2)} =1$ provided that
$\cM^{(2)}$ does not contain terms $\sim \lr{\cF^{(1)}}^2$. This is guaranteed not to happen because
$\cM^{(2)}$ is the two-loop contribution to the four-particle amplitude and, due to the unitarity
conditions, it is given by a sum of irreducible two-loop integrals.

\subsubsection{Rung rule}\label{rr} 
 
As mentioned in Sect.~\ref{sec:rung-rule}, the two-loop result \p{2.9} can be obtained in a very straightforward manner from the one-loop result \p{2.3} (including the precise coefficients) by making use of the amplitude/correlation function duality. This yields a procedure equivalent to the rung-rule procedure for amplitudes, and in this context generates the single $f-$graph $f^{(2)}$ by gluing two one-loop $f-$graphs together. Graphically, one obtains the octahedron in Fig.~\ref{fig:bodies} from gluing two pyramids together. 

The key to this graphical procedure is the duality relation in the form \p{rung-2loop}. On the right-hand side we observe two terms, one is factorized into lower-loop terms, $\frac12 \lr{\cF^{(1)}}^2$, and the other is irreducible (a genuine two-loop graph), $\cM^{(2)}$. We want to consider the $f-$graphs corresponding to the two $\cF-$terms in \p{rung-2loop}. Schematically, the relation between  the two types of terms is given by 
\begin{align}\label{Ff}
f^{(1)} = \frac{\cF^{(1)}}{x^2_{12}x^2_{23}x^2_{34}x^2_{41} (x^2_{13}x^2_{24})^2}\,,
\end{align}
where we ignore for the time being the light-like limit $x^2_{i,i+1}\to 0$. The completion of the one-loop $\cF-$graph to the $f^{(1)}-$graph is shown in Fig.~\ref{fF}. The denominator in \p{Ff} can be viewed as a frame of solid lines connecting the four external points $1,\ldots,4$. Superposing this frame onto the $\cF^{(1)}-$graph, there is partial cancellation of solid (denominator) versus dashed (numerator) lines.

\begin{figure}[h!]
\psfrag{F1}[cc][cc]{$\cF^{(1)}$}
\psfrag{f1}[cc][cc]{$f^{(1)}$}
  \centering
  \includegraphics[width=0.6\linewidth]{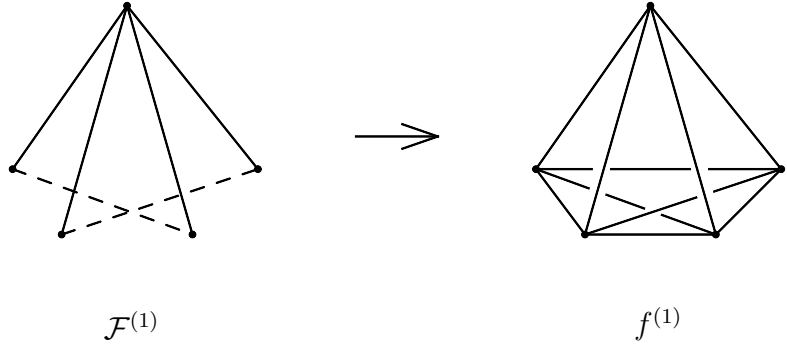}
  \caption{  Obtaining the graph $f^{(1)}$ from  the graph $\cF^{(1)}$. The apex of  the pyramid $\cF^{(1)}$ corresponds to the integration point, the four legs stand on the external points. The dashed lines are numerator factors canceling the conformal weight of  $\cF^{(1)}$ at the external points. }
  \label{fF}
\end{figure}

Next, let us do the same with the factorized term $\frac12 \lr{\cF^{(1)}}^2$ in the duality relation \p{rung-2loop}. The result is shown in Fig.~\ref{FFf}. Here we have chosen to attribute the entire frame of solid lines to one of the factors $\cF^{(1)}$, thus making it into $f^{(1)}$. The other factor $\cF^{(1)}$ is left in its original form. Gluing the two objects together and deleting the pairs of overlapping solid/dashed lines, we obtain precisely the octahedron representing the  $f^{(2)}-$graph. What we have seen here is an example of iteration of the one-loop $f-$function to two loops. 

\begin{figure}[h!]
\psfrag{FF1}[cc][cc]{$(\cF^{(1)})^2$}
\psfrag{Ff1}[cc][cc]{$\cF^{(1)} f^{(1)}$}
\psfrag{f2}[cc][cc]{$f^{(2)}$}
  \centering
  \includegraphics[width=0.8\linewidth]{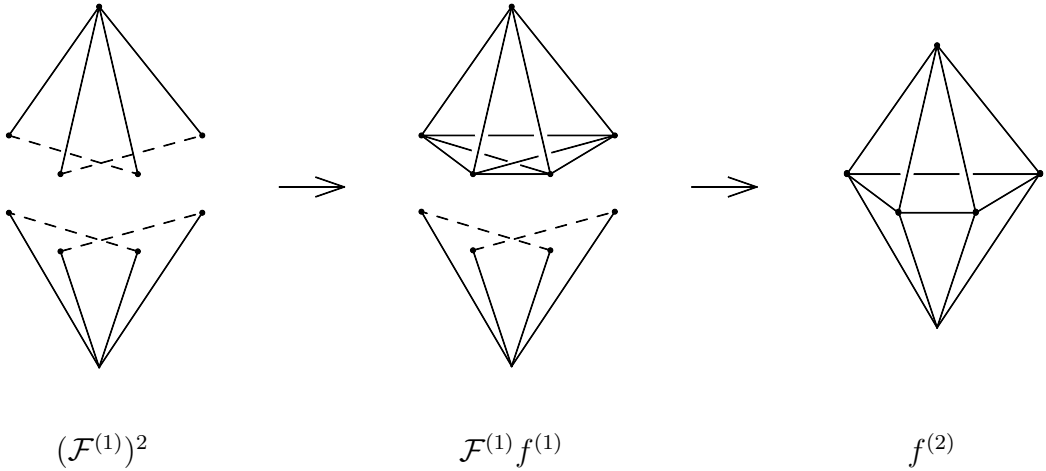}
  \caption{  Graphical representation of the  term $\frac12 \lr{\cF^{(1)}}^2$ and of the corresponding $f-$graph.  }
  \label{FFf}
\end{figure}

This very simple example shows how to proceed to higher loops. We start by recasting the duality relation  \p{eq:2} in the form
\begin{align}\label{duFF}
\cF^{(\ell)} = \cF^{(1)}\times \cF^{(\ell-1)} + \ldots\,,
\end{align}
which is achieved by replacing all lower-loop amplitude $\cM-$terms by $\cF-$terms, according to the inverse relations \p{M-F}.  We want to use the ``gluing" of the one-loop factor $\cF^{(1)}$ to the $(\ell-1)-$loop  factor $\cF^{(\ell-1)}$ as a means of generating new $\ell-$loop topologies $f^{(\ell)}$. At this point we have to remember  that the relation between the functions $f^{(\ell)}$ and $\cF^{(\ell)}$ involves  the light-cone limit $x_{12}^2,x_{23}^2,x_{34}^2,x_{41}^2\to 0$. According to \p{integg}, in this limit $\cF^{(\ell)}$ defined in \p{F-hat}  receives a non-vanishing
contribution only from the terms in $f^{(\ell)}$ which include $1/(x_{12}^2 x_{23}^2x_{34}^2x_{41}^2)$. This denominator factor is needed to cancel out the vanishing prefactor on the right-hand side of \p{integg}. Inspecting the cross-term shown in \p{duFF}, we see that this denominator factor should come from  the part of $f^{(\ell-1)}$ whose restriction to the light cone appears in ${\cF^{(\ell-1)}}$.
In terms of graphs, this means that the relevant $f^{(\ell-1)}-$graphs should have a closed path of length four 
going through the external points $1,2,3,4$. For planar $f-$graphs (which we can view as the edges of a polyhedron) such a 4-cycle has the interpretation of the boundary of  either a rectangular face or of two triangular faces glued along the common edge.\footnote{There are also closed paths of length four lying inside the polyhedron, that is, which split the polyhedron into two pieces. These correspond to product amplitudes contained in $\cF^{(\ell-1)}$ but which cancel in $\cM^{(\ell-1)}$ (see~(\ref{F-M}), (\ref{M-F})).}

Next, we need to upgrade the $f^{(\ell-1)}-$graphs with such faces to
$f^{(\ell)}-$graphs. This is done by gluing to them a square pyramid with two dashed line diagonals at its base, corresponding to the factor  ${\cF^{(1)}}$ in \p{duFF},  along the common path of length four. Four of the pyramid's vertices correspond
to the external points and the fifth one to the new integration point
$x_\ell$ of the $f^{(\ell)}-$graph. The  two diagonals
at the base of the ${\cF^{(1)}}-$graph are numerator factors which keep the balance of
conformal weights at the external  (gluing) points.     The two possible configurations are illustrated in the top line of
Fig.~\ref{fig:rungrule}. In the first case, the base of the pyramid is glued to two triangular faces with a common edge. Here one of the dashed lines of the ${\cF^{(1)}}-$graph deletes this common edge, the other remains as a numerator factor.  In the second case the pyramid is glued to a rectangular face, so both numerator factors from ${\cF^{(1)}}$ survive. The latter case occurs for the first time in the iteration from four to five loops, see Sect.~\ref{sec:five-looops}.

\begin{figure}[h!]
\psfrag{a}[cc][cc]{}
\psfrag{b}[cc][cc]{}
  \centering
  \includegraphics[width=.6\linewidth]{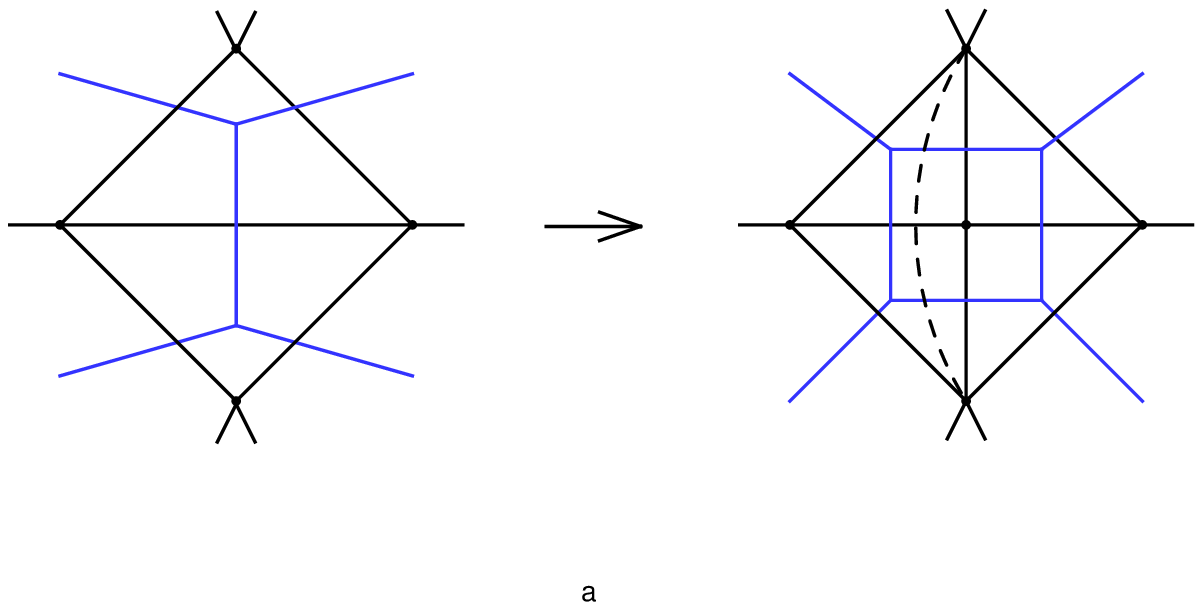}\\[-5mm]
  \includegraphics[width=.6\linewidth]{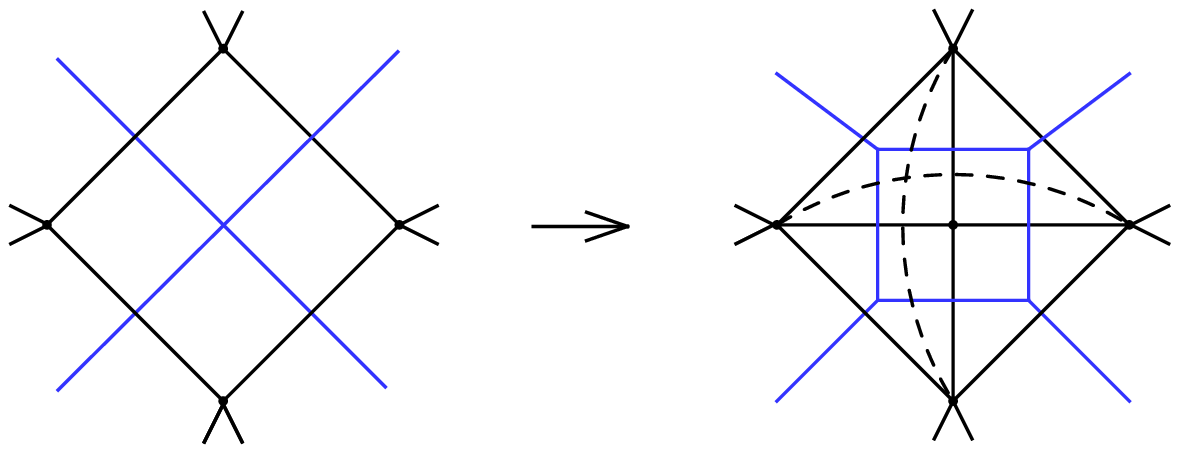}
  \caption{Figures illustrating the rung rule in both position space (black solid lines) and in momentum space (blue lines). In position space, for any square one inserts a
    vertex with four edges coming from it together with two numerator (dashed)
    lines. In the first case one of these numerator lines simply
    cancels the diagonal propagator line.  The momentum space graphs
    give  the well-known rung rule for obtaining higher loop amplitude
    integrands from lower ones. }
  \label{fig:rungrule}
\end{figure} 
 
Fig.~\ref{fig:rungrule} illustrates another important aspect of our graphical iteration procedure. As mentioned in Sect.~\ref{sec:rung-rule}, it is equivalent to the well-known ``rung rule" \cite{Bern:1997nh} for iteration of planar loop integrals 
for the four-particle amplitude. To see this in more detail, let us again use the simplest iteration from one to two loops. In Fig.~\ref{12boxes} we recall the familiar dual space description of the one- and two-loop scalar boxes. In fact, the two pictures here are incomplete, we have deleted the leftmost parts of the momentum space graph (the boxes are not closed on the left) and of the dual space graph (point 4 is missing). The rung rule in momentum space consists in adding an extra vertical line (``rung") and thus creating a new box. In dual space this amounts  to gluing a figure of the type of the  ${\cF^{(1)}}-$graph above to the points $1,2,3,5$ of the one-loop dual graph. This is what happens  in the top row of Fig.~\ref{fig:rungrule}, with the diagonal in the left-hand side figure being the analog of the segment $[x_2,x_5]$ in Fig.~\ref{12boxes}. The only difference between the two figures is the absence of the external  lines $[x_1,x_2]$ and $[x_2,x_3]$  in Fig.~\ref{12boxes}. The reason is that the dual graphs depicted  in Fig.~\ref{12boxes} are contributions to the $\cF-$terms, while in   Fig.~\ref{fig:rungrule} we have drawn $f-$graphs. The dashed line present after the iteration corresponds to the numerator factor known from the standard rung rule of Ref.~\cite{Bern:1997nh}. 

The configuration depicted in the bottom row of Fig.~\ref{fig:rungrule} appears for the first time in the iteration from four to five loops (see Sect.~\ref{sec:five-looops}) and upwards. The new feature here is the  square face of the polyhedron in the left-hand side figure.  It is clear that such a face cannot be obtained by the rung-rule procedure, i.e. by gluing a pyramid to a lower-loop polyhedron with triangular faces. At four loops, for the first time, we have a non-rung-rule topology with square faces. Upgrading it to a five-loop graph, we obtain the right-hand side figure in the bottom row of Fig.~\ref{fig:rungrule}. This is a generalization of the rung rule,
which has also been considered previously in the amplitude case, and
(for obvious reasons) is known as the ``square
insertion''~\cite{Bern:2007ct}.

\begin{figure}[h!]
\psfrag{x1}[cc][cc]{$x_1$}\psfrag{x2}[cc][cc]{$x_2$}\psfrag{x3}[cc][cc]{$x_3$}
\psfrag{x5}[cc][cc]{$x_5$}\psfrag{x6}[cc][cc]{$x_6$}
\psfrag{p1}[cc][cc]{$p_1$}\psfrag{p2}[cc][cc]{$p_2$}
  \centering
  \includegraphics[width=0.6\linewidth]{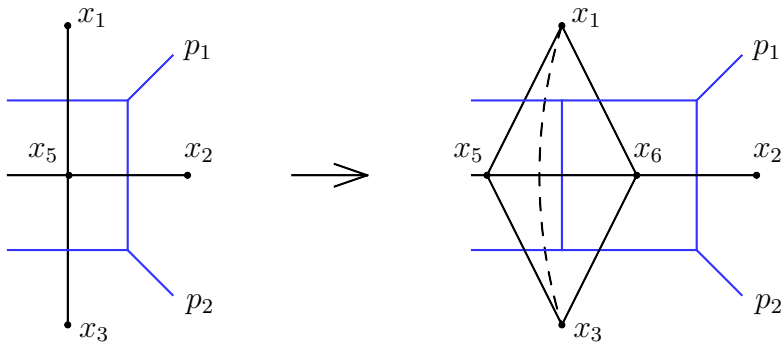}
  \caption{  The rung rule from one to two loops. We show both the momentum $p-$space (in blue) and the dual $x-$space (in black) graphs.}
  \label{12boxes}
\end{figure}

It is by now clear how to continue this iteration procedure and induce higher-loop $f-$terms from the known lower-loop ones. Further applications of this rule are shown in the following sections. We would like to emphasize that our procedure not only generates new $f-$graph topologies, but also fixes their coefficients. They are inherited from the lower-loop amplitude/correlation function via the cross-term in \p{duFF}.

One might wonder if we can consider further cross-terms appearing
in~(\ref{duFF}) in a similar way to obtain further higher-loop
terms. For example, from four loops on we see terms such as
$\cF^{(\ell)} \sim \cF^{(\ell-2)} \times \cF^{(2)}$. Geometrically this
would correspond to  gluing the two-loop octahedron $f-$graph to an
$(\ell -2)$-loop $f-$graph. However, we can see that the same
polyhedron could be obtained by consecutively gluing two pyramids to the $(\ell-2)$-loop $f-$graph. So, we find that this term will be picked up by the rung-rule procedure. In other words, considering further cross-terms gives us nothing new.

Another comment concerns the possibility  of having planar $f-$function topologies which could completely vanish in the light-cone limit and would thus not be detectable via the duality  $G=\cA^2$. For this to happen, the topology should not contain a 4-cycle, i.e. a subset of four vertices connected with solid lines. If such a 4-cycle is absent, the prefactor in \p{integg} will make this contribution vanish in the light-cone limit. An $f-$graph with no 4-cycle occurs for the first time at 11 loops, however this $f-$graph is non-planar. We have checked up to 15  loops and found no example of a planar $f-$graph which contains no 4-cycle. Of course, this does not prove that such graphs may not exist at even higher loop orders.

{As a final comment, we note that  imposing   the constraint \p{faco}
and its higher-loop counterparts only for $\alpha=0$ determines
precisely that part of the correlator which is also determined by the
rung rule up to five loops. At six loops it also enforces the
vanishing of certain non-rung-rule coefficients.  }

\section{Correlation function at three loops}\label{cftl}

In this section we reconstruct the three-loop correction to $G_4$ from Ref.~\cite{Eden:2011we}. According to Table ~\ref{Table1}, the three-loop integrand receives contributions from $f-$graphs of four different topologies. 
As was shown in \cite{Eden:2011we}, only one topology survives in the planar limit and its coefficient was fixed  there  
via the duality with the scattering amplitude. We show in this section that the three-loop
correlation function can be fully constructed without any reference to the amplitude. Instead, in order to fix the coefficients, we use the singular limits 
\p{lim1} and \p{lim2} discussed above. Starting from three loops we encounter a new feature: three of the $f-$graph 
topologies are non-planar and they may result in non-planar contributions to the correlation function.
However, the careful analysis of the singular limits shows that the non-planar contribution is actually vanishing at three loops. 

\subsection{The three-loop integrand}

According to \p{integg}, the three-loop correction to the correlation function takes the form
\begin{align}\label{F-3loop} 
  F^{(3)}(x_1,x_2,x_3,x_4)={x_{12}^2 x_{13}^2 x_{14}^2 x_{23}^2 
  x_{24}^2 x_{34}^2\over 3!\,(-4\pi^2)^3} \int d^4x_5   d^4x_6 d^4x_7\,  f^{(3)}(x_1,  \dots , x_{7})\,,
\end{align}
where the $S_7$ symmetric function $f^{(3)}$ can be read off  form \p{f-sum}:
\begin{align}\label{f3}
f^{(3)}(x_1,\ldots,x_{7}) = \sum_{\alpha=1}^{4} c^{(3)}_{\alpha} \, f_\alpha^{(3)}(x_1,\ldots,x_{7}) 
=  \sum_{\alpha=1}^{4} c^{(3)}_{\alpha}\frac{ P^{(3)}_\alpha (x_1,\ldots,x_{7})}{\prod_{1\le i< j \le 7} x_{ij}^2} \,.
\end{align}
Here the sum runs over $n_{\ell=3}=4$ different $P-$graph topologies and $c^{(3)}_{\alpha}$
are arbitrary coefficients.

According to the definition \p{P-cond}, the three-loop $P-$graphs are multigraphs with $7$ vertices of degree $2$.
These graphs are listed in Fig.~\ref{Fgr}. The corresponding expressions for $P^{(3)}_\alpha$
are
\begin{align}
  & P^{(3)}_1= \ft1{14} x_{12}^2 x_{23}^2 x_{34}^2
  x_{45}^2 x_{56}^2 x_{67}^2  x_{71}^2  \ +\ {S_7\ \mathrm{permutations}} , \nonumber
 \\[2mm]
  & P^{(3)}_2= \ft1{20}(x_{12}^2)^2( x_{34}^2 x_{45}^2 x_{56}^2 x_{67}^2 x_{73}^2)  \ +\ {S_7\ \mathrm{permutations}} , 
 \nonumber
 \\[2mm]
   & P^{(3)}_3= \ft1{48} (x_{12}^2  x_{23}^2 x_{31}^2) ( x_{45}^2
  x_{56}^2 x_{67}^2 x_{74}^2 ) \ +\ {S_7\ \mathrm{permutations}} ,  
  \nonumber
  \\[2mm]
  & P^{(3)}_4= \ft1{48} (x_{12}^2)^2( x_{34}^2)^2( 
   x_{56}^2 x_{67}^2 x_{75 }^2)  \ +\ {S_7\ \mathrm{permutations}} , \label{eq:11}
\end{align}
where the sum runs over the $S_7$ permutations of the indices $1,\ldots,7$ and the coefficients ensure that each distinct term appears only once in the sum. 

To obtain the most general integrand \p{f3}, we consider a linear combination of the four polynomials $P^{(3)}_\alpha$ with arbitrary coefficients  and then  divide it by the
product of all distances between the seven points. We observe that each factor in the expression for
the polynomials $P^{(3)}_1$ and $P^{(3)}_3$ cancels against a similar factor in the denominator leading
to the integrands $f^{(3)}_1$ and $f^{(3)}_3$  with numerators equal to 1. 
For the polynomials  $P^{(3)}_2$ and $P^{(3)}_4$, due to the presence of $(x_{\sigma_1\sigma_2}^2)^2$, the numerators of the corresponding integrands $f^{(3)}_2$ and $f^{(3)}_4$ contain an additional factor of $x_{\sigma_1\sigma_2}^2$.
The four $f-$graphs are obtained by discarding the numerators and focusing on the denominators only, with the result shown in Fig.~\ref{Fgr}.   For $f^{(3)}_1$ and $f^{(3)}_3$, all the vertices of the corresponding $f-$graphs have  
degree $4$, whereas for $f^{(3)}_2$ and $f^{(3)}_4$ some vertices have degree $5$. Such vertices have 
an excess of conformal weight, which is compensated by factors of 
$x_{\sigma_1\sigma_2}^2$ in the numerator for the integrand $f^{(3)}$. 

Examining the three-loop graphs shown in Fig.~\ref{Fgr}, we notice that only the graph $f^{(3)}_2$ is planar (i.e., has genus 0) whereas
the remaining three graphs have genus 1. One might wonder whether their contribution to the three-loop 
correlation function \p{F-3loop} could become planar after multiplication by the prefactor on
the right-hand side of  \p{F-3loop},  similarly to what happened at
one loop. Close examination shows that the three-loop non-planar
$f-$graphs might result in non-planar corrections to the correlation
function $F^{(3)}$. In fact, as we show below, at three loops this does not happen. The additional conditions on the singular behavior of the correlation function simply rule out the entire non-planar sector. We will have to wait till four loops for the first non-planar contributions to appear. 

\subsection{Fixing the coefficients from the logarithmic singularities}\label{3lcondi}

In this subsection we apply the criteria on the singular behavior of the correlation function, as explained in Sect.~\ref{ficolo}. Similarly to the two-loop case, we are able to completely fix the freedom in \p{f3}.

In the double short-distance limit \p{lim1}, expanding the log in the left-hand side of \p{SD}, we find  at level $a^3$
\begin{align}\label{3.1}
  x^4_{13} \left[F^{(3)} - 6 x^4_{13} F^{(1)} F^{(2)} +12 x^8_{13} (F^{(1)})^3\right]
  \ \stackrel{u\to 0\atop v\to 1}{\longrightarrow}
  \
  \frac1{12}\gamma_{\mathcal K}^{(3)}  \ln u +O(u^0)\,.
\end{align}
Then, we substitute the known expressions for  $F^{(1)}$ and $F^{(2)}$, Eqs.~\p{F-1loop} and \p{F-2loop}, and
replace $F^{(3)}$ by its general expression \p{F-3loop} with $f^{(3)}$ given by \p{f3} to get for $x_2\to x_1$ and
$x_4\to x_3$
\begin{align}\label{p-3loop}
\text{l.h.s. of \p{3.1}}= x^4_{13} \int  \frac{d^4 x_5 d^4 x_6 d^4 x_7 \ Q(x_i)}{x_{56}^2 x_{67}^2x_{75}^2(x_{15}^4 x_{16}^4x_{17}^4) (x_{35}^4 x_{36}^4x_{37}^4) } \,.
\end{align}
Here  $Q(x_i)$ is a homogeneous polynomial,  symmetric under the exchange of the
integration points $x_5,x_6,x_7$ and linear in the constants $c^{(3)}_\a$. 
To save space, we do not present its explicit expression here. As before, we remark the presence of factors 
like $x_{15}^4$, etc., in the denominator. When $x_5 \to x_1$, these factors cause a higher-order log singularity of 
the integral. {In order to soften the singularity, as required by \p{3.1}, the numerator of the integrand on the right-hand side of 
\p{p-3loop} should vanish in this limit,
\begin{align}\label{Q0}
\lim_{x_5 \to x_1} Q(x_i) = 0 \,,
\end{align}
with $x_6,\, x_7$~fixed.  }
Unlike  the two-loop case, this condition does not have a unique solution:
\begin{align}\label{3.2}
c_1^{(3)}\equiv c(N_c) \,,\qquad c_2^{(3)}=1- c(N_c)  \,,\qquad c_3^{(3)}=- 2 c(N_c)  \,,\qquad c_4^{(3)}=c(N_c)    \,,
\end{align}
where $c(N_c)$ is an arbitrary color-dependent constant.

{The same result can be obtained by considering the other singular regime, the light-like Minkowski limit \p{lim2}. This time the relevant term in the expansion \p{2.13} is
\begin{align}\label{4.8}
\cF^{(3)} -2\cF^{(1)}\cF^{(2)}+\frac43  (\cF^{(1)} )^3  \ 
 \stackrel{u,v\to 0}{\sim} \ (\ln u)^3 +(\ln v)^3 \,,
\end{align}
in accord with \p{LC}. Thus, the analog of \p{p-3loop} now is
\begin{align}\label{3.1'}
\text{l.h.s. of \p{4.8}} 
=  x_{13}^2x_{24}^2   \int  \frac{d^4 x_5 d^4 x_6 d^4 x_7 \ Q'(x_i)}{ x_{56}^2x_{67}^2x_{75}^2\prod_{i=1}^4 x_{i5}^2 x_{i6}^2x_{i7}^2} \,,
\end{align}
with a different polynomial $Q'(x_i)$. When one of the integration points approaches
a light-like segments, say, $x_5^\mu \to  (1-\alpha) x_1^\mu + \alpha x_2^\mu$, the integral in \p{3.1'} in general develops a stronger singularity than what is required in \p{4.8}.   The  condition for softening the singularity is
\begin{align}\label{Q'0}
\lim_{x_5  \to  (1-\alpha) x_1  + \alpha x_2 } Q'(x_i) = 0 \,, 
\end{align}
with $x_6,\, x_7$~fixed. This condition has the same general solution as in \p{3.2}.}

At this stage  the three-loop correction $F^{(3)}$ to the correlation function takes the following form:
\begin{align}\label{FP}
 F^{(3)}(x_i) ={1\over 3!\,(-4\pi^2)^3} \int \frac{d^4x_5   d^4x_6 d^4x_7\,  P(x_i)}{x_{56}^2x_{67}^2x_{75}^2\prod_{i=1}^4 x_{i5}^2 x_{i6}^2x_{i7}^2}\,,
\end{align}
where $P(x_i)$ is given by a linear combination of
the $S_7-$symmetric polynomials \p{eq:11} with coefficients as defined in \p{3.2},
\begin{align}\label{P-nonpl1}
 P(x_i) =P_2^{(3)} + c(N_c)  \left[ P_1^{(3)}-P_2^{(3)}-2P_3^{(3)}+P_4^{(3)}\right] \,.
\end{align}
It may seem that we are unable to completely determine the form of the correlation function, because of the arbitrary parameter $c(N_c)$ left in \p{P-nonpl1}. However, we have not yet taken into account the Gram determinant conditions on the vectors entering in  \p{P-nonpl1}.  

By definition, $P(x_i)$ is a conformally covariant polynomial 
depending on seven four-dimensional vectors $x_1,\ldots,x_7$. Notice that any five vectors
are linearly dependent in four dimensions. In general, the corresponding Gram determinant 
conditions $\det \| x_i\cdot x_j\|=0$ do not respect the conformal symmetry and, therefore, 
they are not consistent with the properties of the polynomial $P(x_i)$. However, 
we can construct linear combinations of Gram determinants which are conformally 
covariant (see Appendix \ref{Gram} for details). Such combinations necessarily involve seven points 
and, therefore, appear for the first time at three loops. In this case there exists only one conformally covariant Gram determinant condition and
it takes the form
\begin{align}\label{Gc3}
P_1^{(3)}-P_2^{(3)}-2P_3^{(3)}+P_4^{(3)}=0\,.
\end{align}
Comparing with 
\p{P-nonpl1}, we immediately arrive at
\begin{align}\label{PP}
P(x_i) =P_2^{(3)}\,.
\end{align}
We conclude that  the entire three-loop correction \p{FP} is given by the single planar topology:
\begin{align}\label{FP2}
 F^{(3)}(x_i) ={1\over 3!\,(-4\pi^2)^3} \int \frac{d^4x_5   d^4x_6 d^4x_7\,  P^{(3)}_2(x_i)}{x_{56}^2x_{67}^2x_{75}^2\prod_{i=1}^4 x_{i5}^2 x_{i6}^2x_{i7}^2}\,.
\end{align} 

Remarkably,  among the four three-loop topologies listed in Fig.~\ref{Fgr} only the planar one, $f^{(3)}_2$, contributes to the numerator \p{PP}.  Moreover, it comes with a fixed {\it color-independent} coefficient (the only color dependence is in the universal color factor in \p{intriLoops}). This implies that the three-loop correlation function \p{FP2} {\it  receives no
non-planar corrections}, neither in terms of non-planar integral topologies, nor in terms of subleading color-dependent coefficients.  
As a consequence, we derive from \p{3.1} that the three-loop Konishi anomalous dimension $\gamma_{\mathcal K}^{(3)}$ is color exact. The same is true for the anomalous dimensions of the twist-two local
operators that appear in the OPE expansion of the four-point correlation 
function.~\footnote{Since all twist-two operators  with a given (super)conformal spin
form an irreducible $\mathcal{N}=4$ supermultiplet \cite{Belitsky:2003sh}, the same property holds for {\it all}  twist-two operators in $\mathcal{N}=4$ SYM.} 
This result, obtained here without any Feynman graphs, is in agreement with the known explicit perturbative calculations of anomalous dimensions \cite{Bianchi:2000hn,Dolan:2001tt,Eden:2004ua}.~\footnote{It is likely that the absence of non-planar contributions to the three-loop correlation function is due to the vanishing of the color factors of the non-planar three-loop Feynman graphs, for symmetry reasons. This point deserves further investigation.}

{We would like to make the following important comment on the singular behavior of the correlation function that we have constructed. In Eqs.~\p{3.1} and \p{4.8} we have shown the expected behavior  of the logarithm of the correlation function in the two singular regimes we have considered.  Then, in Eqs.~\p{Q0} and \p{Q'0} we have formulated natural conditions for softening the singularities of the generic three-loop integrals. However, it is not obvious that these conditions are {\it sufficient} to obtain exactly the behavior  \p{3.1} and \p{4.8} - the residual divergences of our integrals may still be stronger.  In our case, we can justify this {\it a posteriori}, by the fact that we have found a {\it unique} form of the correlation function. Being unique, it has to behave as required by  \p{3.1} and \p{4.8}. }

The role played by the  Gram determinant condition in deriving the three-loop correlation function illustrates the fact that our graphical construction of the integrand results, in general, in an over-complete basis of integrands. This phenomenon persists to higher loops, where the number of independent Gram determinant conditions rapidly increases (see Appendix~\ref{Gram}). {We note however that these conditions mix together planar and non-planar topologies, as we can see in \p{Gc3}. As a consequence, the Gram determinant conditions become
irrelevant in the planar sector, but they have to be taken into account in the non-planar sector.}

In conclusion, we have found the unique form of the three-loop integrand, Eqs.~\p{FP} and \p{PP}. 
Next, we can expand the three-loop  correction to the correlation function in terms of {\it only planar} three-loop scalar integrals (the details can be found in Ref.~\cite{Eden:2011we})  
\begin{align}\notag
F^{(3)} =&  \big[T(1,3;2,4)+ 11 \mbox{ perms}\big]
+ 
\big[E(1,3;2,4)+ 11 \mbox{ perms}\big]
\\[2mm] \notag +& \big[L(1,3;2,4)+ 5 \mbox{
  perms}\big] 
+ 
  \big[({g\times h})(1,3;2,4)+  5 \mbox{ perms}\big]
 \nonumber 
 \\[2mm] +&  
 {\textstyle \frac12}\big[  H(1,3;2,4)+ 11 \mbox{
  perms}\big],
  \label{eq:15} 
\end{align}
where ``$+11$ perms'' etc. denotes a sum over the remaining distinct
 $S_4$ permutations of external points $1,2,3,4$. The three-loop integrals  in \p{eq:15} are defined below:
\begin{eqnarray} \label{eq:14}
 T(1,2;3,4)&=&-{x_{34}^2\over (4\pi^2)^3}\int 
  \frac{ d^4x_5 d^4x_6 d^4x_7 \ x_{17}^2 }{(x_{15}^2 x_{35}^2) (x_{16}^2  x_{46}^2) (x_{37}^2 x_{27}^2  x_{47}^2)
    x_{56}^2 x_{57}^2 x_{67}^2}\ ,\nonumber \\
E(1,2;3,4) & = & - \frac{x^2_{23} x^2_{24}}{(4 \pi^2)^3}
\int \frac{d^4x_5 \, d^4x_6 \, d^4x_7 \ x^2_{16}}{(x_{15}^2 x_{25}^2 x_{35}^2)
x_{56}^2 (x_{26}^2 x_{36}^2 x^2_{46}) x^2_{67} (x_{17}^2 x_{27}^2 x_{47}^2)}
\, , \nonumber \\
L(1,2;3,4) & = & - \frac{x^4_{34}}{(4 \pi^2)^3}
\int \frac{d^4x_5 \, d^4x_6 \, d^4x_7}{(x_{15}^2 x_{35}^2 x_{45}^2) x_{56}^2
(x_{36}^2 x_{46}^2) x^2_{67} (x_{27}^2 x_{37}^2 x_{47}^2)} \, , \nonumber \\
  {({g\times h})}(1,2;3,4) &=& - {x_{12}^2 x_{34}^4\over (4\pi^2)^3}\int 
   \frac{d^4x_5d^4x_6 d^4x_7}{(x_{15}^2  x_{25}^2  x_{35}^2 x_{45}^2) 
 (x_{16}^2  x_{36}^2x_{46}^2) 
    (x_{27}^2  x_{37}^2   x_{47}^2) x_{67}^2} \ ,\nonumber\\
H(1,2;3,4) & = & - \frac{x_{41}^2 x_{23}^2
    x_{34}^2 
}{(4 \pi^2)^3}
\int \frac{d^4x_5 \, d^4x_6 \, d^4x_7 \ x^2_{57}}{(x_{15}^2 x_{25}^2 x_{35}^2
x^2_{45}) x_{56}^2 (x_{36}^2 x^2_{46}) x^2_{67} (x_{17}^2 x_{27}^2 x^2_{37}
x_{47}^2)} \, .
\end{eqnarray}

We would like to emphasize that in \cite{Eden:2011we}   the coefficient $c^{(3)}_2$ of the planar three-loop topology  was fixed using input from the three-loop amplitude via the duality relation \p{M-F}. As we have shown in this subsection, this is not necessary, the intrinsic properties of the correlation function are sufficient to fully determine this, as well as the remaining coefficients of the non-planar topologies. 

\subsection{Relation to the three-loop amplitude}

In the preceding subsection we have shown that the  three-loop correlation function is given by {\it planar} Feynman integrals only. This allows us to apply the duality relation \p{M-F} to 
predict the precise form of the integrand of the three-loop {\it planar} amplitude:
\begin{align}\label{3.3}
\cM^{(3)}=\cF^{(3)} - \cF^{(1)}\cF^{(2)} +\frac12 \big(\cF^{(1)}\big)^3=\cF^{(3)} -\cM^{(1)}\cM^{(2)}\,,
\end{align}
where in the last relation we made use of \p{spec1}.
We recall that $\cF^{(\ell)}$ stands for the light-cone limit \p{F-hat} of the integrand 
of the correlation function in the planar
approximation. 

Inspecting the list of integrals \p{eq:14} appearing in $\cF^{(3)}$, we observe that two of them, $E$ and $H$, 
vanish in the light-cone limit \p{lim2} due to the prefactors on the right-hand side of \p{eq:14}. Two other
integrals, $T$ and $L$, coincide in the light-cone limit with the so-called ``tennis court" and ``ladder" integrals if
expressed in terms of the dual momenta $p_i=x_i-x_{i+1}$. Similarly, the remaining integral $g\times h$ is simply 
the product of the one- and two-loop momentum box integrals, and this contribution is exactly canceled  by the cross-term 
$\cM^{(1)}\times \cM^{(2)}$ on the right-hand side of \p{3.3}. In this way, we finally obtain
\begin{align}\notag
\cM^{(3)}= x_{13}^2 x_{24}^2 &  \big[L(1,3;2,4) +L(2,4;1,3)  
  \\[2mm] \label{M3}
   &   + T(1,3;2,4)+ T(1,3;4,2)+
  T(2,4;1,3)+T(2,4;3,1)\big]. 
\end{align}
Once again, this relation should be understood at the level of integrands. Eq.~\p{M3} agrees with
the known result for the three-loop amplitude in planar $\cN=4$ SYM \cite{Bern:2005iz}. We would like to stress that the relative
coefficient between the $L-$ and $T-$integrals in \p{M3} follows
from the permutation symmetry of the $f-$function, while the overall coefficient was fixed from the singular  limit behavior of the correlation function.

\subsection{Rung rule at three loops}

As at two loops, we find that at three loops we could have obtained the full planar answer using the rung-rule procedure described in detail in section~\ref{rr}. At three  loops this amounts to gluing a pyramid to the two-loop $f-$graph (an octahedron) across two adjacent triangular faces (see Fig.~\ref{fig:bodies}). The resulting decahedron is precisely that of the three-loop $f-$graph $f_2^{(3)}$, the sole contribution to the planar three-loop correlator and the coefficient is also determined by this rung rule.

\section{Correlation function at four loops}

The construction of the four-loop correlation function
\begin{align}\label{F-4loop} 
  F^{(4)}(x_1,x_2,x_3,x_4)={x_{12}^2 x_{13}^2 x_{14}^2 x_{23}^2 
  x_{24}^2 x_{34}^2\over 4!\,(-4\pi^2)^4} \int d^4x_5   d^4x_6 d^4x_7 d^4x_8\,  f^{(4)}(x_1,  \dots , x_{8})\,,
\end{align}
goes along the same lines as
in the previous section. We first identify all relevant four-loop $f-$graphs,
classify them into planar and non-planar ones and then try to fix their contribution
from the requirement that $F^{(4)}$ has the correct asymptotic behavior
in the singular limits \p{lim1} and \p{lim2}.

The main difference from three loops is that the non-planar sector is much bigger at four loops.
As we show in this section, using the general properties formulated above, we are able to fully 
determine the correlation function in the planar sector and to considerably
restrict the freedom in the non-planar sector, reducing it to four arbitrary coefficients.
 
According to Table~\ref{Table1}, at four loops there are 32 different $f-$graphs. 
Each of them is a connected graph with 8 vertices of degree $\ge 4$ (see Eq.~\p{f-cond}). 
So, the most general $f-$function in \p{F-4loop} is a linear combination of the 32 topologies:
\begin{align}\label{f4l}
f^{(4)}(x_1,\ldots,x_{8}) = \sum_{\alpha=1}^{32} c^{(4)}_{\alpha} \, f_\alpha^{(4)}(x_1,\ldots,x_{8}) \,.
\end{align}
At the first step, we have to classify the $f-$graphs according to their 
genus. This proves to be a non-trivial task due to the large number of edges and vertices
in the graphs. To compute the genus of the $f-$graphs, we used the open-source mathematics 
software system {\it Sage} \cite{Sage}. We found that only
three $f-$graphs have genus 0 and the rest have genus 1. The planar graphs
are shown in Fig.~\ref{Fig:4loop} (to save space we do not show the non-planar $f-$graphs).
\begin{figure}[h!]
  \begin{center}
   \psfrag{1}[cc][cc]{$\scriptstyle f_1^{(4)}$}\psfrag{2}[cc][cc]{$\scriptstyle f_2^{(4)}$}\psfrag{3}[tc][cc]{$\scriptstyle f_3^{(4)}$}
   \includegraphics[width=.8\linewidth]{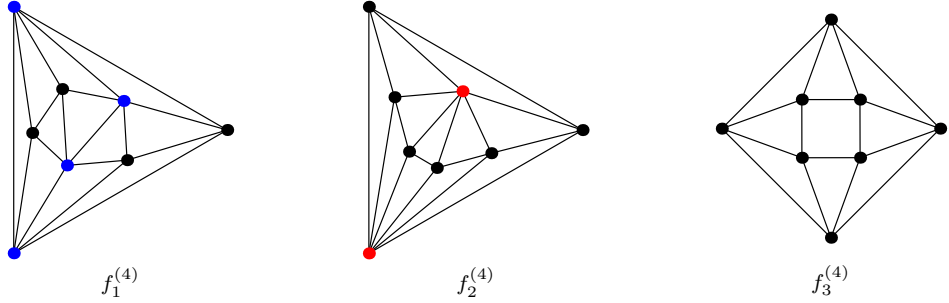}
  \end{center}
  \caption{Planar four-loop $f-$graphs. Black vertices are of degree 4
    (4 edges ending there),  blue vertices  are of degree 5 and red
    vertices are of degree 6. There is thus a dashed line (not shown)
    coming from each blue vertex and two dashed lines from each red
    vertex to ensure the total conformal weight is 4 at each
    vertex. These numerator lines are unambiguously fixed at four and
    five loops. 
}
  \label{Fig:4loop}
\end{figure}

Our next task is to try to fix the coefficients $c^{(4)}_{\alpha}$ in \p{f4l}.
Let us first discuss in more detail their color dependence. We recall that for the Feynman diagrams contributing to the correlation function there exists
a one-to-one correspondence between the genus g of the graph and the leading large $N_c$ scaling of the color factor associated with the diagram, $\sim 1/(N_c^2)^{\rm g}$. The $f-$graphs that we use as our main building blocks are, of course, not Feynman diagrams. Nevertheless, the Feynman integrals with (non-)planar topology that we can obtain from the $f-$graphs are ultimately made of (non-)planar Feynman diagrams and should come with coefficients with the appropriate color scaling. This suggests to assign a color scaling to the coefficients $c_\alpha^{(4)}$ in accord with the genus of the corresponding $f-$graphs: 
\begin{align}\label{color}\notag
& {\rm g}=0: \quad  c_\a^{(4)} = O(N_c^0)\,, \quad \a=1,2,3\,;
\\[2mm]
& {\rm g}=1: \quad   c_{\a}^{(4)} = O(1/N_c^2)\,, \quad \a=4,\ldots,32  \,.
\end{align} 
We would like to emphasize that, in general, the color factors are given by  series in $1/N_c^2$, so that 
a planar $f-$graph can also contribute subleading $O(1/N_c^2)$ color corrections to $F^{(4)}$.
This allows us to write $c_\a^{(4)} = c_{0;\a}^{(4)} + c_{1;\a}^{(4)}/N_c^2$  for $\alpha=1,2,3$ and $c_{\a}^{(4)}= c_{1;\a}^{(4)}/N_c^2$  for $\alpha=4,\ldots,32$. Then we can split the four-loop integrand   into a sum of planar and non-planar contributions,
\begin{align}\label{f-3loop-pl}
f^{(4)} = f^{(4)}_{\rm g=0}(x_i) + \frac1{N_c^2} f^{(4)}_{\rm g=1}(x_i)\,.
\end{align}
Here $f^{(4)}_{\rm g=0}$ is made only from planar $f-$graphs, whereas
$f^{(4)}_{\rm g=1}$ is a sum over all (planar and non-planar) $f-$graphs: 
\begin{align}\label{c-hat}
f^{(4)}_{\rm g=0} = \sum_{\alpha=1}^{3}c^{(4)}_{0;\a} f_\a^{(4)}(x_i) \,,\qqqquad  f^{(4)}_{\rm g=1} = \sum_{\a=1}^{32} 
 c_{1;\a}^{(4)}f_\a^{(4)} (x_i)\,,
\end{align}
and the coefficients $c^{(4)}_{0;\a}$ and $c^{(4)}_{1;\a}$ scale as $O(N_c^0)$.

{We should keep in mind that the connection between the genus of the $f-$graphs and that of the Feynman graphs which make them up, is far from obvious. One might imagine a situation in which the sum of a set of  non-planar Feynman graphs of, say, $\rm g=1$, may turn out to have $\rm g=0$. If the corresponding $f-$graphs do not appear elsewhere in the sector of the planar Feynman graphs, then some of the leading color coefficients might vanish, $c^{(\ell)}_{0;\a}=0$ for some values of $\a$. At four and at five loops this does not happen: all leading color coefficients $c^{(4)}_{0;\a}\neq0$, $c^{(5)}_{0;\a}\neq0$, see \p{c-4loop} and \p{c-5loop}, respectively. However, at six loops we find that some $c^{(6)}_{0;\a}$ do indeed vanish, see \p{c-6loop}. The statement we argue for in appendix~\ref{app:genus} is that the coefficient of $f-$graph with genus $g$ may contain corrections $O(1/N_c^{2h})$ for all $h\geq g$.}  
 
Substituting 
\p{f-3loop-pl} into \p{F-4loop}, we find that the four-loop correction to the correlation
function $F^{(4)}$ takes the general form \p{top},
\begin{align}
F^{(4)}(x_i)=F^{(4)}_{\rm g=0} + \frac1{N_c^2}F^{(4)}_{\rm g=1}\,.
\end{align}
Here  
\begin{align}\label{F4-g}
 F^{(4)}_{\rm g}(x_i) ={1\over 4!\,(-4\pi^2)^4} \int \frac{d^4x_5   d^4x_6 d^4x_7d^4x_8\,  P^{(4)}_{\rm g}(x_i)}{x_{56}^2x_{57}^2x_{58}^2x_{67}^2x_{68}^2x_{78}^2\prod_{i=1}^4 x_{i5}^2 x_{i6}^2x_{i7}^2x_{i8}^2}
\end{align}
is defined by the $S_8-$symmetric polynomials $P^{(4)}_{\rm g=0}(x_i)$ or $P^{(4)}_{\rm g=1}(x_i)$  given by
\begin{align}\label{P4-sum}
P^{(4)}_{\rm g=0}(x_i) = \sum_{\a=1}^3 c_{0;\a}^{(4)} P^{(4)}_\a(x_i)\,,\qqqquad
P^{(4)}_{\rm g=1}(x_i) = \sum_{\a=1}^{32}  c_{1;\a}^{(4)} P^{(4)}_\a(x_i)\,.
\end{align}

Below we present the numerator polynomials $P_\alpha^{(4)}$ corresponding to all 32 graphs and try to fix their coefficients, first in the planar sector and then in the non-planar.

\subsection{Planar sector}\label{Pls}

The planar part of the four-loop correlation function $F^{(4)}_{\rm g=0}$ receives contributions from the three $f-$graphs
shown in Fig.~\ref{Fig:4loop}. The corresponding polynomials take 
the form
\begin{align} \notag \label{P-4loop-p}
P^{(4)}_1(x_1, \dots, x_8)&=\ft1{24} x_{12}^2 x_{13}^2 x_{16}^2 x_{23}^2 x_{25}^2 x_{34}^2 x_{45}^2 x_{46}^2 x_{56}^2 x_{78}^6+\text{$S_8$ permutations}\,,  \\ \notag
P^{(4)}_2(x_1, \dots, x_8)&=\ft18x_{12}^2 x_{13}^2 x_{16}^2
   x_{24}^2 x_{27}^2 x_{34}^2 x_{38}^2 x_{45}^2 x_{56}^4 x_{78}^4+\text{$S_8$ permutations}\,,\\
P^{(4)}_3(x_1, \dots, x_8)&=\ft1{16} x_{12}^2 x_{15}^2 x_{18}^2 x_{23}^2 x_{26}^2 x_{34}^2
   x_{37}^2 x_{45}^2 x_{48}^2 x_{56}^2 x_{67}^2 x_{78}^2+\text{$S_8$ permutations}\,.
 \end{align}
According to \p{P4-sum} and \p{F4-g}, the planar four-loop
correlation function $F^{(4)}_{\rm g=0}$ is determined by the three coefficients $c_{0;\a}^{(4)}$ (with $\a=1,2,3$).
 
As before, to fix the coefficients  we require that the four-loop correlation function should have
the correct asymptotic behavior \p{SD} and \p{LC} in the singular limits \p{lim1} and \p{lim2}, respectively,  to all orders in $1/N_c$. Replacing $F(x_i)$ in \p{SD} and \p{LC} by its general expression \p{top} and expanding
both sides in powers of $1/N_c$, we find that the planar correction $F^{(\ell)}_{\rm g=0}$
satisfies the same relations \p{SD} and \p{LC}. Further, expanding both sides of \p{SD} to order $O(a^4)$ we find in the planar limit  
\begin{align}\label{log4loop}
 \widehat F_{\rm g=0}^{(4)} - 6 \widehat F^{(1)} \widehat F^{(3)} 
- 3  (\widehat F^{(2)})^2+36   (\widehat F^{(1)})^2\widehat F^{(2)}-54  (\widehat F^{(1)})^4 
\stackrel{u\to 0\atop v\to 1}{\sim}  \ln u + O(u^0) \,,
\end{align}
where $\widehat F^{(\ell)}\equiv x^4_{13}F^{(\ell)}(x_i)$. Replacing $F^{(1)}$, $F^{(2)}$ and $F^{(3)}$ by their expressions obtained in the preceding sections, and $F_{\rm g=0}^{(4)}$ by a linear combination of the three planar structures (see \p{P4-sum}), we can write the left-hand side 
of \p{log4loop} in a form similar to \p{p-3loop}. Then, the
relation \p{log4loop} translates into the  condition that the resulting numerator polynomial $Q$ should vanish for $x_5\to x_1$, with  the remaining integration points $x_6,x_7,x_8$ in general positions (recall \p{Q0}). Equating to zero the coefficients
in front of the different terms depending on $x_6,x_7,x_8$, we obtain an overdetermined system of equations for
the coefficients $c_{0;\a}^{(4)}$ (with $\a=1,2,3$). We found that it has the general solution
\begin{align}\label{co4l}
c_{0;1}^{(4)} = - c_{0;3}^{(4)} \,,\qqquad c_{0;2}^{(4)}=1\,,
\end{align}
so that the value of $c_{0;1}^{(4)}$ remains undetermined. Thus, the double short-distance limit is not powerful enough to fix all the planar coefficients at four loops.

{Let us now repeat the  analysis in the light-cone limit.   Expanding the left-hand side of \p{LC} to order $O(a^4)$, 
we expect that for $u, v\to 0$ the leading singularity of the following combination of integrals:  
\begin{align}\label{log4loop1}
 \cF_{\rm g=0}^{(4)} -2\cF^{(1)}\cF_{\rm g=0}^{(3)}-(\cF^{(2)})^2+4(\cF^{(1)})^2\cF^{(2)} -2 (\cF^{(1)} )^4\ 
 \stackrel{u,v\to 0}{\sim} \ (\ln u)^4 +(\ln v)^4 \,,
\end{align}
should be much softer than that of the correlation function itself (see \p{Sud}). 
We put the terms in the left-hand side of this relation  under a common denominator and identify the corresponding $S_8-$symmetric polynomial in the numerator. The softening of the light-cone singularity required by \p{log4loop1} implies the condition that this polynomial should vanish when the integration point $x_5$ approaches one of the light-like edges, say, $x_5\to (1-\alpha)x_1+\alpha x_2$, and the remaining integration points $x_6,x_7,x_8$ are in general positions. } This again yields an
overdetermined system of equations, which this time has the unique solution \footnote{This is an example where we exploit the full power of the light-cone condition, requiring it to hold for  $\a$ {\it arbitrary}.}
\begin{align}\label{c-4loop}
c_{0;1}^{(4)} = c_{0;2}^{(4)}= - c_{0;3}^{(4)} =1\,.
\end{align}
Thus, the light-cone limit allows us to fix all planar coefficients leading to the following result for the
four-loop correction to the correlation function in the planar limit:
\begin{align}\label{F4-g=0}
 F^{(4)}_{\rm g=0}(x_i) ={1\over 4!\,(-4\pi^2)^4} \int \frac{d^4x_5   d^4x_6 d^4x_7d^4x_8\, \left[ P^{(4)}_1(x_i)+P^{(4)}_2(x_i)-P^{(4)}_3(x_i)\right]}{x_{56}^2x_{57}^2x_{58}^2x_{67}^2x_{68}^2x_{78}^2\prod_{i=1}^4 x_{i5}^2 x_{i6}^2x_{i7}^2x_{i8}^2}\,,
\end{align}
with the $P-$polynomials defined in \p{P-4loop-p}.  
 
\subsection{Rung rule at four loops}
 
At four loops we can also implement the rung-rule procedure described in
Sect.~\ref{rr} to obtain parts of the planar result. This
procedure amounts to gluing pyramids to the decahedron describing
the planar three-loop graph $f^{(3)}_2$, thus obtaining four-loop $f-$graphs (see Fig.~\ref{fig:bodies}). Implementing
this procedure we find there are two inequivalent ways of gluing the
pyramid to the graph $f_2^{(3)}$ and this gives two out of
the three planar four-loop $f-$graphs, namely $f_1^{(4)}$ and $f_2^{(4)}$. This is illustrated graphically in
Fig.~\ref{fig:bodies}, where the right-most two polyhedra correspond
to $f_1^{(4)}$ and $f_2^{(4)}$.  We recall that the rung rule not only predicts the new higher loop topologies, but also fixes the coefficients of their contributions.

The remaining planar graph $f_3^{(4)}$
is not generated by the rung rule, so we have to identify it as the other member of the class of planar $f-$graph topologies.~\footnote{A characteristic feature of the new  non-rung-rule topology is the presence of rectangular faces. All planar $f-$graphs so far have had only triangular faces, and the rung rule (gluing a pyramid by its square base) cannot produce a rectangular face. } We recall that the weaker Euclidean limit  \p{log4loop} could not fix all coefficients (see \p{co4l}). However, if we now use the additional information that the rung rule fixes $c_{0;1}^{(4)} =1$, then the first relation in \p{co4l} allows us to also determine the coefficient of the non-rung-rule topology. 

\subsection{Relation to the four-loop amplitude} \label{Sect-4loop}

We can use the obtained expression for the four-loop planar correlation function \p{F4-g=0} to predict 
the integrand for the planar four-loop four-gluon amplitude $\cM^{(4)}$ with the help of the duality relation \p{M-F}. 
We recall that the duality relation only holds in the planar limit and it is not valid beyond it.

Restricting the loop corrections to the correlation function $F^{(\ell)}$ to their light-cone limits $\cF^{(\ell)}=\lim_{x_{i,i+1}^2\to 0} x_{13}^2 x_{24}^2 F^{(\ell)}$, from the duality relation \p{M-F} we obtain the expression for the integrand of the planar four-loop amplitude,
\begin{align}\label{M4}
\cM^{(4)} = \cF^{(4)} - \cM^{(1)}\cM^{(3)}- \frac12\lr{\cM^{(2)}}^{2} \,,
\end{align}
where $\cM^{(1,2,3)}$ are defined in Eqs.~\p{spec2} and \p{3.3}, and $\cF^{(4)}$ is given by a sum over the various conformal $x-$integrals originating from the three different $f-$graph
topologies shown in Fig.~\ref{Fig:4loop}. When translated into the dual
momentum space, $p_i=x_i-x_{i+1}$, these integrals become dual conformal momentum scalar integrals. 

Notice that some of the integrals
in $\cF^{(4)}$ are reducible, that is they can be factored out into a product of lower loop integrals. At the same time, such integrals cannot appear in the four-loop amplitude
$\cM^{(4)}$ due to the unitarity condition. This implies that,  on the right-hand side of \p{M4} all reducible integrals inside $\cF^{(4)}$ should cancel against the products of integrals
generated by the cross-terms $\cM^{(1)}\cM^{(3)}+ \frac12\lr{\cM^{(2)}}^{2}$. 
Indeed, we verified that 
the resulting expression for $\cM^{(4)}$ is given by a sum over 15 different irreducible conformal $x-$integrals.
 The coefficients of these integrals are uniquely fixed by the three coefficients
defined in \p{c-4loop}.
 We also verified that the expression for  $\cM^{(4)}$ obtained in this way is in agreement with the results of Ref.~\cite{Bern:2006ew} on the four-loop planar four-gluon amplitude. Namely, the two rung-rule topologies $f_1^{(4)}$  and  $f_2^{(4)}$ shown in Fig.~\ref{Fig:4loop} give rise to the 12 rung-rule type dual momentum integrals listed in \cite{Bern:2006ew}. Similarly, the third topology $f_3^{(4)}$ in Fig.~\ref{Fig:4loop}  produces the three non-rung-rule integrals in  \cite{Bern:2006ew}, with relative factors $(-1)$, as follows from \p{F4-g=0}.

\subsection{Non-planar sector}\label{Nps}

At four loops, there are 29 non-isomorphic non-planar $f-$graphs of genus 1. The corresponding basis $P-$polynomials are listed in Appendix \ref{Np4} (see Eq.~\p{P-4loop-np}). 

The non-planar correction to the correlation function $F_{\rm g=1}^{(4)}$ is given by 
\p{F4-g} with the polynomial $P_{\rm g=1}^{(4)}$ equal to the linear combination \p{P4-sum}
of all 32 polynomials defined in
Eqs.~\p{P-4loop-p} and \p{P-4loop-np} with arbitrary coefficients $c_{1;\a}^{(4)}$. As before, we may try to fix the coefficients by requiring that $F_{\rm g=1}^{(4)}$ should have the correct asymptotic
behavior \p{SD} and \p{LC} to all orders in $1/N_c$.   

Expanding both sides of \p{SD}  in  $1/N_c$ and keeping only the leading non-planar terms, we obtain  the following condition:
\begin{align}\label{SD-nonpl}
  \frac{x_{13}^4 F_{\rm g=1} (x_i)}{1+6x_{13}^4 F_{\rm g=0} (x_i)} & \
\stackrel{u\to 0\atop v\to 1}{\longrightarrow} \  \frac1{12} \gamma_{{\mathcal K},{\rm g=1}}(a) \ln u+ O(u^0)  \,,
\end{align}
where $\gamma_{{\mathcal K},{\rm g=1}}(a)$  denotes the non-planar
$1/N_c^2$ correction to the Konishi anomalous dimension.  We recall that, in virtue of \p{FP2},
the correlation function receives non-planar corrections starting from four loops. Then, the relation  
\p{SD-nonpl} leads to 
\begin{align}\label{SD-nonpl3}
   { x_{13}^4 F^{(4)}_{\rm g=1} (x_i)} & \
\stackrel{u\to 0\atop v\to 1}{\longrightarrow} \  \frac1{12} \gamma^{(4)}_{{\mathcal K},{\rm g=1}}  \ln u+ O(u^0)\,.
\end{align}
Repeating the same analysis in the light-cone limit, from \p{LC} we find the following condition
on $F_{\rm g=1}^{(4)}$:
\begin{align}
\label{LC-nonpl3}
   { x_{13}^2 x_{24}^2  F^{(4)}_{\rm g=1}(x_i)} & \
\stackrel{u, v\to 0}{\sim} \ 0\times \left[ (\ln u)^4 + (\ln v)^4 \right] +\ldots\,.
\end{align}
{Here we multiplied the leading term on the right-hand side by zero to recall that the non-planar corrections at $\ell$ loops involve $\ln u$ and $\ln v$ to 
the total power  $\leq \ell-1$.

As in Sect.~\ref{Pls}, the relations \p{SD-nonpl3} and \p{LC-nonpl3} can be translated into the condition that the polynomial $P_{\rm g=1}(x_i)$ in \p{P4-sum} should vanish in the singular regimes:  
(i) for $x_5\to x_1$ in the short-distance limit \p{lim1} and (ii) for $x_5\to (1-\alpha) x_1 + \alpha x_2$ in the light-cone limit \p{lim2}, with the three remaining integration points $x_6, x_7$ and $x_8$ in general positions.  Since the light-cone limit \p{lim2}
turned out to be more restrictive in the planar sector, we employ the same limit in the non-planar
sector as well. Namely, we require that for $x_{12}^2,x_{23}^2,x_{34}^2,x_{41}^2= 0$
\begin{align}\label{nonpl-eqs}
 P_{\rm g=1}^{(4)}(x_1,\ldots,x_8) \to 0\,, \qquad
\text{as \ $x_5\to (1-\alpha) x_1+\alpha x_2$}\,.
\end{align} 
Equating to zero the coefficients in front of the different distinct terms and of the powers of $\alpha$ in
the left-hand side of this relation, we obtain a system of equations for the non-planar coefficients  
$c_{1;\a}^{(4)}$. We found that it fixes all but 7 coefficients. This means that the general
solution to \p{nonpl-eqs} is a sum of 7 terms with arbitrary coefficients each given by a linear 
combination of 32 polynomials with definite coefficients.

So far we have not taken into account the conformal Gram determinant condition (see Appendix~\ref{Gram}). 
We have shown in Sect.~\ref{3lcondi} that this condition completely eliminates the non-planar correction
at three loops.  At four loops there are three independent conformal Gram determinant conditions. Taking them into account, we find that the number of independent solutions to 
\p{nonpl-eqs} reduces from 7 to 4.

The explicit form of the conformal Gram determinant conditions can be
found by the method explained in Appendix~\ref{Gram}.  There is however a much simpler way to identify the same conditions. We start
with a linear combination of the 32 polynomials $\sum_{i=1}^{32} a_i P_i^{(4)}(x_i)$ and generate
32 random kinematical configurations of 8 four-dimensional vectors $(x_1,\ldots,x_8)$. Requiring
this linear combination to vanish for all configurations, we arrive at a system of 32 equations for
the coefficients $a_i$. As expected, it has three linearly independent solutions:  
\begin{align}\label{Gram-conf}
\sum_{i=1}^{32}  a_{k,i} P_i^{(4)}(x_1,\ldots,x_8) = 0 \,,\qquad (k=1,2,3)\,,
\end{align}
where $a_1, a_2, a_3$ are three lists of 32 coefficients:
\begin{align}\notag
a_1=&{\scriptstyle  (16, 6, -8, 8, -10, 24, 0, -4, 8, 6, -2, 4, -4, -6, 3, -9, 0, 3, 4, 
-5, -2, -18, -2, 3, -3, 1, 0, 0, 0, 0, 0, 0)} \\ \notag
a_2=&{\scriptstyle (-18, -9, 12, -8, 12, -24, 0, 4, -7, -6, 0, -2, 4, 6, -3, 9, 0, -2, 
-5, 5, -2, 18, 2, -3, 3, 0, -24, -2, 0, 2, 0, 0)} \\
a_3=&{\scriptstyle (4, -1, 2, 4, -2, 12, 0, -2, 5, 2, -4, 4, -2, -2, 1, -3, -1, 2, 1, -1,  
-6, -6, 0, 1, 0, 0, -36, 0, 1, 0, -1, 1)}\ .
\end{align}
Notice that the left-hand side of \p{Gram-conf} contains a sum over the polynomials corresponding to 
both planar and non-planar $f-$graphs. Then, making use of \p{Gram-conf} we can express the
three planar polynomials, $P_1^{(4)}$, $P_2^{(4)}$ and $P_3^{(4)}$, in terms of the non-planar ones.

Finally, the general solution to \p{nonpl-eqs}  has the form
\begin{align}\label{P=Q}
P_{\rm g=1}^{(4)}(x_i) = c_1^{(4)} Q_1(x_i) + c_2^{(4)} Q_2(x_i) + c_3^{(4)} Q_3(x_i) + c_4^{(4)} Q_4(x_i) \,,
\end{align}
where $c_k^{(4)}$ are arbitrary and the polynomials $Q_k$ are given by linear combinations
of the polynomials  \p{P-4loop-p} and \p{P-4loop-np}
\begin{align}\label{Qk}
Q_k = \sum_{i=1}^{32} q_{k,i} P_i^{(4)}(x_1,\ldots,x_8)\,,\qquad (k=1,2,3,4)\,,
\end{align}
and $q_{k,i}$ are four lists of 32 integer coefficients:
\begin{align}\notag
& q_1= {\scriptstyle  (0, 0, 0, 0, 0, 0, 0, 0, 0, 0, 0, 0, 0, 0, 0, 0, 0, 0, 0, 0, 0, 0, 0, 
0, 0, 0, 1, 0, 0, 0, 0, 0)}\,,
\\ \notag
& q_2= {\scriptstyle  (-2, 0, 2, -8, 0, -6, 0, 0, 0, 0, 2, -2, 1, 0, 0, 0, 0, -2, -1, 1, 6, 
0, 0, 0, 0, 0, 0, 0, 0, 0, 0, 0)} \,,
\\ \notag
& q_3= {\scriptstyle  (-2, 0, -3, 4, 1, -6, -2, 2, -4, 0, 2, -2, 1, 1, -1, 2, 1, 0, 0, 0, 0, 
0, 0, 0, 0, 0, 0, 0, 0, 0, 0, 0)} \,,
\\
& q_4=  {\scriptstyle   (-14, -8, 10, -8, 8, -18, 0, 3, -4, -4, 0, -2, 3, 4, -2, 6, 0, -2, -4, 
4, 0, 12, 2, -2, 2, 0, 0, 0, 0, 0, 0, 0)} \,.
\end{align}
The first solution is particularly interesting, 
\begin{align}
Q_1=P_{27}^{(4)}(x_i)\,,
\end{align}
implying that the polynomial $P_{27}^{(4)}$ satisfies the condition \p{nonpl-eqs}.
The corresponding $f-$graph has a high symmetry as indicated by the large combinatorial factor in the denominator in the definition of $P_{27}^{(4)}$, Eq.~\p{P-4loop-np}. The remaining three solutions, $Q_2, Q_3, Q_4$, 
involve  contributions from the three planar graphs. However, applying the conformal Gram determinant conditions
\p{Gram-conf} we can eliminate the planar graphs from $Q_i$, at the expense of increasing the number of
contributing (non-planar) polynomials.

Our derivation of \p{P=Q} was based on the light-cone limit. We can verify that the obtained expression for the
non-planar correction \p{P=Q} has the correct asymptotic behavior in the double short-distance limit, $x_1\to x_2$
and $x_3\to x_4$. The latter condition implies that each polynomial $Q_k$, Eq.~\p{Qk}, in this limit should vanish for $x_5\to x_1$. Indeed, we have examined the expressions for the polynomials $Q_k$ and have found that, surprisingly, for $x_1=x_2$ and $x_3=x_4$ they are all proportional to each other leading to
\begin{align} 
\lim_{x_1\to x_2\atop x_3\to x_4} P_{\rm g=1}^{(4)}(x_i) & =  \lr{ c_1^{(4)}-120 c_2^{(4)} +68 c_3^{(4)} -24 c_4^{(4)}}
\lim_{x_1\to x_2\atop x_3\to x_4} Q_1(x_i)\,,
\end{align}
with
\begin{align}
\lim_{x_1\to x_2\atop x_3\to x_4} Q_1(x_i) = \frac16 x_{13}^4 
\lr{x_{56}^2x_{78}^2+ x_{57}^2x_{68}^2+x_{58}^2x_{67}^2}
\prod_{i=5,6,7,8} x_{1i}^2 x_{3i}^2\,.
\end{align}
As expected, this expression vanishes when the integration points $x_5,x_6,x_7,x_8$ approach one
of the external points, $x_1$ and $x_3$.
 
To summarize,  in this subsection we have constructed the integrand of the non-planar four-loop correction
to the correlation function and demonstrated that it is fixed up to four arbitrary constants as
\begin{align}\label{F4-g=1}
 F^{(4)}_{\rm g=1}(x_i) =\sum_{k=1}^4   {c_k^{(4)} \over 4!\,(-4\pi^2)^4} \int \frac{d^4x_5   d^4x_6 d^4x_7d^4x_8\, \, Q_k(x_1,\ldots,x_8) }{x_{56}^2x_{57}^2x_{58}^2x_{67}^2x_{68}^2x_{78}^2\prod_{i=1}^4 x_{i5}^2 x_{i6}^2x_{i7}^2x_{i8}^2}\,,
\end{align}
with the polynomials $Q_k$ defined in \p{Qk}. To fix the values of the coefficients $c_k^{(4)}$ in  \p{F4-g=1}
we would need more detailed information about the asymptotic behavior of the correlation function $F^{(4)}_{\rm g=1}(x_i) $ in the singular limits defined above.

\section{Five loops}
\label{sec:five-looops}

We can read from Table~\ref{Table1} that the five-loop integrand $f^{(5)}$, Eq.~\p{f-sum}, receives   
contributions from 930 different $f-$graphs. The general arguments
given in the previous sections together with the assumption that there
is a one-to-one correspondence between the genus of the $f-$graphs and
the large $N_c$ scaling of their contribution to the correlation
function, allows us to classify all $f-$graphs into planar and non-planar ones.

We find that among the 930 five-loop $f-$graphs only 7 are planar. They yield the following expression for the five-loop  $S_9-$symmetric numerator polynomial in the planar sector:
\begin{equation}
  \label{eq:5}
  P_{\rm g=0}^{(5)} =  \sum_{\a=1}^7 c^{(5)}_\a
  P_\a^{(5)}(x_1,\ldots,x_9)\,,
\end{equation}
where $c^{(5)}_\a$ are arbitrary coefficients~\footnote{To simplify the notation, we drop the leading-color label, so here $c^{(5)}_\a \equiv c^{(5)}_{0;\a}$.} and
\begin{equation}\label{P5}
\begin{array}{l}
 P_1^{\text{(5)}}=\frac{1}{2} x_{13}^2 x_{16}^2 x_{18}^2 x_{19}^2 x_{24}^4 x_{26}^2 x_{29}^2 x_{37}^2 x_{38}^2 x_{39}^2 x_{47}^2 x_{48}^2
   x_{56}^2 x_{57}^2 x_{58}^2 x_{59}^2 x_{67}^2 +\ldots\,,
   \\
 P_2^{\text{(5)}}=\frac{1}{4} x_{13}^2 x_{16}^2 x_{18}^2 x_{19}^2 x_{24}^4 x_{26}^2 x_{29}^2 x_{37}^4 x_{39}^2 x_{48}^4 x_{56}^2 x_{57}^2
   x_{58}^2 x_{59}^2 x_{67}^2 +\ldots\,,\\
 P_3^{\text{(5)}}=\frac{1}{4} x_{13}^4 x_{17}^2 x_{19}^2 x_{24}^2 x_{26}^2 x_{27}^2 x_{29}^2 x_{36}^2 x_{39}^2 x_{48}^6 x_{56}^2 x_{57}^2
   x_{58}^2 x_{59}^2 x_{67}^2+\ldots\,, \\
 P_4^{\text{(5)}}=\frac{1}{6} x_{13}^2 x_{16}^2 x_{19}^4 x_{24}^4 x_{28}^2 x_{29}^2 x_{37}^4 x_{38}^2 x_{46}^2 x_{47}^2 x_{56}^2 x_{57}^2
   x_{58}^2 x_{59}^2 x_{68}^2+\ldots\,, \\
 P_5^{\text{(5)}}=\frac{1}{8} x_{13}^4 x_{16}^2 x_{18}^2 x_{24}^4 x_{28}^2 x_{29}^2 x_{37}^2 x_{39}^2 x_{46}^2 x_{47}^2 x_{56}^2 x_{57}^2
   x_{58}^2 x_{59}^2 x_{69}^2 x_{78}^2+\ldots\,, \\
 P_6^{\text{(5)}}=\frac{1}{28} x_{13}^2 x_{17}^2 x_{18}^2 x_{19}^2 x_{24}^8 x_{36}^2 x_{38}^2 x_{39}^2 x_{56}^2 x_{57}^2 x_{58}^2 x_{59}^2
   x_{67}^2 x_{69}^2 x_{78}^2 +\ldots\,,\\
 P_7^{\text{(5)}}=\frac{1}{12} x_{13}^2 x_{16}^2 x_{17}^2 x_{19}^2 x_{26}^2 x_{27}^2 x_{28}^2 x_{29}^2 x_{35}^2 x_{38}^2 x_{39}^2 x_{45}^2
   x_{46}^2 x_{47}^2 x_{49}^2 x_{57}^2 x_{58}^2 x_{68}^2+\ldots\,.
\end{array}
\end{equation}
The planar $f-$graphs corresponding to the seven polynomials in \p{P5} are shown
in Fig.~\ref{Fig:5loops}.  
\begin{figure}
  \begin{center}
  \psfrag{1}[tc][cc]{$\scriptstyle f_1^{(5)}$}\psfrag{2}[cc][cc]{$\scriptstyle f_2^{(5)}$}\psfrag{3}[cc][cc]{$\scriptstyle f_3^{(5)}$}
  \psfrag{4}[cc][cc]{$\scriptstyle f_4^{(5)}$}\psfrag{5}[tc][cc]{$\scriptstyle f_5^{(5)}$}\psfrag{6}[cc][cc]{$\scriptstyle f_6^{(5)}$}\psfrag{7}[tc][cc]{$\scriptstyle f_7^{(5)}$}
    \includegraphics[width=\linewidth]{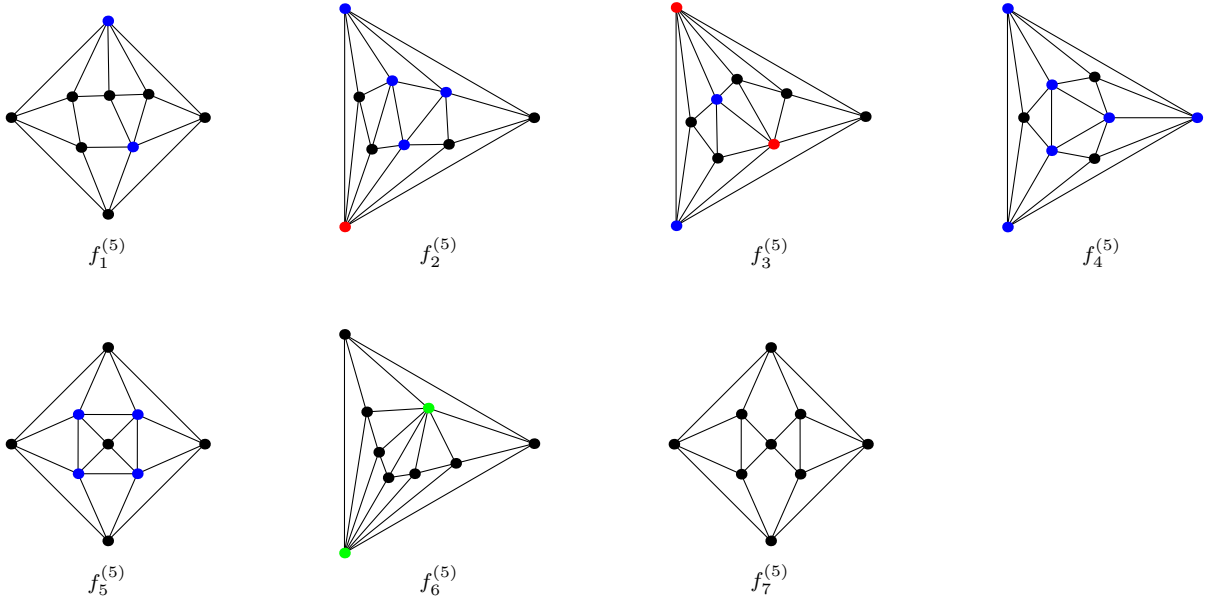}
  \end{center}
    \caption{Planar $f-$graphs at  five loops. 
  The definition of black, blue and red vertices is the same as in
 Fig.~\ref{Fig:4loop}, the green vertices have degree 7.
    The graphs $f_1^{(5)}$ and $f_5^{(5)}$ can be obtained from the four-loop graph $f_3^{(4)}$ in Fig.~\ref{Fig:4loop} by the rung rule. The same rule relates the graphs  $f_2^{(5)},f_3^{(5)},f_4^{(5)}$ with $f_1^{(4)}$ and $f_6^{(5)}$ with $f_2^{(4)}$. The only non-rung-rule five-loop graph is $f_7^{(5)}$.}
    \label{Fig:5loops}
\end{figure}

As before, the ellipses in \p{P5}  denote terms with $S_9-$permutations of indices.  In general,  in the sum over permutations, the same term
could be repeated multiple times due to the symmetry of the corresponding graph. Indeed, for a symmetric graph there is
a set of permutations which map the graph back to itself while preserving the edge-vertex connectivity. This set forms the automorphism group of the corresponding numerator polynomial. In
order to take this symmetry into account, we simply divide each polynomial in \p{P5} by the dimension of 
its automorphism group (which can be computed, for example, by the {\it Combinatorica} package of {\it Mathematica}). 

To fix the coefficients $c^{(5)}_{\a}$ in \p{eq:5} we require that the five-loop correction to the correlation
function should satisfy the Minkowski singular limit relation \p{LC}. At five loops, this relation leads to (for $u, v\to 0$)
\begin{align}\label{log5loop1}\notag
 \cF_{\rm g=0}^{(5)}  -2\cF^{(1)}\cF_{\rm g=0}^{(4)} & -2\cF^{(2)}\cF_{\rm g=0}^{(3)}
+4(\cF^{(1)})^2\cF_{\rm g=0}^{(3)} \\
& +4\cF^{(1)}(\cF^{(2)})^2-8(\cF^{(1)})^3\cF^{(2)} +\ft{16}5 (\cF^{(1)} )^5 \ \ \stackrel{u,v\to 0}{\sim} \ \ (\ln u)^5+ (\ln v)^5 \,.
\end{align}
Repeating the analysis along the same lines as at four loops we find
that this relation fixes the coefficients to be \footnote{Rather
  surprisingly, at five loops the Euclidean short-distance limit also
  unambiguously fixes all the 7 coefficients of the planar topologies,
  unlike what happened at four loops. }
\begin{align}\label{c-5loop}
-c^{(5)}_{1}=c^{(5)}_{2}=c^{(5)}_{3}=c^{(5)}_{4}=-c^{(5)}_{5}=c^{(5)}_{6}=c^{(5)}_{7}=1 \,.
\end{align}
We observe that all topologies in Fig.~\ref{Fig:5loops} except $f_7^{(5)}$ can be obtained from the planar
four-loop graphs in Fig.~\ref{Fig:4loop} via the rung rule. As a consequence, the values of their
coefficients $c_\a^{(5)}$ coincide with the corresponding four-loop coefficients \p{c-4loop}. Thus, we are left
with only one coefficient, that of the extra $f_7^{(5)}-$topology, not induced from lower loops. Notice that unlike at four loops, this non-rung-rule coefficient is positive, $c^{(5)}_{7}=+1$ (to be compared with  $c^{(4)}_{3}=-1$).~\footnote{An attempt to explain the alternation of signs for rung-rule and non-rung-rule topologies was made in Ref.~\cite{Cachazo:2008dx}. The latter were interpreted as correction terms needed to cancel some spurious singularities in the rung-rule sector (hence the negative sign). However, this interpretation is known to fail at six loops \cite{CarrascoJohansson}.} 
 
Combining together Eqs.~\p{c-5loop}, \p{eq:5},  \p{integg} and \p{eq:10}, we finally obtain the five-loop correlation
function in the planar limit as
\begin{align}\label{F5-g=0}
 F^{(5)}_{\rm g=0}(x_i) ={1\over 5!\,(-4\pi^2)^5} \int \frac{\prod_{i=5}^9 d^4x_i \left[ -P^{(5)}_1+P^{(5)}_2+P^{(5)}_3+P^{(5)}_4-P^{(5)}_5+P^{(5)}_6+P^{(5)}_7\right]}{x_{56}^2x_{57}^2x_{58}^2x_{59}^2 
 x_{67}^2x_{68}^2x_{69}^2x_{78}^2x_{79}^2x_{89}^2
 \prod_{i=1}^4 x_{i5}^2 x_{i6}^2x_{i7}^2x_{i8}^2x_{i9}^2}\,,
\end{align}
with the $P^{(5)}-$polynomials defined in \p{P5}. 

What about the non-planar corrections to the five-loop correlation
function? These have the following form
\begin{align}
F^{(5)}(x_i)= F^{(5)}_{\rm g=0} +\frac1{N_c^2} F^{(5)}_{\rm g=1} +\frac1{N_c^4} F^{(5)}_{\rm g=2} \,,
\end{align}
where $F^{(5)}_{\rm g}$ receives the contribution from five-loop
$f-$graphs of genus  $\le \rm g$. 
  
Such graphs can be easily generated using the graph-theoretical methods described in the previous
sections. As before, to fix the coefficients accompanying these graphs we shall require that the
correlation function
should have the correct asymptotic behavior, Eqs.~\p{SD} and \p{LC}, in the singular (Minkowski light-cone and Euclidean short-distance) 
limits to order $1/N_c^2$ and $1/N_c^4$. We recall that non-planar corrections first
appear at four loops and they depend on four coefficients, see Eq.~\p{F4-g=1}. Then, expanding both sides of
\p{LC}, we find that the non-planar $1/N_c^2$ corrections at four and five loops are related to each other in the
Minkowski light-cone limit. In close analogy with the similar iterative structure in the planar sector, the $1/N_c^2$ corrections at five loops 
{may well} be uniquely fixed in terms of the four non-planar four-loop coefficients \p{F4-g=1}. At the same time, $1/N_c^4$ corrections first appear at five loops and, like the 
$1/N_c^2$ corrections at four loops, we expect that the corresponding coefficients will not be fixed completely 
in the Minkowski light-cone limit.


As in Sect.~\ref{Sect-4loop}, we can use the obtained expression for the five-loop correlation function \p{F5-g=0} to predict the integrand for the five-loop planar amplitude from the
duality relation \p{F-M}
\begin{align}\label{5loop-amp}
\cM^{(5)} = \cF^{(5)}-\cM^{(1)}\cM^{(4)}-\cM^{(2)}\cM^{(3)}\,,
\end{align}
with $\cF^{(5)}=\lim_{x_{i,i+1}^2\to 0} x_{13}^2 x_{24}^2 F_{\rm g=0}^{(5)}$. Like 
four loops, all reducible integrals inside
$\cF^{(5)}$ cancel against similar integrals produced by the last two terms on the
right-hand side of \p{5loop-amp}. The resulting expression for  $\cM^{(5)}$ is
given by the sum of 34 irreducible distinct conformal integrals generated by the $f-$graphs of
7 topologies shown in Fig.~\ref{Fig:5loops}. The relative coefficients of these
integrals are uniquely fixed by the 7 coefficients in the numerator of \p{F5-g=0}.
Going to the dual momenta $p_i=x_i-x_{i+1}$, we verified that our result for $\cM^{(5)}$
is in agreement with the ansatz for the planar five-loop four-point amplitude proposed
in Refs.~\cite{Bern:2007ct,Carrasco:2011mn}.

\section{Six loops}\label{Sl}

As follows from Table~\ref{Table1}, at six loops there are 189341 different $f-$graphs. Most of them 
are non-planar, however. Restricting our analysis to the planar sector, we find that there are only
36 planar $f-$graphs.  As a result, the six-loop correction to the correlation function in the planar
limit, $F_{\rm g=0}^{(6)}(x_i)$, is determined by the  numerator polynomial $P_{\rm g=0}^{(6)}$ 
given by
\begin{equation}\label{P-6loop}
  P_{\rm g=0}^{(6)} =  \sum_{\a=1}^{36} c_\a^{(6)}
   P_\a^{(6)}(x_1,\ldots,x_{10})\,,
\end{equation}
with the basis polynomials $ P_\a^{(6)}$ listed in Appendix~\ref{6loopP}.    

We can further split all planar $f-$graphs into those induced from five loops via the rung rule, 23 graphs in total (shown in Fig.~\ref{fig:sixinduced}) and the remaining 13 non-rung-rule graphs (shown in Fig.~\ref{fig:sixextra}). The corresponding  $S_{10}-$symmetric polynomials are listed in \p{eq:3} and \p{eq:4}, respectively.
\begin{figure}[t!]
  \psfrag{1}[tc][cc]{$\scriptstyle f_1^{(6)}$}\psfrag{2}[cc][cc]{$\scriptstyle f_2^{(6)}$}\psfrag{3}[cc][cc]{$\scriptstyle f_3^{(6)}$}
  \psfrag{4}[cc][cc]{$\scriptstyle f_4^{(6)}$}\psfrag{5}[tc][cc]{$\scriptstyle f_5^{(6)}$}\psfrag{6}[tc][cc]{$\scriptstyle f_6^{(6)}$}
  \psfrag{7}[tc][cc]{$\scriptstyle f_7^{(6)}$}\psfrag{8}[tt][cc]{$\scriptstyle f_8^{(6)}$}\psfrag{9}[t][c]{${\phantom{A}}_{\scriptstyle f_9^{(6)}}$} \psfrag{10}[tc][cc]{$\scriptstyle f_{10}^{(6)}$}
  \psfrag{11}[tc][cc]{$\scriptstyle f_{11}^{(6)}$}\psfrag{12}[tc][cc]{$\scriptstyle f_{12}^{(6)}$}\psfrag{13}[cc][cc]{$\scriptstyle f_{13}^{(6)}$}
  \psfrag{14}[cc][cc]{$\scriptstyle f_{14}^{(6)}$}\psfrag{15}[cc][cc]{$\scriptstyle f_{15}^{(6)}$}\psfrag{16}[tc][cc]{$\scriptstyle f_{16}^{(6)}$}
  \psfrag{17}[cc][cc]{$\scriptstyle f_{17}^{(6)}$}\psfrag{18}[tc][cc]{$\scriptstyle f_{18}^{(6)}$}\psfrag{19}[tc][cc]{$\scriptstyle f_{19}^{(6)}$}
   \psfrag{20}[cc][cc]{$\scriptstyle f_{20}^{(6)}$}
  \psfrag{21}[tc][cc]{$\scriptstyle f_{21}^{(6)}$}\psfrag{22}[tc][cc]{$\scriptstyle f_{22}^{(6)}$}\psfrag{23}[t][cc]{$\scriptstyle f_{23}^{(6)}$}
  \begin{center}
    \includegraphics[width=\linewidth]{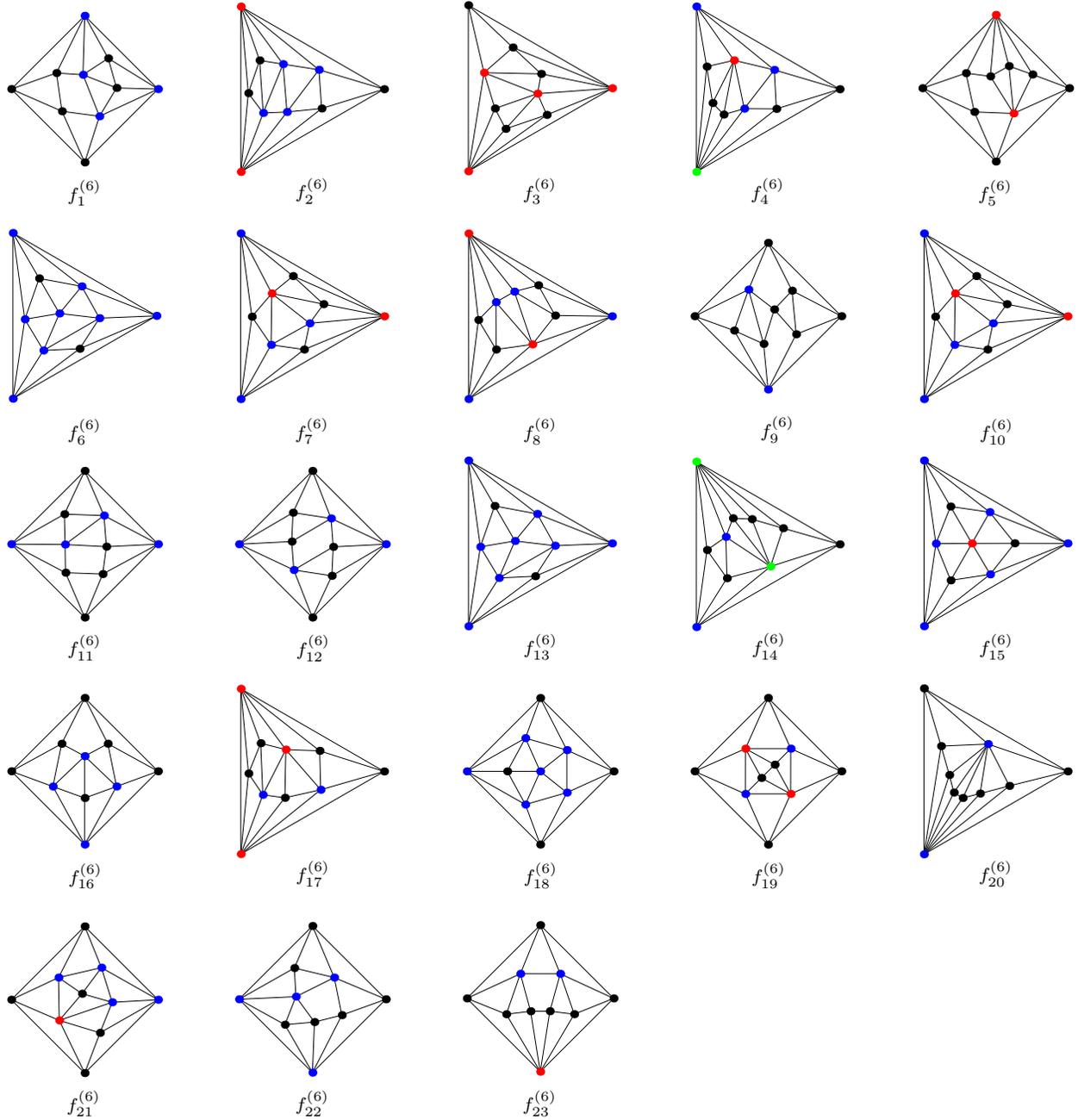}
  \end{center}
  \caption{Six-loop $f-$graphs induced from five loops by the rung rule.  The definition of black, blue and red vertices is the same as in
 Fig.~\ref{Fig:4loop}, the green vertices have degree 7.}
\label{fig:sixinduced}
\end{figure}

\begin{figure}[t!]
 \psfrag{24}[tc][cc]{$\scriptstyle f_{24}^{(6)}$}\psfrag{25}[tc][cc]{$\scriptstyle f_{25}^{(6)}$}\psfrag{26}[tc][cc]{$\scriptstyle f_{26}^{(6)}$}
  \psfrag{27}[tc][cc]{$\scriptstyle f_{27}^{(6)}$}\psfrag{28}[tc][cc]{$\scriptstyle f_{28}^{(6)}$}\psfrag{29}[tc][cc]{$\scriptstyle f_{29}^{(6)}$}
   \psfrag{30}[tc][cc]{$\scriptstyle f_{30}^{(6)}$} \psfrag{31}[tc][cc]{$\scriptstyle f_{31}^{(6)}$} \psfrag{32}[tc][cc]{$\scriptstyle f_{32}^{(6)}$} \psfrag{33}[tc][cc]{$\scriptstyle f_{33}^{(6)}$}\psfrag{34}[tc][cc]{$\scriptstyle f_{34}^{(6)}$}\psfrag{35}[c][c]{$\scriptstyle f_{35}^{(6)}$}\psfrag{36}[c][c]{$\scriptstyle f_{36}^{(6)}$}
  \begin{center}
    \includegraphics[width=\linewidth]{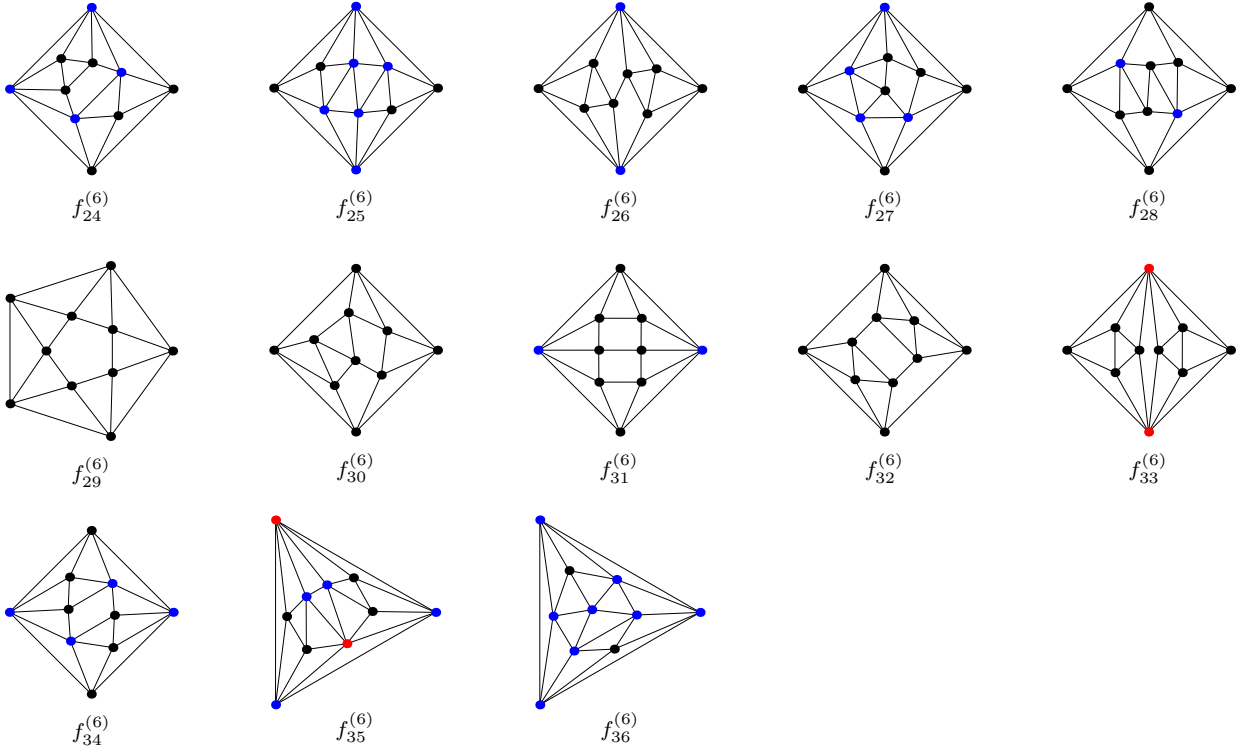}
  \end{center}
\caption{Potential additional planar six-loop $f-$graphs. In fact, only
  $f^{(6)}_{28}, f^{(6)}_{29}$ and $f^{(6)}_{31}$ contribute.}
\label{fig:sixextra}
\end{figure}
We recall that the $f-$graph only depicts the denominator of the integrand. The numerator is needed
to ensure that the overall conformal weight equals four at each vertex of the $f-$graph. 
If we depict a factor of $x_{ij}^2$ in the numerator as a dashed line attached to the vertices $i$ and $j$, then
there should be a dashed
line ending at each vertex of degree 5 (blue dots), two dashed lines ending at
each vertex of degree 6 (red dots) and three dashed lines ending at each vertex of degree 7
(green dots). Vertices of degree 4 (black dots) have no dashed
lines.  

For all but 6 of the 36 six-loop $f-$graphs
shown in Figs.~\ref{fig:sixinduced} and~\ref{fig:sixextra}, the
numerators are unambiguously fixed (given that dashed lines cannot
duplicate and thus cancel existing propagator lines, thus changing the edge-vertex connectivity of the 
graph). The 6 graphs whose numerator edges are not fixed uniquely are
$f_6^{(6)},f_8^{(6)},f_{12}^{(6)},f_{34}^{(6)},f_{35}^{(6)}$ and $f_{36}^{(6)}$. 
More precisely, three pairs of integrands, $f_6^{(6)}$ and $f_{36}^{(6)}$, $f_8^{(6)}$ 
and $f_{35}^{(6)}$, $f_{12}^{(6)}$ and $f_{34}^{(6)}$ have the same denominator 
(and, therefore, are depicted by the same $f-$graph in Figs.~\ref{fig:sixinduced} and
\ref{fig:sixextra}) but differ by the numerator. We display these 6 graphs with the 
numerator (dashed) lines in Fig.~\ref{fig:6loopnum}.  
\begin{figure}[h!]
\psfrag{f}[c][c]{}
\psfrag{6}[c][c]{$\scriptstyle f_6^{(6)}$}
\psfrag{8}[c][c]{$\scriptstyle f_8^{(6)}$}
\psfrag{12}[tc][c]{$\scriptstyle f_{12}^{(6)}$}
\psfrag{36}[c][c]{$\scriptstyle f_{36}^{(6)}$}
\psfrag{35}[c][c]{$\scriptstyle f_{35}^{(6)}$}
\psfrag{34}[tc][c]{$\scriptstyle f_{34}^{(6)}$}
  \centering
    \includegraphics[width=.6\linewidth]{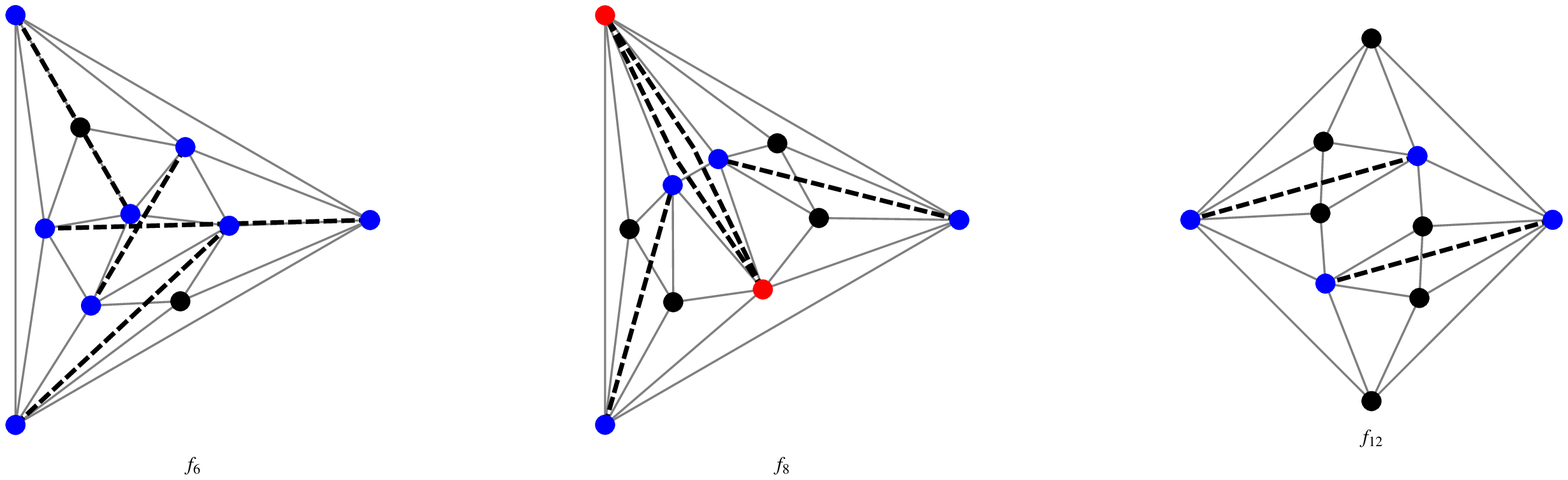}
    \includegraphics[width=.6\linewidth]{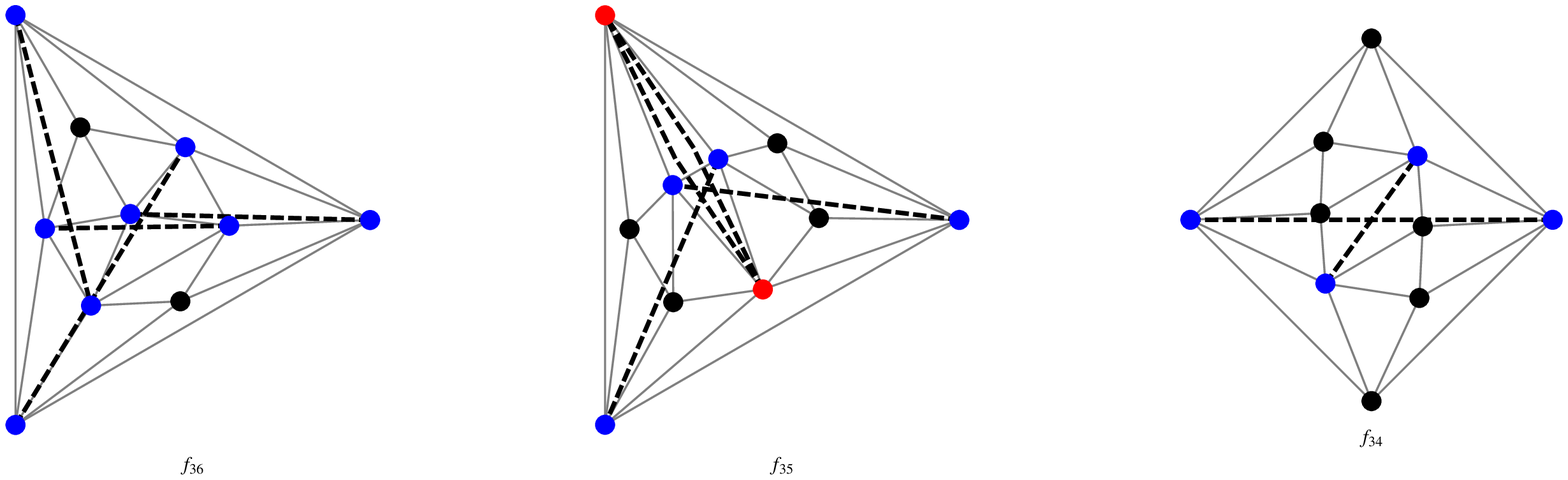}  
  \caption{The six-loop $f-$graphs (solid lines) whose numerators (dashed lines) are not uniquely fixed. Notice they come in three pairs, the three graphs in the first line are induced from five-loop graphs, whereas the 
remaining  three graphs in the second line do not contribute to the correlation function (they have vanishing coefficient, see Eq.~\p{c-6loop}).}
  \label{fig:6loopnum}
\end{figure}

Let us now turn to fixing the coefficients $c_\a^{(6)}$ in \p{P-6loop}. We shall use the light-cone limit \p{lim2}
and impose the condition \p{LC}.  At six loops, expanding the left-hand side of \p{LC} to order $O(a^6)$ we
get  (for $u, v\to 0$)
\begin{align}\label{log6loop1}\notag 
 \cF_{\rm g=0}^{(6)}  -2\cF^{(1)}&\cF_{\rm g=0}^{(5)} -2\cF^{(2)}\cF_{\rm g=0}^{(4)}
+4(\cF^{(1)})^2\cF_{\rm g=0}^{(4)} 
  -(\cF_{\rm g=0}^{(3)} )^2+8\cF^{(1)}\cF^{(2)}\cF_{\rm g=0}^{(3)} 
  -8(\cF^{(1)})^3\cF_{\rm g=0}^{(3)} 
\\[2mm]  
&+\ft43(\cF^{(2)})^3
-12(\cF^{(1)})^2(\cF^{(2)})^2 
  +16 (\cF^{(1)})^4 \cF^{(2)}-\ft{16}3 (\cF^{(1)} )^6    \stackrel{u,v\to 0}{\sim} (\ln u)^6+(\ln v)^6 \,.
\end{align}
We replace the five-loop terms $\cF^{(\ell)}$ (with $\ell=1,\ldots,5$) by their expressions obtained in the previous 
sections and substitute $\cF_{\rm g=0}^{(6)}= \lim_{x_{i,i+1}^2\to 0} x_{13}^2x_{24}^2 F_{\rm g=0}^{(6)}$ with its expression 
in terms of the polynomial \p{P-6loop}. Then, relation \p{log6loop1} translates into the condition that the 
numerator of the integrand in the left-hand side of \p{log6loop1} should vanish for $x_5\to (1-\alpha) x_1+\alpha 
x_2$ and $x_6,\ldots,x_{10}$ in general position.  
We found that this condition allows us to fix all the coefficients: 
\begin{align} \notag \label{c-6loop}
&
   c_1^{\text{(6)}}=c_5^{\text{(6)}}=c_{11}^{\text{(6)}}=c_{12}^{\text{(6)}}=c_{13}^{\text{(6)}}=c_{18}^{\text{(6)}}=c_{19}^{\text{(6)}}=c_{2
   1}^{\text{(6)}}=c_{22}^{\text{(6)}}=c_{23}^{\text{(6)}}=-1 \,,
\\[2mm] \notag
&
   c_2^{\text{(6)}}=c_3^{\text{(6)}}=c_4^{\text{(6)}}=c_6^{\text{(6)}}=c_7^{\text{(6)}}=c_8^{\text{(6)}}=c_9^{\text{(6)}}=c_{10}^{\text{(6)}}
   =c_{14}^{\text{(6)}}=c_{15}^{\text{(6)}}=c_{16}^{\text{(6)}}=c_{17}^{\text{(6)}}=c_{20}^{\text{(6)}}=c_{28}^{\text{(6)}}=c_{31}^{\text{(6)
   }}=1 \,,
\\[2mm] \notag
&
   c_{24}^{\text{(6)}}=c_{25}^{\text{(6)}}=c_{26}^{\text{(6)}}=c_{27}^{\text{(6)}}=c_{30}^{\text{(6)}}=c_{32}^{\text{(6)}}=c_{33}^{\text{(6)}
   }=c_{34}^{\text{(6)}}=c_{35}^{\text{(6)}}=c_{36}^{\text{(6)}}=0 \,,
\\[2mm]  
&  c_{29}^{\text{(6)}}=2\,.
\end{align}
The following comments are in order.

Our derivation of \p{c-6loop} was based on the analysis of the correlation function in the light-cone limit \p{LC}
and did not rely on the rung rule.%
\footnote{We verified that the obtained expression for the six-loop correlation function also has the correct
short-distance behavior \p{SD}.}
 Nevertheless, from \p{c-6loop} we see that the coefficients $c_1^{(6)},\ldots,
c_{23}^{(6)}$ take values $(\pm 1)$ in perfect agreement with the rung-rule prediction. 
So, had we used the rung rule, we could have obtained these coefficients from the five-loop coefficients \p{c-5loop}
without any calculations. This fact justifies the applicability of the rung rule up to six loops. 

The rung rule does not fix however the remaining 13 coefficients $c_{24}^{(6)},\ldots,c_{36}^{(6)}$. 
According to \p{c-6loop},  ten of these coefficients vanish and the only 
non-zero ones are $c_{28}^{\text{(6)}}=c_{31}^{\text{(6)}}=1$ and $c_{29}^{\text{(6)}}=2$.\footnote{Once the coefficients of the 23 rung-rule terms have been found, the weaker Euclidean condition  \p{SD} turns out to be sufficient to fix the remaining 13 coefficients.} This means
that the six-loop correlation function receives contributions from
three planar non-rung-rule  topologies only. It is an interesting
question to find out what role these three non-rung-rule topologies
play in the correlation function. {Another question is why the
 other ten non-rung-rule topologies have vanishing coefficients. In
 fact, eight out of these ten topologies
 ($f_{24}^{(6)},f_{25}^{(6)},f_{26}^{(6)}$ and $f_{32}^{(6)},\dots,
f_{36}^{(6)}$) lead to six-loop amplitude graphs
 which have a product structure. So these have to have vanishing
 coefficient in the planar theory, if we assume the
correlator/amplitude duality and the
 fact that the amplitude can not have such product integrands.
Interestingly, imposing   the six-loop counterpart to the Minkowski
constraint \p{faco} with $\alpha=0$ (recall that this determines
precisely that part of the correlator which is also determined by the
rung rule up to five loops) enforces the vanishing of the same eight
topologies, in addition to the giving the rung-rule coefficients.
We are
 thus left with only two unexplained zero coefficients, namely those
 of $f_{27}^{(6)}$ and $f_{30}^{(6)}$.
We recall also that Eq.~\p{c-6loop} displays only the leading color
coefficients $c^{(6)}_{0;\a}$. So, these ten topologies may be made up
of Feynman graphs of higher genus, as evoked in the comment after
Eq.~\p{c-hat}. }

It is worth mentioning that the total number of coefficients in \p{c-6loop} is a few orders of magnitude smaller compared to the total number of $f-$graphs at six loops (see Table~\ref{Table1}). This indicates in turn that the number of non-rung-rule coefficients
grows at higher loop orders much more slowly than  the total number of $f-$graphs. Following the strategy of systematic use of the rung rule  makes it possible
to extend the analysis of this section to higher loops.  
 
We can now apply the obtained result for the six-loop correlation function to predict the integrand
of the six-loop four-particle amplitude. At six loop the duality relation \p{M-F} reads
\begin{align}\label{6loop}
\cM^{(6)}=\cF^{(6)} - \cM^{(2)}\cM^{(4)}-\frac12 (\cM^{(2)})^{2}-\cM^{(1)}\cM^{(6)}\,,
\end{align}
where  $\cF^{(6)}$ stands for light-cone limit of the six-loop integrand
$\cF^{(6)}=\lim_{x_{i,i+1}^2\to 0} x_{13}^2 x_{24}^2 F_{\rm g=0}^{(6)}$. As follows from
our analysis, $\cF^{(6)}$ receives non-zero contributions from $f-$graphs of 26 topologies
only. Among them there are 23 rung-rule-induced topologies shown
in Fig.~\ref{fig:sixinduced} and 3 non-rung-rule topologies shown
in Fig.~\ref{fig:sixextra}. In this way, we find from \p{6loop} that the six-loop planar amplitude $\cM^{(6)}$ is given by the sum of 229 distinct integrals. We would like
to emphasize that despite this large number of integrands, the total number of independent
integrands is much smaller and equals 26, that is the total number of relevant $f-$topologies. The remaining 203 integrands can be obtained from them by making use of the permutation symmetry explained above.
We verified that our expression for $\cM^{(6)}$ is in agreement with the result for
the planar six-loop four-particle amplitude obtained by two different methods in
Refs.~\cite{CarrascoJohansson} and \cite{Bourjaily:2011hi}.

\section*{Acknowledgements}

{We are grateful to James Drummond, Henrik Johansson, David Kosower, Volodya Smirnov and Christian Vergu  for a
number of enlightening discussions. Special acknowledgments are due to John Joseph Carrasco
and Henrik Johansson and to
Mark Spradlin  and Jacob Bourjaily  for sharing with us their results
on the six-loop planar four-gluon amplitude.  GK and ES acknowledge partial support by the French National Agency
for Research (ANR) under contract StrongInt
(BLANC-SIMI-4-2011). BE is supported by the Deutsche Forschungsgemeinschaft (DFG), Sachbeihilfe ED 78/4-1. PH acknowledges support from an
STFC Consolidated Grant number ST/J000426/1.}

\appendix
\makeatletter
\def\@seccntformat#1{Appendix\ \csname the#1\endcsname\quad}
\makeatother  

\section{The genus of the $f-$graphs}
\label{app:genus}

A particularly nice feature of the $f-$graphs, which is especially useful at five and
six loops in obtaining a much reduced basis,  is that in the planar limit the correlation
function $G_4$ at  $\ell\ge 2$ loops receives the contribution from the planar $f-$graphs only.
Furthermore, we claim that for the $SU(N_c)$ gauge group:
\begin{align*}
\textit{\framebox{The $(1/N_c^2)^g$ corrections to $G_4$  are
given by $f-$graphs of genus $\leq g$.  }} 
\end{align*}
 The single
exception to this rule is the one-loop case, $\ell=1$, whose $f-$graph, shown in 
Fig.~\ref{Fgr}, is non-planar (genus one) whilst the correlation function  is obviously 
planar (there exist no non-planar one-loop Feynman graphs).  
Here we give  a detailed
step-by-step argument of this statement, explaining the failure for $\ell=1$.  

The starting point of our argument is the following assumption:
\textit{All $SO(6)$ components of the correlation function at
$O(1/N_c^{2g})$ can be defined in terms of conformal graphs of genus
$\leq g$.}
The statement is certainly true for the individual Feynman graphs contributing to $G_4$, the assumption here is that this property is not spoiled  when simplifying and summing Feynman graphs into conformal graphs.

Given this assumption, let us focus on a particular $SO(6)$ component of the correlation
function \p{cor4loop} corresponding to the following choice of the harmonic variables: $y_1=y_3$ and $y_2=y_4$. This component is easily obtained from \p{intriLoops} as   
\begin{align}
    G_4^{(\ell)}(1,2,3,4) \sim & \frac{2 \,
      (N_c^2-1)}{(4\pi^2)^{4}} \times {y_{12}^8 x_{13}^2 x_{24}^2 \over
      x_{12}^2 x_{23}^2 x_{34}^2 x_{14}^2}   \times  F^{(\ell)}(x_i)
    \nonumber\\
= &\frac{2 \,
      (N_c^2-1)}{(-4\pi^2)^{4+\ell}\ell!} \times y_{12}^8 \int d^4x_5 \dots
  d^4x_{4+\ell} \, \times \, {x_{13}^4 x_{24}^4 }   \times  f^{(\ell)}(x_1,  \dots , x_{4+\ell})\,, \label{eq:6}
  \end{align}
where the second line is obtained using~(\ref{integg}). Thus, given our assumption, we have that
\begin{align}
\textit{All $O(1/N_c^{2g})$ corrections to ${x_{13}^4 x_{24}^4 }   \times  f^{(\ell)}(x_1,  \dots ,
   x_{4+\ell})$  are given by genus $\leq g$ graphs.} 
   \label{item:2}
\end{align}
It remains therefore to show that this implies the same property for the $f-$graphs
themselves. 

Firstly
recall that for every $f-$graph there is
a corresponding $P-$graph related
via (\ref{eq:10}). Then, it is easy to see graphically, that  for all $\ell >1$, and for all $P-$graphs there
  is a term  containing
   $x_{13}^2x_{24}^2$.
 Recall that the $P-$graphs are loop-less
multigraphs of order $(4+\ell)$ and of degree  $(\ell-1)$.
 Now simply choose an arbitrary
   point of the $P-$graph and label it $x_1$,  then choose an adjacent
   point and label it $x_3$. Such a point always exists for $\ell>1$ since the
   vertex has non-zero degree and 
   thus must have at least one edge attached to it, and this edge  must
   end on a different vertex (since the graph is loop-less). Then 
choose yet another 
   different point for $x_2$ and an adjacent point for $x_4$. Clearly, this
   fails for $\ell = 1$ since the graph has degree 0 and  $P^{(1)}=1$.
   
The fact that all $P-$graphs produce terms  containing $x_{13}^2x_{24}^2$ implies 
 that for all  $\ell>1$ and for all $f-$graphs there is a term without $x_{13}^2$ or $x_{24}^2$ in its denominator.
   The graph corresponding to this term is identical (ignoring numerators)  to the graph of  $x_{13}^4
   x_{24}^4 \times f^{(\ell)}_\alpha$ (since the $x_{13}^4
   x_{24}^4$ terms do not kill any propagator lines). 
 Then, it follows from \p{item:2} that for $\ell>1$,
  at leading order in $1/N_c$, $f-$graphs are planar
  and $1/N_c^{2g}$ corrections come from graphs of genus $\leq g$.

So given our assumption about the relation between the genus of conformal correlator graphs and $1/N_c$ corrections we see the analogous connection between these corrections and $f-$graphs, with the single exception of one loop.
In particular this means that beyond one loop in the planar theory we may focus exclusively on planar $f-$graphs.

\section{Conformal Gram determinant}\label{Gram}

When considering the numerator  polynomials $P^{(\ell)}$ we should take into account the Gram determinant condition, which expresses the fact that any five vectors are linearly dependent in
$D=4$ dimensions
\begin{align}
\sum_{i=1}^5 c_i\, x_i^\mu = 0 \,.
\end{align}
Projecting both sides of this relation onto $x_j$, we get a system of five linear
equations for $c_i$ with the consistency condition
\begin{align}
  {\rm det} \| x_i\cdot x_j \| = 0\,,\qquad (i,j=1,\ldots,5)\,.
\end{align}  
At $\ell$ loops we have four external vectors and $\ell$ internal ones. Due to translation invariance the integrand
depends on the coordinate differences $x_{i1}$, $x_{i2}$, $x_{i3}$ and $x_{i4}$ with $i=5,\ldots,4+\ell$.
As a result, the Gram determinant condition becomes relevant starting from $\ell=2$, where we have five independent differences. However, we should not forget that 
the polynomial  $P^{(\ell)}$ is conformally covariant, while the Gram determinant
does not respect this symmetry in general. So, we need to work out Gram
determinant conditions consistent with the conformal symmetry. This can be done by embedding the four-dimensional Minkowski space into a six-dimensional one, where the $D=4$ conformal symmetry is linearly realized 
as the $SO(4,2)$ rotation symmetry of six-dimensional null vectors
$X^A_i=(1,x_i^\mu,x_i^2)$, so that $X^2_i=0$. Any set of seven vectors of this type are linearly dependent  in six dimensions,
\begin{align}
\sum_{i=1}^7 a_i \, X_i^A = 0  \,.
\end{align}
Projecting both sides onto $X_j^A$ we get
\begin{align}
  {\rm det} \| X_i\cdot X_j \| = 0   \,,\qquad (i,j=1,\ldots,7)\,,
\end{align} 
with $X_i\cdot X_j =(x_i-x_j)^2$.
By construction, the conformal Gram determinant  depends on $7=4+3$ points and, therefore,
it becomes relevant starting from three loops. There the integrand depends on 
six independent differences $x_{ij}^\mu$ and, therefore, there are two conventional Gram
determinant conditions. As follows from our analysis, only one of them is consistent 
with conformal symmetry. Its precise form was discussed in Sect.~\ref{3lcondi}. 

Let us now consider the general case of $\ell$ loops. We have $4+\ell$ six-dimensional
vectors $X_i^A$ out of which we construct the Gram matrix
\begin{align}
G_{ij} = X_i\cdot X_j\,.
\end{align}
This matrix has dimension $4+\ell$ and rank $6$. This means that all minors of size
$7\times 7$ should vanish. Let us choose the minors made of the first 6 rows and columns of  the Gram matrix, to which we add the $i$th row and the $j$th column (with $7\le i,j \le 4+\ell$).
This leads to $(\ell-2)^2$ conditions that we shall denote as
\begin{align}\label{Gr}
{\rm Gram}(1,2,3,4,5,6,i,j)=0\,,\qquad (7\le i,j \le 4+\ell)\,.
\end{align}
However, taking into account that the matrix
${\rm Gram}(1,2,3,4,5,6,i,j)$ is symmetric in $i$ and $j$, the number of independent conditions \p{Gr} reduces to 
\footnote{We are grateful to Christian Vergu for a discussion on this point.}
\begin{align}
\frac12 (\ell-2)(\ell-1) = 1\,, 3\,, 6\,,10\,,15\,,\ldots \,, \qquad \mbox{for $\ell=3,4,5,6,7,\ldots$}\ .
\end{align}
Notice that the conditions \p{Gr} are not  $S_{4+l}$ symmetric. In addition, ${\rm Gram}(1,2,3,4,5,6,i,j)$ has
the conformal weight $(-2)$ at the points $1,\ldots,6$ and the conformal weight $(-1)$ at the points $i$ and $j$.
Therefore, in order to contribute to the $S_{4+l}$ symmetric conformal polynomial $P^{(\ell)}(x_i)$, Eq.~\p{eq:10}, the Gram
determinant \p{Gr} has to be accompanied by another polynomial $P_{ij}$ which compensates a mismatch in the conformal weight 
\begin{align}
 {\rm Gram}(1,2,3,4,5,6,i,j) P_{ij}(x_1,\ldots,x_{4+\ell}) + \text{$S_{4+l}$ permutations}=0\,.
\end{align}
Here $P_{ij}$ should have the conformal weight $-(\ell-3)$ at the points $1,\ldots,6$, the weight
$-(\ell-2)$ at the points $i,j$ and the weight $-(\ell-1)$ at the remaining points. 
 
\section{Non-planar four-loop topologies}\label{Np4}

At four loops, there are 29 non-planar $f-$graphs of genus 1. To save space we
do not display them here but present instead the corresponding $S_8-$symmetric polynomials:
\begin{align} \notag
 P_4^{\text{(4)}}&=\ft{1}{16} x_{12}^2 x_{13}^2 x_{18}^2 x_{23}^2 x_{24}^2 x_{34}^2 x_{45}^2 x_{56}^2 x_{57}^2 x_{67}^2 x_{68}^2 x_{78}^2 +\ldots\,, \\ \notag
 P_5^{\text{(4)}}&=\ft{1}{4} x_{12}^2 x_{13}^2 x_{18}^2 x_{23}^2 x_{26}^2 x_{34}^2 x_{45}^2 x_{48}^2 x_{56}^2 x_{57}^2 x_{67}^2 x_{78}^2 +\ldots\,, \\ \notag
 P_6^{\text{(4)}}&=\ft{1}{12} x_{12}^2 x_{13}^2 x_{18}^2 x_{23}^2 x_{26}^2 x_{34}^2 x_{45}^2 x_{47}^2 x_{56}^2 x_{58}^2 x_{67}^2 x_{78}^2 +\ldots\,, \\ \notag
 P_7^{\text{(4)}}&=\ft{1}{48} x_{12}^2 x_{16}^2 x_{18}^2 x_{23}^2 x_{25}^2 x_{34}^2 x_{38}^2 x_{45}^2 x_{47}^2 x_{56}^2 x_{67}^2 x_{78}^2 +\ldots\,, \\ \notag
 P_8^{\text{(4)}}&=\ft{1}{2} x_{12}^2 x_{13}^2 x_{16}^2 x_{23}^2 x_{25}^2 x_{34}^2 x_{45}^2 x_{46}^2 x_{57}^2 x_{68}^2 x_{78}^4 +\ldots\,, \\ \notag
 P_9^{\text{(4)}}&=\ft{1}{4} x_{12}^2 x_{13}^2 x_{16}^2 x_{23}^2 x_{27}^2 x_{34}^2 x_{45}^2 x_{46}^2 x_{56}^2 x_{58}^2 x_{78}^4 +\ldots\,, \\ \notag
 P_{10}^{\text{(4)}}&=\ft{1}{8} x_{12}^2 x_{14}^2 x_{16}^2 x_{23}^2 x_{25}^2 x_{34}^2 x_{36}^2 x_{45}^2 x_{57}^2 x_{68}^2 x_{78}^4 +\ldots\,, \\ \notag
 P_{11}^{\text{(4)}}&=\ft{1}{4} x_{12}^2 x_{13}^2 x_{14}^2 x_{23}^2 x_{24}^2 x_{35}^2 x_{46}^2 x_{56}^2 x_{57}^2 x_{68}^2 x_{78}^4 +\ldots\,, \\ \notag
 P_{12}^{\text{(4)}}&=\ft{1}{4} x_{12}^2 x_{13}^2 x_{14}^2 x_{23}^2 x_{24}^2 x_{35}^2 x_{48}^2 x_{56}^4 x_{67}^2 x_{78}^4 +\ldots\,, \\ \notag
 P_{13}^{\text{(4)}}&=\ft{1}{2} x_{12}^2 x_{13}^2 x_{14}^2 x_{23}^2 x_{27}^2 x_{35}^2 x_{46}^2 x_{48}^2 x_{56}^4 x_{78}^4 +\ldots\,, \\ \notag
 P_{14}^{\text{(4)}}&=\ft{1}{4} x_{12}^4 x_{13}^2 x_{24}^2 x_{34}^2 x_{35}^2 x_{46}^2 x_{56}^2 x_{57}^2 x_{68}^2 x_{78}^4 +\ldots\,, \\ \notag
 P_{15}^{\text{(4)}}&=\ft{1}{2} x_{12}^4 x_{13}^2 x_{24}^2 x_{34}^2 x_{35}^2 x_{48}^2 x_{56}^4 x_{67}^2 x_{78}^4 +\ldots\,, \\ \notag
 P_{16}^{\text{(4)}}&=\ft{1}{12} x_{12}^4 x_{13}^2 x_{24}^2 x_{35}^2 x_{37}^2 x_{46}^2 x_{48}^2 x_{56}^4 x_{78}^4 +\ldots\,, \\ \notag 
 P_{17}^{\text{(4)}}&=\ft{1}{8} x_{12}^4 x_{13}^2 x_{25}^2 x_{34}^4 x_{48}^2 x_{56}^4 x_{67}^2 x_{78}^4 +\ldots\,, \\ \notag
 P_{18}^{\text{(4)}}&=\ft{1}{8} x_{12}^4 x_{15}^2 x_{25}^2 x_{34}^4 x_{36}^2 x_{46}^2 x_{57}^2 x_{68}^2 x_{78}^4 +\ldots\,, 
 \end{align}\begin{align} \notag
 P_{19}^{\text{(4)}}&=\ft{1}{4} x_{12}^4 x_{15}^2 x_{25}^2 x_{34}^2 x_{36}^2 x_{37}^2 x_{46}^2 x_{48}^2 x_{56}^2 x_{78}^4 +\ldots\,, \\ \notag
 P_{20}^{\text{(4)}}&=\ft{1}{4} x_{12}^4 x_{15}^2 x_{28}^2 x_{34}^4 x_{36}^2 x_{46}^2 x_{56}^2 x_{57}^2 x_{78}^4 +\ldots\,, \\ \notag
 P_{21}^{\text{(4)}}&=\ft{1}{8} x_{12}^4 x_{13}^2 x_{23}^2 x_{34}^2 x_{45}^2 x_{46}^2 x_{57}^2 x_{58}^2 x_{67}^2 x_{68}^2 x_{78}^2 +\ldots\,,  \\ \notag
 P_{22}^{\text{(4)}}&=\ft{1}{144} x_{12}^2 x_{14}^2 x_{16}^2 x_{23}^2 x_{25}^2 x_{34}^2 x_{36}^2 x_{45}^2 x_{56}^2 x_{78}^6 +\ldots\,, \\ \notag
 P_{23}^{\text{(4)}}&=\ft{1}{8} x_{12}^2 x_{13}^2 x_{14}^2 x_{23}^2 x_{24}^2 x_{35}^2 x_{46}^2 x_{56}^4 x_{78}^6 +\ldots\,, \\ \notag
 P_{24}^{\text{(4)}}&=\ft{1}{8} x_{12}^4 x_{13}^2 x_{24}^2 x_{34}^2 x_{35}^2 x_{46}^2 x_{56}^4 x_{78}^6 +\ldots\,, \\ \notag
 P_{25}^{\text{(4)}}&=\ft{1}{12} x_{12}^4 x_{13}^2 x_{25}^2 x_{34}^4 x_{46}^2 x_{56}^4 x_{78}^6 +\ldots\,, \\ \notag
 P_{26}^{\text{(4)}}&=\ft{1}{16} x_{12}^4 x_{15}^2 x_{25}^2 x_{34}^4 x_{36}^2 x_{46}^2 x_{56}^2 x_{78}^6 +\ldots\,, \\ \notag
 P_{27}^{\text{(4)}}&=\ft1{1152}{x_{12}^2 x_{13}^2 x_{14}^2 x_{23}^2 x_{24}^2 x_{34}^2 x_{56}^2 x_{57}^2 x_{58}^2 x_{67}^2 x_{68}^2 x_{78}^2}
   +\ldots\,, \\ \notag
 P_{28}^{\text{(4)}}&=\ft{1}{96} x_{12}^2 x_{13}^2 x_{14}^2 x_{23}^2 x_{24}^2 x_{34}^2 x_{56}^4 x_{57}^2 x_{68}^2 x_{78}^4 +\ldots\,, \\ \notag
 P_{29}^{\text{(4)}}&=\ft{1}{32} x_{12}^4 x_{13}^2 x_{24}^2 x_{34}^4 x_{56}^4 x_{57}^2 x_{68}^2 x_{78}^4 +\ldots\,, \\ \notag
 P_{30}^{\text{(4)}}&=\ft{1}{192} x_{12}^2 x_{13}^2 x_{14}^2 x_{23}^2 x_{24}^2 x_{34}^2 x_{56}^6 x_{78}^6 +\ldots\,, \\ \notag
 P_{31}^{\text{(4)}}&=\ft{1}{32} x_{12}^4 x_{13}^2 x_{24}^2 x_{34}^4 x_{56}^6 x_{78}^6 +\ldots\,, \\ \label{P-4loop-np} 
 P_{32}^{\text{(4)}}&=\ft{1}{384} x_{12}^6 x_{34}^6 x_{56}^6 x_{78}^6+\ldots\,,
\end{align}
where the ellipses denote the terms with all possible $S_8-$permutation of indices.

\section{Planar six-loop topologies}\label{6loopP}

Here we list the 23 numerator polynomials for the six-loop $f-$graphs of the rung-rule type:
\begin{align}\notag
 P_1^{\text{(6)}}&=\ft{1}{2} x_{13}^4 x_{18}^2 x_{19}^2 x_{24}^2 x_{27}^2 x_{28}^2 x_{29}^2 x_{37}^2 x_{39}^2 x_{46}^4 x_{47}^2 x_{56}^2
   x_{57}^2 x_{58}^2 x_{59}^2 x_{68}^2 x_{69}^2 x_{78}^2 x_{1,10}^2 x_{2,10}^2 x_{3,10}^2 x_{4,10}^2 x_{5,10}^2 +\ldots,\\ \notag
 P_2^{\text{(6)}}&=\ft{1}{2} x_{13}^4 x_{18}^4 x_{24}^2 x_{27}^2 x_{28}^2 x_{29}^2 x_{37}^2 x_{39}^2 x_{46}^4 x_{47}^2 x_{56}^2 x_{57}^2
   x_{58}^2 x_{59}^2 x_{69}^4 x_{78}^2 x_{1,10}^2 x_{2,10}^2 x_{3,10}^2 x_{4,10}^2 x_{5,10}^2 +\ldots,\\ \notag
 P_3^{\text{(6)}}&=\ft{1}{8} x_{13}^2 x_{17}^2 x_{18}^2 x_{19}^2 x_{24}^6 x_{28}^2 x_{36}^2 x_{37}^2 x_{38}^2 x_{47}^2 x_{56}^2 x_{57}^2
   x_{58}^2 x_{59}^2 x_{69}^6 x_{78}^2 x_{1,10}^2 x_{2,10}^2 x_{3,10}^2 x_{4,10}^2 x_{5,10}^2 +\ldots,\\ \notag
 P_4^{\text{(6)}}&=\ft{1}{2} x_{13}^2 x_{16}^4 x_{18}^2 x_{24}^4 x_{27}^2 x_{28}^2 x_{37}^2 x_{38}^2 x_{39}^2 x_{49}^6 x_{56}^2 x_{57}^2
   x_{58}^2 x_{59}^2 x_{67}^2 x_{78}^2 x_{1,10}^2 x_{2,10}^2 x_{3,10}^2 x_{5,10}^2 x_{6,10}^2 +\ldots,\\ \notag
 P_5^{\text{(6)}}&=\ft{1}{2} x_{13}^2 x_{17}^2 x_{18}^2 x_{19}^2 x_{24}^6 x_{27}^2 x_{36}^2 x_{38}^2 x_{39}^2 x_{48}^2 x_{49}^2 x_{56}^2
   x_{57}^2 x_{58}^2 x_{59}^2 x_{67}^2 x_{69}^2 x_{78}^2 x_{1,10}^2 x_{2,10}^2 x_{3,10}^2 x_{5,10}^2 x_{6,10}^2 +\ldots,\\ \notag
 P_6^{\text{(6)}}&=\ft{1}{4} x_{13}^2 x_{17}^4 x_{19}^2 x_{24}^4 x_{28}^2 x_{29}^2 x_{36}^2 x_{38}^4 x_{39}^2 x_{46}^2 x_{47}^2 x_{49}^2
   x_{56}^2 x_{57}^2 x_{58}^2 x_{59}^2 x_{78}^2 x_{1,10}^2 x_{2,10}^2 x_{5,10}^2 x_{6,10}^4 +\ldots,\\ \notag
 P_7^{\text{(6)}}&=\ft{1}{2} x_{13}^2 x_{17}^2 x_{18}^2 x_{19}^2 x_{24}^4 x_{27}^2 x_{36}^4 x_{39}^2 x_{48}^4 x_{49}^2 x_{56}^2 x_{57}^2
   x_{58}^2 x_{59}^2 x_{67}^2 x_{68}^2 x_{79}^2 x_{1,10}^2 x_{2,10}^4 x_{3,10}^2 x_{5,10}^2 +\ldots,\\ \notag
 P_8^{\text{(6)}}&=\ft{1}{8} x_{13}^2 x_{16}^2 x_{17}^2 x_{18}^2 x_{24}^4 x_{27}^2 x_{28}^2 x_{39}^6 x_{46}^2 x_{47}^2 x_{56}^2 x_{57}^2
   x_{58}^2 x_{59}^2 x_{68}^4 x_{79}^2 x_{1,10}^2 x_{2,10}^2 x_{3,10}^2 x_{4,10}^2 x_{5,10}^2 +\ldots,\\ \notag
 P_9^{\text{(6)}}&=\ft{1}{2} x_{13}^2 x_{17}^2 x_{18}^2 x_{19}^2 x_{24}^4 x_{26}^2 x_{29}^2 x_{36}^2 x_{37}^2 x_{38}^2 x_{47}^2 x_{48}^2
   x_{56}^2 x_{57}^2 x_{58}^2 x_{59}^2 x_{68}^2 x_{69}^2 x_{79}^2 x_{1,10}^2 x_{2,10}^2 x_{3,10}^2 x_{4,10}^2 x_{5,10}^2 +\ldots,\\ \notag
 P_{,10}^{\text{(6)}}&=\ft{1}{2} x_{13}^2 x_{16}^2 x_{17}^2 x_{24}^6 x_{29}^2 x_{37}^2 x_{38}^2 x_{39}^2 x_{47}^2 x_{48}^2 x_{56}^2 x_{57}^2
   x_{58}^2 x_{59}^2 x_{68}^4 x_{69}^2 x_{79}^2 x_{1,10}^4 x_{2,10}^2 x_{3,10}^2 x_{5,10}^2 +\ldots,\\ \notag
 P_{11}^{\text{(6)}}&=x_{13}^2 x_{16}^2 x_{18}^2 x_{19}^2 x_{24}^4 x_{28}^2 x_{29}^2 x_{37}^4 x_{38}^2 x_{46}^2 x_{47}^2 x_{48}^2 x_{56}^2
   x_{57}^2 x_{58}^2 x_{59}^2 x_{69}^2 x_{79}^2 x_{1,10}^2 x_{2,10}^2 x_{3,10}^2 x_{5,10}^2 x_{6,10}^2 +\ldots,\\ \notag
 P_{12}^{\text{(6)}}&=\ft{1}{4} x_{13}^2 x_{16}^2 x_{18}^2 x_{19}^2 x_{24}^4 x_{28}^2 x_{29}^2 x_{36}^2 x_{37}^2 x_{38}^2 x_{46}^2 x_{47}^2
   x_{56}^2 x_{57}^2 x_{58}^2 x_{59}^2 x_{68}^2 x_{79}^4 x_{1,10}^2 x_{2,10}^2 x_{3,10}^2 x_{4,10}^2 x_{5,10}^2 +\ldots,\\ \notag
 P_{13}^{\text{(6)}}&=\ft{1}{16} x_{13}^4 x_{16}^2 x_{19}^2 x_{24}^4 x_{28}^2 x_{29}^2 x_{37}^2 x_{38}^2 x_{46}^2 x_{47}^2 x_{56}^2 x_{57}^2
   x_{58}^2 x_{59}^2 x_{68}^4 x_{79}^4 x_{1,10}^2 x_{2,10}^2 x_{3,10}^2 x_{4,10}^2 x_{5,10}^2 +\ldots,\\ \notag
 P_{14}^{\text{(6)}}&=\ft{1}{4} x_{13}^2 x_{16}^2 x_{17}^2 x_{18}^2 x_{24}^8 x_{35}^2 x_{38}^2 x_{39}^2 x_{48}^2 x_{57}^2 x_{58}^2 x_{59}^2
   x_{67}^2 x_{68}^2 x_{69}^2 x_{79}^4 x_{1,10}^2 x_{2,10}^2 x_{3,10}^2 x_{5,10}^2 x_{6,10}^2 +\ldots, 
    \end{align}\begin{align}\notag   
 P_{15}^{\text{(6)}}&=\ft{1}{2} x_{13}^2 x_{16}^2 x_{18}^4 x_{24}^4 x_{28}^2 x_{29}^2 x_{37}^4 x_{39}^4 x_{46}^2 x_{47}^2 x_{48}^2 x_{56}^2
   x_{57}^2 x_{58}^2 x_{59}^2 x_{69}^2 x_{1,10}^2 x_{2,10}^2 x_{5,10}^2 x_{6,10}^2 x_{7,10}^2 +\ldots,  \\ \notag
 P_{16}^{\text{(6)}}&=\ft{1}{4} x_{13}^4 x_{17}^2 x_{18}^2 x_{19}^2 x_{24}^4 x_{28}^2 x_{29}^2 x_{36}^2 x_{39}^2 x_{46}^2 x_{47}^2 x_{49}^2
   x_{56}^2 x_{57}^2 x_{58}^2 x_{59}^2 x_{68}^2 x_{78}^2 x_{2,10}^2 x_{3,10}^2 x_{5,10}^2 x_{6,10}^2 x_{7,10}^2 +\ldots,\\ \notag
 P_{17}^{\text{(6)}}&=\ft{1}{4} x_{13}^2 x_{15}^2 x_{17}^2 x_{18}^2 x_{24}^6 x_{27}^2 x_{36}^4 x_{39}^4 x_{45}^2 x_{48}^2 x_{58}^2 x_{59}^2
   x_{68}^2 x_{69}^2 x_{78}^2 x_{79}^2 x_{1,10}^2 x_{2,10}^2 x_{5,10}^2 x_{6,10}^2 x_{7,10}^2 +\ldots,\\ \notag
 P_{18}^{\text{(6)}}&=\ft{1}{2} x_{13}^2 x_{17}^4 x_{19}^2 x_{24}^4 x_{28}^2 x_{29}^2 x_{36}^4 x_{38}^2 x_{46}^2 x_{47}^2 x_{56}^2 x_{57}^2
   x_{58}^2 x_{59}^2 x_{69}^2 x_{78}^2 x_{89}^2 x_{1,10}^2 x_{2,10}^2 x_{3,10}^2 x_{4,10}^2 x_{5,10}^2 +\ldots,\\ \notag
 P_{19}^{\text{(6)}}&=\ft{1}{4} x_{13}^2 x_{16}^4 x_{19}^2 x_{24}^6 x_{28}^2 x_{37}^2 x_{38}^2 x_{39}^2 x_{47}^2 x_{49}^2 x_{56}^2 x_{57}^2
   x_{58}^2 x_{59}^2 x_{67}^2 x_{68}^2 x_{89}^2 x_{1,10}^2 x_{2,10}^2 x_{3,10}^2 x_{5,10}^2 x_{7,10}^2 +\ldots,
\\ \notag
 P_{20}^{\text{(6)}}&=\ft{1}{32} x_{13}^2 x_{17}^2 x_{18}^2 x_{19}^2 x_{24}^{10} x_{36}^2 x_{38}^2 x_{39}^2 x_{56}^2 x_{57}^2 x_{58}^2
   x_{59}^2 x_{67}^2 x_{69}^2 x_{78}^2 x_{89}^2 x_{1,10}^2 x_{3,10}^2 x_{5,10}^2 x_{6,10}^2 x_{7,10}^2 +\ldots,\\ \notag
 P_{21}^{\text{(6)}}&=x_{13}^2 x_{17}^4 x_{19}^2 x_{24}^4 x_{26}^4 x_{36}^2 x_{37}^2 x_{38}^2 x_{47}^2 x_{48}^2 x_{49}^2 x_{56}^2 x_{57}^2
   x_{58}^2 x_{59}^2 x_{69}^2 x_{89}^2 x_{1,10}^2 x_{2,10}^2 x_{3,10}^2 x_{5,10}^2 x_{8,10}^2 +\ldots,
 \\ \notag
 P_{22}^{\text{(6)}}&=\ft{1}{2} x_{13}^2 x_{16}^2 x_{17}^2 x_{18}^2 x_{19}^2 x_{24}^4 x_{26}^2 x_{29}^2 x_{37}^4 x_{39}^2 x_{46}^2 x_{47}^2
   x_{48}^2 x_{56}^2 x_{57}^2 x_{58}^2 x_{59}^2 x_{89}^2 x_{2,10}^2 x_{3,10}^2 x_{5,10}^2 x_{6,10}^2 x_{8,10}^2 +\ldots,\\  
 P_{23}^{\text{(6)}}&=\ft{1}{2} x_{13}^2 x_{16}^2 x_{17}^2 x_{18}^2 x_{19}^2 x_{24}^4 x_{28}^4 x_{37}^2 x_{38}^2 x_{39}^2 x_{46}^2 x_{47}^2
   x_{49}^2 x_{56}^2 x_{57}^2 x_{58}^2 x_{59}^2 x_{67}^2 x_{2,10}^2 x_{3,10}^2 x_{5,10}^2 x_{6,10}^2 x_{9,10}^2
   +\ldots\,,   \label{eq:3}
\end{align}
where the ellipses denote the terms with all possible $S_{10}$ permutations of indices.

The remaining 13 polynomials of the non-rung-rule type are:
\begin{align} \notag
 P_{24}^{\text{(6)}}&=\ft{1}{4} x_{12}^2 x_{13}^2 x_{16}^2 x_{18}^2 x_{19}^2 x_{23}^2 x_{24}^2 x_{34}^2 x_{35}^2 x_{37}^2 x_{45}^2 x_{46}^2
   x_{48}^2 x_{56}^2 x_{59}^4 x_{67}^2 x_{78}^2 x_{79}^2 x_{89}^2 x_{2,10}^4 x_{6,10}^2 x_{7,10}^2 x_{8,10}^2 +\ldots,\\ \notag
 P_{25}^{\text{(6)}}&=\ft{1}{2} x_{12}^2 x_{13}^2 x_{16}^2 x_{18}^2 x_{19}^2 x_{23}^2 x_{24}^2 x_{34}^2 x_{35}^2 x_{37}^2 x_{45}^2 x_{46}^2
   x_{48}^2 x_{56}^2 x_{59}^4 x_{67}^2 x_{78}^4 x_{89}^2 x_{2,10}^4 x_{6,10}^2 x_{7,10}^2 x_{9,10}^2 +\ldots,\\ \notag
 P_{26}^{\text{(6)}}&=\ft{1}{2} x_{12}^2 x_{13}^2 x_{16}^2 x_{18}^2 x_{19}^2 x_{23}^2 x_{24}^2 x_{34}^2 x_{35}^2 x_{37}^2 x_{45}^2 x_{46}^2
   x_{48}^2 x_{56}^2 x_{58}^2 x_{59}^2 x_{67}^2 x_{78}^2 x_{79}^2 x_{89}^2 x_{2,10}^4 x_{6,10}^2 x_{7,10}^2 x_{9,10}^2 +\ldots,\\ \notag
 P_{27}^{\text{(6)}}&=\ft{1}{2} x_{12}^2 x_{13}^2 x_{16}^2 x_{23}^2 x_{24}^2 x_{27}^2 x_{29}^2 x_{34}^2 x_{35}^2 x_{38}^2 x_{45}^2 x_{46}^2
   x_{49}^2 x_{56}^2 x_{57}^2 x_{67}^2 x_{68}^2 x_{78}^2 x_{89}^4 x_{1,10}^4 x_{5,10}^2 x_{7,10}^2 x_{9,10}^2 +\ldots,\\ \notag
 P_{28}^{\text{(6)}}&=\ft{1}{2} x_{12}^2 x_{13}^2 x_{16}^2 x_{23}^2 x_{24}^2 x_{27}^2 x_{29}^2 x_{34}^2 x_{35}^2 x_{38}^2 x_{45}^2 x_{46}^2
   x_{49}^2 x_{56}^2 x_{57}^2 x_{67}^2 x_{68}^2 x_{78}^2 x_{79}^2 x_{89}^2 x_{1,10}^4 x_{5,10}^2 x_{8,10}^2 x_{9,10}^2 +\ldots,\\ \notag
 P_{29}^{\text{(6)}}&=\ft{1}{20} x_{12}^2 x_{13}^2 x_{16}^2 x_{19}^2 x_{23}^2 x_{24}^2 x_{27}^2 x_{34}^2 x_{35}^2 x_{38}^2 x_{45}^2 x_{46}^2
   x_{49}^2 x_{56}^2 x_{57}^2 x_{67}^2 x_{68}^2 x_{78}^2 x_{79}^2 x_{89}^2 x_{1,10}^2 x_{2,10}^2 x_{5,10}^2 x_{8,10}^2 x_{9,10}^2 +\ldots,  \\ \notag
 P_{30}^{\text{(6)}}&=\ft{1}{4} x_{12}^2 x_{13}^2 x_{15}^2 x_{17}^2 x_{23}^2 x_{24}^2 x_{27}^2 x_{29}^2 x_{34}^2 x_{35}^2 x_{38}^2 x_{45}^2
   x_{46}^2 x_{56}^2 x_{59}^2 x_{67}^2 x_{68}^2 x_{78}^2 x_{79}^2 x_{89}^2 x_{1,10}^2 x_{4,10}^2 x_{6,10}^2 x_{8,10}^2 x_{9,10}^2 +\ldots,\\ \notag
 P_{31}^{\text{(6)}}&=\ft{1}{4} x_{12}^2 x_{13}^2 x_{15}^2 x_{17}^2 x_{19}^2 x_{23}^2 x_{24}^2 x_{27}^2 x_{28}^2 x_{34}^2 x_{35}^2 x_{45}^2
   x_{46}^2 x_{49}^2 x_{56}^2 x_{58}^2 x_{67}^2 x_{78}^2 x_{79}^2 x_{89}^2 x_{3,10}^2 x_{6,10}^4 x_{8,10}^2 x_{9,10}^2 +\ldots, \\ \notag
 P_{32}^{\text{(6)}}&=\ft{1}{16} x_{12}^2 x_{13}^2 x_{14}^2 x_{16}^2 x_{23}^2 x_{24}^2 x_{26}^2 x_{29}^2 x_{34}^2 x_{35}^2 x_{38}^2 x_{45}^2
   x_{47}^2 x_{56}^2 x_{59}^2 x_{67}^2 x_{68}^2 x_{78}^2 x_{79}^2 x_{89}^2 x_{1,10}^2 x_{5,10}^2 x_{7,10}^2 x_{8,10}^2 x_{9,10}^2 +\ldots,\\ \notag
 P_{33}^{\text{(6)}}&=\ft{1}{16} x_{12}^2 x_{13}^2 x_{14}^2 x_{16}^2 x_{18}^2 x_{23}^2 x_{24}^2 x_{26}^2 x_{28}^2 x_{34}^2 x_{35}^2 x_{37}^2
   x_{45}^2 x_{47}^2 x_{56}^2 x_{58}^2 x_{67}^2 x_{78}^2 x_{79}^2 x_{89}^2 x_{5,10}^2 x_{6,10}^2 x_{9,10}^6 +\ldots,\\ \notag
 P_{34}^{\text{(6)}}&=\ft{1}{4} x_{12}^2 x_{13}^2 x_{14}^2 x_{16}^2 x_{23}^2 x_{24}^2 x_{26}^2 x_{29}^2 x_{34}^2 x_{35}^2 x_{38}^2 x_{45}^2
   x_{47}^2 x_{56}^2 x_{59}^2 x_{67}^2 x_{68}^2 x_{78}^2 x_{89}^4 x_{1,10}^2 x_{5,10}^2 x_{7,10}^4 x_{9,10}^2 +\ldots,\\ \notag
 P_{35}^{\text{(6)}}&=\ft{1}{8} x_{12}^2 x_{13}^2 x_{15}^2 x_{16}^2 x_{23}^2 x_{24}^2 x_{28}^4 x_{34}^2 x_{37}^4 x_{45}^2 x_{46}^2 x_{49}^2
   x_{57}^2 x_{58}^2 x_{67}^2 x_{68}^2 x_{69}^2 x_{78}^2 x_{1,10}^2 x_{5,10}^2 x_{9,10}^6 +\ldots,\\   \label{eq:4}
 P_{36}^{\text{(6)}}&=\ft{1}{8} x_{12}^2 x_{13}^2 x_{16}^2 x_{18}^2 x_{19}^2 x_{23}^2 x_{25}^4 x_{34}^4 x_{37}^2 x_{45}^2 x_{46}^2 x_{49}^2
   x_{56}^2 x_{58}^2 x_{67}^2 x_{78}^4 x_{89}^2 x_{2,10}^2 x_{6,10}^2 x_{7,10}^2 x_{9,10}^4+\ldots\ .  
\end{align}

\end{document}